\documentclass[twocolumn]{aastex61}

\usepackage{amsmath}
\usepackage{morefloats}
\newcommand{\code}[1]{\texttt{#1}}
\newcommand{\mesa}{\code{MESA}}
\newcommand{\MESA}{\mesa}
\newcommand{\MESAWeb}{\code{MESA-Web}}
\newcommand{\ADIPLS}{\code{ADIPLS}}
\newcommand{\GYRE}{\code{GYRE}}

\newcommand{\stella}{\code{STELLA}}
\newcommand{\STELLA}{\stella}

\newcommand{\mesastar}{\mesa\code{star}}

\newcommand{\MESAstar}{\mesastar}

\newcommand{\kap}{\code{kap}}
\newcommand{\eos}{\code{eos}}

\newcommand{\atm}{\code{atm}}

\newcommand{\mlt}{\code{mlt}}
\newcommand{\rates}{\code{rates}}

\newcommand{\neu}{\code{neu}}

\newcommand{\colors}{\code{colors}}

\newcommand{\kB}{\ensuremath{k_\mathrm{B}}} %Boltzmann's constant
\newcommand{\dif}{\ensuremath{\mathrm{d}}}
\newcommand{\Dif}{\ensuremath{\mathrm{D}}}

\newcommand{\ddt}[1]{\frac{\partial #1}{\partial t}} %partial time derivative 
\newcommand{\DDt}[1]{\frac{\Dif #1}{\Dif t}} % Lagrangian time derivative
\newcommand{\ddm}[1]{\frac{\partial #1}{\partial m}} %partial derivative wrt m 
\newcommand{\ddr}[1]{\frac{\partial #1}{\partial r}} %partial derivative wrt r
\newcommand{\dmbar}[1]{\ensuremath{\overline{dm}_{#1}}} % Mass of cell face
\newcommand{\dm}[1]{\ensuremath{dm_{#1}}} % Mass of cell 
\newcommand{\mC}{\ensuremath{m_{\mathrm{C}}}}

\newcommand{\Ncells}{\ensuremath{N_{\mathrm{target}}}}
\newcommand{\timestep}{\ensuremath{\delta t}} % Numerical timestep

\newcommand{\rhoR}{\ensuremath{\rho_{\mathrm{R}}}}
\newcommand{\rhoL}{\ensuremath{\rho_{\mathrm{L}}}}
\newcommand{\PR}{\ensuremath{P_{\mathrm{R}}}}
\newcommand{\PL}{\ensuremath{P_{\mathrm{L}}}}
\newcommand{\uR}{\ensuremath{u_{\mathrm{R}}}}
\newcommand{\uL}{\ensuremath{u_{\mathrm{L}}}}
\newcommand{\Sr}{\ensuremath{\mathcal{S}_{\mathrm{R}}}}
\newcommand{\Sl}{\ensuremath{\mathcal{S}_{\mathrm{L}}}}
\newcommand{\Ss}{\ensuremath{\mathcal{S}_{*}}}
\newcommand{\Ps}{\ensuremath{P_{*}}}
\newcommand{\Pface}{\ensuremath{P_{\mathrm{face}}}}
\newcommand{\uface}{\ensuremath{u_{\mathrm{face}}}}
\newcommand{\dt}{\ensuremath{\delta t}}
\newcommand{\csR}{\ensuremath{c_{{\rm s},\mathrm{R}}}}
\newcommand{\csL}{\ensuremath{c_{{\rm s},\mathrm{L}}}}
\newcommand{\area}{\ensuremath{\mathcal{A}}}
\newcommand{\swave}{\ensuremath{\mathcal{S}}}

\newcommand{\RTI}{\mathcal{R}}
\newcommand{\alphaRTI}[1]{\ensuremath{\alpha_{\RTI{#1}}}}
\newcommand{\etaRTI}[1]{\ensuremath{\eta_{\RTI{#1}}}}
\newcommand{\sigRTI}[1]{\ensuremath{\sigma_{\RTI{#1}}}}
\newcommand{\Sap}{\ensuremath{S_{\alpha}^{+}}}
\newcommand{\Sam}{\ensuremath{S_{\alpha}^{-}}}
\newcommand{\cs}[1]{\ensuremath{c_{\mathrm{s#1}}}}
\newcommand{\ARTI}{\ensuremath{A_{\RTI}}}
\newcommand{\BRTI}{\ensuremath{B_{\RTI}}}
\newcommand{\CRTI}{\ensuremath{C_{\RTI}}}
\newcommand{\DRTI}{\ensuremath{D_{\RTI}}}

\newcommand{\epsnuc}{\ensuremath{\epsilon_{\mathrm{nuc}}}} % Nuclear energy generation rate
\newcommand{\epsgrav}{\ensuremath{\epsilon_{\mathrm{grav}}}} % Gravitational heating rate
\newcommand{\epsnu}{\ensuremath{\epsilon_{\mathrm{\nu}}}} % Neutrino loss rate
\newcommand{\epsextra}{\ensuremath{\epsilon_{\mathrm{extra}}}} % Extra heating or cooling
\newcommand{\eblast}{\ensuremath{E_{\mathrm{blast}}}} % Blast wave energy

\newcommand{\alphaMLT}{\ensuremath{\alpha_{\mathrm{MLT}}}}	% Mixing length parameter
\newcommand{\grada}{\ensuremath{\nabla_{{\rm ad}}}}

\newcommand{\gradL}{\ensuremath{\nabla_{{\rm L}}}}
\newcommand{\gradr}{\ensuremath{\nabla_{{\rm rad}}}}
\newcommand{\gradT}{\ensuremath{\nabla_T}}
\newcommand{\Teff}{\ensuremath{T_{\rm eff}}}	% Effective temperature
\newcommand{\fov}{\ensuremath{f_{\mathrm{ov}}}} % Convective overshoot parameter
\newcommand{\BV}{Brunt-V\"{a}is\"{a}l\"{a}}
\newcommand{\Mzams}{\ensuremath{M_{\rm ZAMS}}}
\newcommand{\Xhyd}{\ensuremath{X}}
\newcommand{\Yhel}{\ensuremath{Y}}
\newcommand{\Zmet}{\ensuremath{Z}}
\newcommand{\qe}{\ensuremath{q_{\rm e}}}

\newcommand{\gammaone}{\ensuremath{\Gamma_1}}
\newcommand{\mue}{\ensuremath{\mu_{\mathrm{e}}}} % Electron chemical potential

\newcommand{\logT}{\ensuremath{\log(T/\mathrm{K})}} % logT
\newcommand{\logRho}{\ensuremath{\log(\rho/\grampercc)}} % logRho
\newcommand{\unitspace}{\ensuremath{\,}}
\newcommand{\usp}{\unitspace}

\newcommand{\unitstyle}[1]{\ensuremath{\mathrm{#1}}}
\newcommand{\power}[2]{\ensuremath{{#1}^{#2}}}

\newcommand{\centi}{\unitstyle{c}}

\newcommand{\Mega}{\unitstyle{M}}
\newcommand{\Giga}{\unitstyle{G}}

\newcommand{\meter}{\unitstyle{m}}

\newcommand{\second}{\unitstyle{s}}

\newcommand{\Kelvin}{\unitstyle{K}}
\newcommand{\K}{\Kelvin}  %degrees Kelvin

\newcommand{\cm}{\centi\meter}
\newcommand{\gram}{\unitstyle{g}}

\newcommand{\grampercc}{\gram\usp\power{\cm}{-3}} %mass density
\newcommand{\erg}{\unitstyle{erg}} %erg 

\newcommand{\cmpersecondSc}{\cm\usp\power{\second}{-2}} %acceleration
\newcommand{\eV}{\unitstyle{eV}}        %eV
\newcommand{\Msun}{\ensuremath{\unitstyle{M}_\odot}}
\newcommand{\Lsun}{\ensuremath{\unitstyle{L}_{\odot}}}
\newcommand{\Rsun}{\ensuremath{\unitstyle{R}_{\odot}}}

\newcommand{\Myr}{\Mega\yr}
\newcommand{\Gyr}{\Giga\yr}

\newcommand{\yr}{\unitstyle{yr}}        %year
\RequirePackage{bm}
\newcommand{\mesaone}{Paper~I}  % the first mesa paper
\newcommand{\mesatwo}{Paper~II} % the second mesa paper
\newcommand{\mesathree}{Paper~III} % the third mesa paper

\newlength{\apjcolwidth}
\newlength{\figwidth}
\newlength{\doublewide}
\newcommand\duffellinprep{P.~Duffell et al.~(2017, in preparation)}
\newcommand\duffellinprepalt{P.~Duffell et al.~2017, in preparation}

\begin{document}
\title{Modules for Experiments in Stellar Astrophysics (MESA): Convective Boundaries, Element Diffusion, and Massive Star Explosions}
\author{Bill Paxton}
\affiliation{Kavli Institute for Theoretical Physics, University of California, Santa Barbara, CA 93106, USA}

\author[0000-0002-4870-8855]{Josiah Schwab}
\altaffiliation{Hubble Fellow}
\affiliation{Department of Astronomy and Astrophysics, University of California, Santa Cruz, CA 95064, USA}
\correspondingauthor{Josiah Schwab}
\email{jwschwab@ucsc.edu}

\author[0000-0002-4791-6724]{Evan B. Bauer}
\affiliation{Department of Physics, University of California, Santa Barbara, CA 93106, USA}

\author{Lars Bildsten}
\affiliation{Kavli Institute for Theoretical Physics, University of California, Santa Barbara, CA 93106, USA}
\affiliation{Department of Physics, University of California, Santa Barbara, CA 93106, USA}

\author{Sergei Blinnikov}
\affiliation{M.V. Lomonosov Moscow State University, Sternberg Astronomical Institute, Universitetsky~pr.,~13,~Moscow, 119234, Russia}
\affiliation{NIC Kurchatov Institute -- Institute for Theoretical and Experimental Physics (ITEP), 117218 Moscow, Russia}
\affiliation{Kavli Institute for the Physics and Mathematics of the Universe (WPI), The University of Tokyo, Kashiwa, Chiba 277-8583, Japan}

\author{Paul Duffell}
\affiliation{Department of Astronomy and Theoretical Astrophysics Center, University of California, Berkeley, CA 94720, USA}

\author[0000-0003-3441-7624]{R. Farmer}
\affiliation{Anton Pannenkoek Institute for Astronomy, University of Amsterdam, NL-1090 GE Amsterdam, the Netherlands}
\affiliation{School of Earth and Space Exploration, Arizona State University, Tempe, AZ 85287, USA}

\author[0000-0003-1012-3031]{Jared A. Goldberg}
\affiliation{Department of Physics, University of California, Santa Barbara, CA 93106, USA}

\author{Pablo Marchant}
\affiliation{Department of Physics and Astronomy, Northwestern University, 2145 Sheridan Road, Evanston, IL 60208, USA}

\author{Elena Sorokina}
\affiliation{M.V. Lomonosov Moscow State University, Sternberg Astronomical Institute, Universitetsky~pr.,~13,~Moscow, 119234, Russia}
\affiliation{NIC Kurchatov Institute -- Institute for Theoretical and Experimental Physics (ITEP), 117218 Moscow, Russia}

\author{Anne Thoul}
\affiliation{Space sciences, Technologies and Astrophysics Research (STAR) Institute, Universit\'e de Li\`ege, All\'ee du 6 Ao$\hat{u}$t 19C, Bat. B5C, 4000 Li\`ege, Belgium}

\author[0000-0002-2522-8605]{Richard H. D. Townsend}
\affiliation{Department of Astronomy, University of Wisconsin-Madison, Madison, WI 53706, USA}

\author[0000-0002-0474-159X]{F. X. Timmes}
\affiliation{School of Earth and Space Exploration, Arizona State University, Tempe, AZ 85287, USA}

% 250 word maximum
\begin{abstract}
We update the capabilities of the software instrument Modules for
Experiments in Stellar Astrophysics (\MESA) and enhance its ease of
use and availability. Our new approach to locating convective
boundaries is consistent with the physics of convection, and yields
reliable values of the convective core mass during both hydrogen and
helium burning phases. Stars with $M<8\,\Msun$ become white dwarfs and
cool to the point where the electrons are degenerate and the ions are
strongly coupled, a realm now available to study with \MESA\ due to
improved treatments of element diffusion, latent heat release, and
blending of equations of state. Studies of the final fates of massive
stars are extended in \MESA\ by our addition of an approximate Riemann
solver that captures shocks and conserves energy to high accuracy
during dynamic epochs. We also introduce a 1D capability for modeling
the effects of Rayleigh-Taylor instabilities that, in combination with
the coupling to a public version of the \STELLA\ radiation transfer
instrument, creates new avenues for exploring Type II supernovae
properties. These capabilities are exhibited with exploratory models
of pair-instability supernova, pulsational pair-instability supernova,
and the formation of stellar mass black holes. The applicability of
\MESA\ is now widened by the capability of importing multi-dimensional
hydrodynamic models into \MESA.  We close by introducing software
modules for handling floating point exceptions and stellar model
optimization, and four new software tools $-$ \MESAWeb, \MESA-Docker,
\texttt{pyMESA}, and mesastar.org $-$ to enhance \MESA's education and
research impact.
\end{abstract}

% keywords must be chosen from the approved set at http://journals.aas.org/authors/keywords2013.html
% no more than six (6) keywords are allowed
% keywords must be in alphabetical order

\tableofcontents
% !TEX root = ./paper.tex

\section{Introduction}\label{s.introduction}

Over the next decade multi-messenger astronomy will probe the rich
stellar astrophysics of transient phenomena in the sky, including
gravitational waves from the mergers of neutron stars and black
holes, light curves and spectra from core-collapse
supernovae, and the oscillation modes of stars.
On the observational side of this new era, the 
{\it Laser Interferometer Gravitational-Wave Observatory} 
has demonstrated the existence of binary stellar-mass black hole systems 
\citep{abbott_2016_aa,abbott_2016_ab,abbott_2016_ac, abbott_2017_aa, abbott_2017_ab}
and continues to
monitor the sky with broadband detectors for gravitational waves from
compact binary inspirals and asymmetrical exploding massive stars
\citep{fryer_2002_aa,gossan_2016_aa,abbott_2016_ad,abbott_2016_ae,abbott_2016_af,abbott_2017_aa}.
The {\it Gaia} Data Release~1, containing about one billion stars, 
begins the process of converting the spectrophotometric measurements 
to distances, proper motions, luminosities, effective temperatures, surface gravities,
 and elemental compositions \citep{gaia-collaboration_2016_aa,gaia-collaboration_2016_ab}.
This stellar census will provide the observational data to tackle a
range of questions related to the origin, structure, and evolutionary history 
of stars in the Milky Way \citep{creevey_2015_aa,sacco_2015_aa,lindegren_2016_aa,van-leeuwen_2017_aa}.
The {\it Neutron star Interior Composition Explorer} mission,
delivered to the International Space Station in June 2017, will provide
rotation-resolved spectroscopy of the thermal and non-thermal
emissions of neutron stars in the soft X-ray band 
with over 15 million seconds of exposures 
\citep{gendreau_2012_aa,arzoumanian_2014_aa,gendreau_2016_aa}
to open a new window into the interior structure and dynamics
that underlie neutron stars \citep[e.g.,][]{ozel_2016_ac,miller_2016_aa}.
With first light at Palomar Observatory in 2017, the 
{\it Zwicky Transient Facility} \citep{kulkarni_2016_aa} will scan more than
3750 deg$^2\,$hr$^{-1}$ to a depth of about 20 mag to discover young supernovae
less than 24 hours after explosion each night, hunt for
electromagnetic counterparts of gravitational-wave events
\citep{ghosh_2017_ab}, and search for rare and exotic transients.
Repeated imaging of the Northern sky, including the Galactic Plane,
will produce a photometric variability catalog with nearly 300
observations each year \citep{laher_2017_aa} for detailed studies of
variable stars and binary systems.
From its unique high earth orbit,
the {\it Transiting Exoplanet Survey Satellite} aims to survey about 200,000 
nearby G, K and M type stars with apparent magnitudes brighter than about 12 mag
with a 1 minute cadence across a 400 deg$^2$ area of the sky
\citep{ricker_2016_aa,sullivan_2015_aa,sullivan_2017_aa}
to open a new era on stellar variability.
The {\it Large Synoptic Survey Telescope} will image the entire Southern Hemisphere deeply in
multiple optical colors every week with a 3.5 deg$^2$, three billion
pixel digital camera \citep{lsst-science-collaboration_2017_aa} to open new perspectives
on transient objects such as tidal disruption events 
\citep{bade_1996_aa,stern_2004_aa,arcavi_2014_aa,komossa_2015_aa}
and interacting close binary systems
\citep{oluseyi_2012_aa,korol_2017_aa}.
The {\it Jiangmen Underground Neutrino Observatory} will usher in 
a new generation of multipurpose neutrino detectors 
\citep{li_2014_aa,brugiere_2017_aa} designed in part to 
open a new avenue on neutrinos from pre-supernova massive stars 
\citep[e.g.,][]{odrzywolek_2009_aa,misch_2016_aa,patton_2017_aa,patton_2017_ab}
and core-collapse supernova explosions \citep[e.g.,][]{hirata_1987_aa,janka_2017_aa}.

This ongoing explosion of activity in multi-messenger stellar astronomy
powers theoretical and computational developments, in particular the evolution of the
community software instrument Modules for Experiments in Stellar Astrophysics
(\MESA) for research and education. We introduce \MESA\ in \citet[][\mesaone]{paxton_2011_aa}
and significantly expand its range of capabilities in  \citet[][\mesatwo]{paxton_2013_aa}
and \citet[][\mesathree]{paxton_2015_aa}. These  prior papers, as well
as this one, are ``instrument'' papers that describe the capabilities and limitations of \MESA\ while also
comparing to other available numerical or analytic results.
This paper describes the major new advances to \MESA\ for modeling
convective boundaries, element diffusion, implicit shock
hydrodynamics, massive star explosions and light curves, pulsational
pair-instability supernovae, and black hole formation. We do not fully explore these
results and their implications here.  The scientific potential of these
new capabilities will be unlocked in future work via the efforts of the
\MESA\ user community.

The convective regions of stars remain a rich site of fascinating
challenges including the interplay between mixing, composition
gradients, and element diffusion. A convection region transports
energy through the vertical exchange of matter.  The location
where the radial velocity of the bulk motions goes to zero is a natural
way to define the edge of a convection region
\citep{vitense_1953_aa,bohm-vitense_1958_aa}. It is necessary to ensure
that convective boundaries are properly positioned \citep[e.g.,][]{Eggleton1972,Gabriel2014}, because their
exact placement can have a strong influence on the evolution of
the stellar model \citep{Salaris2017}.  An important
new addition to \MESA \ is an improved treatment of convective
boundaries, allowing them to evolve toward a state where the radiative
gradient equals the adiabatic gradient on the convective side of the
boundary. As a consequence, the Schwarzchild and Ledoux criteria now give the same position for
convective boundaries.

Gradients can drive changes in the composition profile of a star.
For example, if gradients occur in the concentrations of chemical
elements, then diffusion tends to smooth out the
differences.  Temperature
gradients can push heavier species towards regions of
higher temperature, while pressure gradients can propel heavier species
to diffuse towards regions of higher pressure
\citep{Thoul94,hansen_2004_aa,Kipp12,michaud_2015_aa}. Treatments of diffusion typically assume
that all diffusing species are ideal gases
\citep[e.g.,][]{Burgers69,Thoul94}. For white dwarf interiors and
neutron star envelopes, degenerate electrons violate this assumption
\citep{DeloyeBildsten,CBA2010}. 
In addition, strong Coulomb coupling in plasmas requires modifications to
the binary scattering formalism for calculating cross-sections used to
obtain diffusion coefficients
\citep{Paquette86coeff,Stanton16,Daligault16,Shaffer17}.
\MESA's extensions of element diffusion for
degenerate and strongly coupled plasmas
 open a pathway into the regime relevant to sedimentation in the
interiors of white dwarfs
\citep{Iben85,Iben92,koester_2009_aa,hollands_2017_aa} and the
surfaces of neutron stars
\citep{chang_2003_aa,chang_2004_aa,beznogov_2016_aa}.

Massive ($M \gtrsim 8\,\Msun$) stars explode when energy from the
collapse of their core to a compact object emerges as an outgoing
shock wave into the outer parts of the star. The outward propagation
of this shock wave generates Rayleigh-Taylor instabilities that can
mix material behind the shock front
\citep{Chevalier1976,Chevalier1978,Weaver1980,Benz1990,Herant1991,Hammer2010,Wongwathanarat2015,Utrobin2017}. The
resulting light curves of Type II supernovae can be sub-divided into
multiple classes but we focus here on Type~IIP supernovae
\citep[e.g.,][]{smartt_2009_aa,smartt_2015_aa,smith_2016_ab}.  Our
improvements to \MESA\ --- implicit shock capturing hydrodynamics,
Rayleigh-Taylor instability modeling in 1D \citep{Duffell2016}, and
radiative transfer using the public version of the \STELLA\ instrument
\citep{Blinnikov2004,Baklanov2005, Blinnikov2006} --- open up new
avenues for researching the diverse set of Type~II supernovae.

Pair-instability leads to a partial collapse, which in turn causes
runaway thermonuclear burning in the carbon-oxygen core
\citep{fowler_1964_aa, rakavy_1967_ab, barkat_1967_aa, rakavy_1967_aa,
  fraley_1968_aa}.  A wide variety of outcomes is possible depending
on the star's mass and rotation. A single energetic burst from
nuclear burning can disrupt the entire star without leaving a black
hole remnant behind to produce a pair-instability supernova
\citep{ober_1983_aa, fryer_2001_aa, scannapieco_2005_aa,
  kasen_2011_aa, chatzopoulos_2013_ab}.  Alternatively, a series of
bursts can trigger a cyclic pattern of nuclear burning, expansion and
contraction, leading to a pulsational pair-instability supernova
that leaves a black hole remnant \citep{barkat_1967_aa,
  woosley_2007_aa, chatzopoulos_2012_aa, Woosley2017,
  limongi_2017_aa}.  Many of these variations can now be explored in
\MESA, as can lower mass progenitors that do not pulse before collapse
to a black hole.

\MESA \ is a community-driven software instrument for stellar astrophysics.  
New directions will be motivated by features useful to the
\MESA \ user community, advances in the physics modules, algorithmic
developments, and architectural evolution. Potential examples for
expanding \MESA's scientific, computational, and educational capabilities include
seamlessly leveraging many-core architectures,
an improved treatment of the equation of state, 
Jupyter/Python notebooks for education,
and continued integration with software instruments 
useful to the astronomy and astrophysics community. Examples include
\ADIPLS \ \citep{christensen-dalsgaard_2008_aa,christensen-dalsgaard_2011_aa}, 
\GYRE \ \citep{townsend_2013_aa}, and 
\STELLA \ \citep{Blinnikov1998,Blinnikov2004,Baklanov2005, Blinnikov2006}.

The paper is organized as follows. 
\deleted{Section~\ref{s.epsgrav} discusses energy accounting in stellar evolution.}
Section~\ref{s.convect} introduces a new treatment of convective boundaries.
In Section~\ref{s.diffusion} we present an implementation of element diffusion 
that accounts for electron degeneracy and strongly coupled interactions.
Section~\ref{s.hydro} describes the Riemann solver for shock capturing in \MESA's new implicit hydrodynamics solver, and 
Section~\ref{s.RTI} presents a model for approximating the 3D effects of the Rayleigh-Taylor instability.
In Section~\ref{s.sneiip} we introduce the coupling of  \MESA \ and an implementation of the \STELLA \ radiative transfer instrument 
to explore the modeling of Type~IIP supernova light curves from post-explosion to post-plateau.
In Section~\ref{s.bhform} we show advances to model pair-instability supernova,
pulsation pair-instability supernova, and black hole formation.
\added{Section~\ref{s.epsgrav} discusses energy accounting in stellar evolution.}

Appendix \ref{s.colors} discusses improvements to estimating a model's absolute magnitude in a chosen color filter,
Appendix \ref{app.relax} offers guidance on importing multi-dimensional models into \MESA, and
Appendix \ref{app.diffusion} details the implementation of element diffusion in \MESA.
Appendix \ref{s.software} introduces two new software modules for handling floating point exceptions and 
stellar model optimization, and four new software tools for education and research: \MESA-Web, \MESA-Docker, \texttt{pyMESA}, and \texttt{mesastar.org}.

Important symbols are defined in Table 1.  Acronyms used are denoted in Table~\ref{tab:acronym}. We denote components of \MESA, such as modules and routines, in typewriter font 
e.g., \texttt{colors}.

% p{2cm}

\startlongtable
\begin{deluxetable}{clr}
  \tablecolumns{3}
  \tablewidth{0.9\apjcolwidth}
  \tablecaption{Important Variables.
    Single character symbols are listed \\
    first, symbols with modifiers are listed second.
    Some symbols may \\ be further subscripted by $\mathrm{c}$ (indicating a central quantity),
    by a \\ cell index $k$, or by an index that runs over species ($i$, $j$, $s$, or $t$).
\label{t.list-of-symbols}}
  \tablehead{\colhead{Name} & \colhead{Description} & \colhead{First Appears}}
  \startdata
  \area          & Area of face                           & \ref{s.hllc}   \\ %ok
$C$            & Concentration                          & \ref{s.degeneracy_errors}  \\ %ok
$e$            & Specific thermal energy                & \ref{s.hllc}           \\ %ok
$F$            & Flux across cell face                  & \ref{s.hllc}               \\ %ok
$\gamma$       & Adiabatic index                        & \ref{s.tests.sedov}        \\ %ok
$\Gamma$       & Plasma coupling parameter              & \ref{sec:latent-heat}      \\ %ok
$K$            & Resistance coefficient                 & \ref{s.degeneracy_errors}  \\ %ok
$\lambda$      & Screening length                       & \ref{s.coefficients}       \\ %ok
$m$            & Baryonic mass coordinate               & \ref{s.growing_core}           \\ %ok
$M$            & Stellar mass                           & \ref{s.growing_core}       \\ %ok
$\mu$          & Chemical potential                     & \ref{s.eps_fund}           \\ %ok
$\Phi$         & Gravitational potential                & \ref{s.implicit_hydro}     \\ %ok
$q$            & Specific heat                          & \ref{s.eps_fund}           \\ %ok
$r$            & Radial coordinate                      & \ref{s.degeneracy_errors}     \\ %ok
$s$            & Specific entropy                       & \ref{s.eps_fund}           \\ %ok
$\swave$       & Wave speed                             & \ref{s.hllc}               \\ %ok
$u$            & Cell-centered velocity                 & \ref{s.hllc}               \\ %ok
$w$            & Diffusion velocity                     & \ref{s.degeneracy_errors}  \\ %ok
$z$            & Resistance coefficient                 & \ref{s.degeneracy_errors}  \\ %ok

%%% Local Variables:
%%% mode: latex
%%% TeX-master: "paper"
%%% End:

  $\bar{A}$         &  Average atomic number                     & \ref{s.eps_fund}          \\ %ok
\alphaMLT         &  Mixing length of MLT                           & \ref{s.howto}             \\ %ok
$c_P$             &  Specific heat at constant pressure             & \ref{sec:implementation}  \\ %ok
$c_{\rm s}$       &  Sound speed                                    & \ref{s.hllc}              \\ %ok
$c_V$             &  Specific heat at constant volume               & \ref{sec:implementation}  \\ %ok
\timestep         &  Numerical timestep                             & \ref{s.timestep_mesh}  \\ %ok
\dm{}             &  Mass of cell                                   & \ref{s.hllc}  \\ %ok
\dmbar{}          &  Mass at cell face                              & \ref{s.hllc}              \\ %ok
\DRTI             &  Rayleigh-Taylor decay coefficient              & \ref{s.RTI_implement}     \\ %ok
$e_{\rm ion}$     & Specific ionization energy                      & \ref{sec:ionization}     \\ %ok
\eblast           &  Blast energy                                   & \ref{s.tests.sedov}       \\ %ok
\epsextra         &  Extra specific heating/cooling rate                      & \ref{s.hllc}              \\ %ok
\epsgrav          &  Gravitational heating rate                     & \ref{s.eps_fund}          \\ %ok
\epsnu            &  Neutrino energy loss rate                      & \ref{s.hllc}              \\ %ok
\epsnuc           &  Nuclear energy generation rate                 & \ref{s.hllc}              \\ %ok
\fov              &  Convective overshoot parameter                 & \ref{s.howto}           \\ %ok
\gammaone         &  First adiabatic index                          & \ref{s.bhform}            \\ %ok
\grada            &  Adiabatic temperature gradient                 & \ref{s.convect}  \\ %ok
\gradL            &  Ledoux temperature gradient                    & \ref{s.predictive-mix}    \\ %ok
\gradr            &  Radiative temperature gradient                 & \ref{s.convect}   \\ %ok
\gradT            &  Temperature gradient from MLT                  & \ref{s.hllc}              \\ %ok
%$k_{\rm B}$       &  Boltzmann constant                             & \ref{sec:ionization}      \\ %ok
%$m_{\rm e}$       &  Electron mass                                  & \ref{s.coefficients}      \\ %ok
%$m_{\rm p}$       &  Proton mass                                    & \ref{sec:ionization}      \\ %ok
%$M_{\rm{bol}}$    &  Absolute bolometric magnitude                  & \ref{s.colors}            \\ %ok
%$n_{\rm e}$       & Electron number density                         & \ref{s.diffusion_methods} \\ %ok
%$N_{\rm A}$       &  Avogadro number                                & \ref{s.eps_fund}          \\ %ok
$N_{\rm B}$       &  Number of baryons                              & \ref{s.eps_fund}          \\ %ok 
$P_{\rm gas}$     &  Gas pressure                                   & \ref{sec:implementation}  \\ %ok 
$P_{\rm rad}$     &  Radiation pressure                             & \ref{sec:implementation}  \\ %ok 
$\overline{P}$    &  Pressure at cell face                          & \ref{s.hllc}              \\ %ok
\qe               &  Electric charge                                & \ref{s.skirting}          \\ %??
$\rho_e$          &  Charge density                                 & \ref{s.degeneracy_errors} \\ %ok
%$t_{\rm exp}$     &  Time since shock breakout                      & \ref{s.sobolev}           \\ %ok
$\overline{T}$    &  Temperature at cell face                       & \ref{s.hllc}              \\ %ok
%$\tau_{\rm c}$    &  Optical depth                                  & \ref{s.sobolev}           \\ %ok
$\tau_{\rm Ros}$  &  Rosseland optical depth                        & \ref{s.sobolev}           \\ %ok
$\tau_{\rm sob}$  &  Sobolev optical depth                          & \ref{s.sobolev}           \\ %ok
$\chi_{\rho}$     &  ($\partial$log$P$/$\partial$log$\rho)|_{T,X}$  & \ref{sec:implementation}  \\ %ok
$\chi_T$          &  ($\partial$log$P$/$\partial$log$T)|_{\rho,X}$  & \ref{sec:implementation}  \\ %ok
$\bar{Z}$         &  Average ion charge                             & \ref{s.eps_fund}          \\ %ok

  \enddata
\end{deluxetable}

% p{2cm}
\startlongtable
\begin{deluxetable}{clr}
  \tablecolumns{3}
  \tablewidth{0.9\apjcolwidth}
  \tablecaption{Acronyms used in this paper.\label{tab:acronym}}
  \tablehead{ \colhead{Acronym} & \colhead{Description} & \colhead{First Appears} }
  \startdata 
      AGB & Asymptotic Giant Branch & \ref{sec:ionization} \\
    
    BC & Bolometric Correction & \ref{s.colors} \\
    BH & Black Hole & \ref{s.bhform} \\

    CFL & Courant-Friedrichs-Lewy & \ref{s:hydro_time} \\
    CHeB & Core Helium Burning & \ref{s.helium_core} \\
    CSM & Circumstellar Material & \ref{s.postbreakout} \\
    
    EOS & Equation of State & \ref{s.epsgrav} \\  
      
    HLLC & Harten-Lax-van Leer-Contact & \ref{s.hydro} \\
    HR & Hertzsprung-Russell & \ref{s.helium_core} \\      
    
    IB & Inner Boundary & \ref{s.corecoll} \\
%    INF & Infinity & \ref{s.nantrap} \\
    
    LTE & Local Thermal Equilibrium & \ref{s.eps_fund} \\ 
   
    MLT & Mixing Length Theory & \ref{s.convect} \\  
    MS & Main Sequence & \ref{s.convect} \\
    
%    NaN & Not A Number & \ref{s.nantrap} \\
%    NS & Neutron Star & \ref{s.bhform}  \\
    
    PPISN & Pulsational Pair-Instability SN & \ref{s.bhform} \\
    PISN & Pair-instability SN & \ref{s.bhform} \\
    
    RTI & Rayleigh-Taylor Instability & \ref{s.RTI} \\
%     RSG & Red Super Giant & only used once \ref{s.specificsne} \\   
      
    SN & Supernova & \ref{s.tests.mmk} \\
    SPH & Smoothed Particle Hydrodynamics & \ref{app.relax} \\
    
    TAMS & Terminal Age Main Sequence & \ref{s.growing_core} \\
    
    WD & White Dwarf & \ref{s.epsgrav} \\
    
    ZAMS & Zero Age Main Sequence & \ref{s.growing_core} \\
    
%     \MESA & Modules for Experiments in Stellar Astrophysics & Title  \\
%     \GYRE & Gives You Random Eigenvalues & \\

  \enddata
\end{deluxetable}

%%% Local Variables:
%%% mode: latex
%%% TeX-master: "paper"
%%% End:

\section{Convective Boundaries}
\label{s.convect}

\citet{Gabriel2014} discuss the correct positioning of
convective boundaries in stellar evolution models.  Following earlier
work \citep[e.g.,][]{Roxburgh1978} they argue that a convective
boundary should be defined as the point where the convective velocity
vanishes. Within local mixing-length theory (MLT), this condition is
equivalent to the requirement $\gradr = \grada$, where $\gradr$ and
$\grada$ are the radiative and adiabatic temperature gradients,
respectively. Critically, this equality must be satisfied on the
\emph{convective} side of the boundary, because the MLT convective
velocity is only well defined there. Moreover, because the fluid on
the convective side is presumed to be well-mixed, the Ledoux
temperature gradient $\gradL = \grada + B$ (Equation 11 of \mesatwo)
can play no part in setting the location of the boundary.

If the chemical composition is continuous across the convective
boundary, then so too are $\gradr$ and $\grada$, and requiring
$\gradr=\grada$ on the convective side of the boundary results in the
same equality on the radiative side. However, a composition
discontinuity produces a jump in density and opacity, and in turn a
discontinuity in $\gradr$ and $\grada$. Hence, it is generally the
case that $\gradr \neq \grada$ on the radiative side of the boundary.

In numerical codes based on discrete grids, the nuance of the
foregoing discussion is often overlooked in favor of a simple approach
for locating convective boundaries based on sign changes in the
discriminant $y = \gradr-\grada$ (or $y=\gradr-\gradL$, if the Ledoux
stability criterion is used).  This approach works well when the
chemical composition remains continuous, but is problematic when the
composition --- and hence $y$ --- is discontinuous at the boundary; it
typically leads to configurations where $\gradr > \grada$ on the
convective side, which is unphysical and ultimately retards the growth
of the convective region.  Previous versions of \MESA\ have taken this
approach; the outcome is evident in Figure 15 of \mesatwo, which shows
the convective core mass as a function of age during the He-burning
evolution of a $3\,\Msun$ star. In the model with no overshoot and the
Schwarzschild stability criterion, the core grows only modestly in
mass before reaching a plateau. Inspection of the model
confirms that $\gradr > \grada$ on the convective side of the core
boundary, signifying that the core growth is being impeded.

\citet{Gabriel2014} highlight a further issue with this simple
sign-change approach, whereby the location of a convective boundary is
not uniquely determined but rather depends on the mixing history near
the boundary.  We have confirmed this issue is present in \MESA\ when
using the sign-change approach.  This manifests itself as a lack of
convergence in some models (e.g., the $3\,\Msun$ He-burning example)
when the resolution is increased and/or the timestep shortened.

To resolve these issues we implement a new ``predictive mixing''
scheme in \MESA. It is inspired both by the ``maximal overshoot''
scheme introduced by \citet{Constantino2015}, and by the procedure
described by \citet{Bossini2015}. In the new scheme, the extent of a
convection region is allowed to expand at each time step until the
boundaries reach the point where $\gradr=\grada$ on their convective
side. We describe the new scheme in detail in the following section
and then present results obtained with this scheme in four scenarios:
a growing convective core in a low-mass star on the main sequence
(MS), a retreating convective core in a high-mass star on the MS,
growing He-burning cores in intermediate- and low-mass stars, and a
surface convective region in a low-mass star on the MS. In all cases,
we assume an initial He mass fraction $\Yhel=0.28$, an initial metal
mass fraction $\Zmet=0.02$, and we neglect rotation and mass loss.

\subsection{Predictive Mixing}
\label{s.predictive-mix}

The \MESA\ predictive mixing scheme initially proceeds in the same
manner as the simple sign-change approach, by finding the cells where
$y > 0$ on one face (convective) and $y < 0$ on the other face
(radiative). For each of these candidate boundary cells, the algorithm
considers how $y$ would change if the cell were completely mixed with
the rest of the adjoining convection region. This prediction involves
re-evaluating opacities, densities and other data throughout the mixed
region, under the assumption that the composition is completely
uniform. If $y$ would become positive on both faces of the candidate
boundary cell, then the adjacent cell in the radiative region becomes
the new candidate boundary cell and a new round of predictive mixing
begins. The process continues iteratively until the candidate cell
after the predictive mixing still has a negative $y$ on the radiative
face. The code reverts to the previous candidate, identifies it as the
final convective boundary cell, recalculates convective diffusivities
and convective velocities using MLT, and writes these into the model
for use in the composition solver (see \mesaone, Section 6.2). No
abundances are directly modified in the model during the predictive
iterations.  Below, we demonstrate that this algorithm leads \MESA\ to
a solution of the stellar structure equations in which $0\le y \ll \grada$
on the convective side of each boundary cell.

The physical justification for our predictive mixing scheme traces
back to a narrative advanced by \citet{Castellani1971}. Focusing on He
core-burning, these authors argue that any gentle mixing outside the
core boundary \replaced{will irreversibly alter the composition there,
  and in turn change the local $\gradr$ from sub-adiabatic to
  super-adiabatic}{irreversibly alters the composition there, and the
  resulting increase in opacity raises the local $\gradr$ from
  sub-adiabatic to super-adiabatic}. The outcome is a `self-driving
mechanism for the extension of the convective region', which continues
until $\gradr = \grada$ on the convective side of the core
boundary. While \citet{Castellani1971} invoked overshoot as the source
of the mixing outside the boundary, \citet{Michaud2007} show that
element diffusion can serve equally well and leads to the same
outcome.

\added{For MS stars with growing convective cores, the extension of
  the core boundary cannot be driven in exactly the same way as the
  He-burning case, because helium has a lower opacity than
  hydrogen. However, gentle mixing outside the core boundary erases
  any composition gradients there, and it is the loss of these
  gradients---and their accompanying stabilizing effect---that drives
  the extension of the convective region until $\gradr=\grada$ on the
  convective side of the core boundary.}

The predictive mixing scheme doesn't specify the nature of the gentle
mixing beyond convective boundaries, instead focusing on its
effects. Tied in with this agnosticism is the presumption that the
mixing-driven expansion of convective boundaries is so rapid that it
can be approximated as instantaneous. This is likely a reasonable
approach during core H and He burning; \citet{Castellani1971} argue
that the growth of the core boundary in the latter case should proceed
on a timescale which is much shorter than the burning
lifetime. However, there may be circumstances where the finite
timescale for boundary growth cannot be neglected.

Because uniform composition is assumed during the predictive mixing
iterations, there is no functional distinction between the
Schwarzschild and Ledoux criteria when evaluating the discriminant
$y$. However, the preliminary search for sign changes in $y$, before
any predictions are made, \emph{does} take into account composition
gradients when the Ledoux criterion is used.  As a result, the initial
candidate boundary cells can differ between the two criteria.  In many
cases this difference is unimportant, with the final location of the
boundaries being insensitive to the choice of criterion. The one
exception is when a region with $\grada < \gradr < \gradL$ is bounded
on both sides by radiative regions; then, it will be completely
overlooked during a preliminary search with the Ledoux criterion. As
we shall demonstrate later, such scenarios arise in our calculations
outside convective cores during MS evolution.

\subsection{Evolution of a Growing Convective Core on the Main
  Sequence}
\label{s.growing_core}

\begin{figure}
\begin{center}
\includegraphics{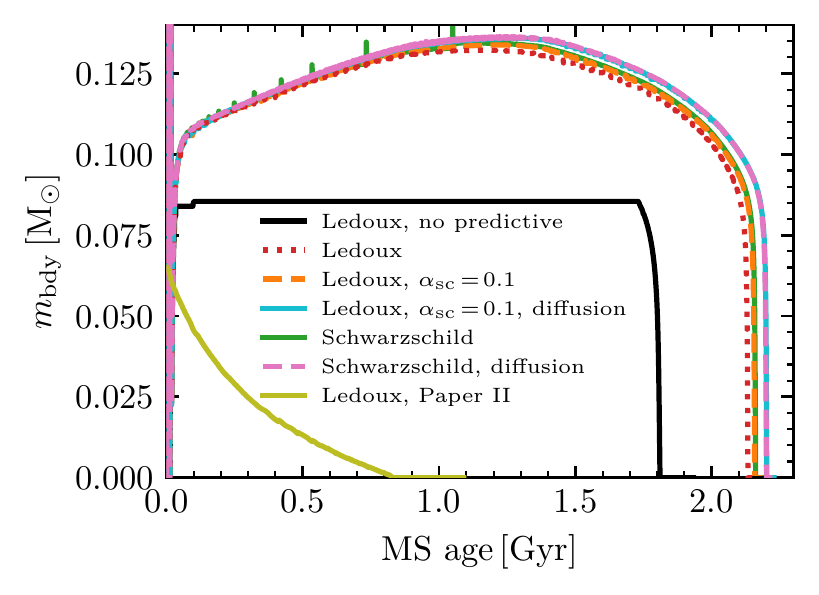}
\caption{The mass coordinate $m_{\rm bdy}$ of the convective-core
  boundary plotted as a function of MS age, for the
  $1.5\,\Msun$ stellar model discussed in
  Section~\ref{s.growing_core}. Different line styles/colors show the
  separate runs described in the text. \deleted{Ledoux models use
  $\alpha_{\rm sc} = 0$ unless otherwise noted.}}
\label{fig:M15-core-bdy}
\end{center}
\end{figure}

\begin{figure*}
\begin{center}
\includegraphics{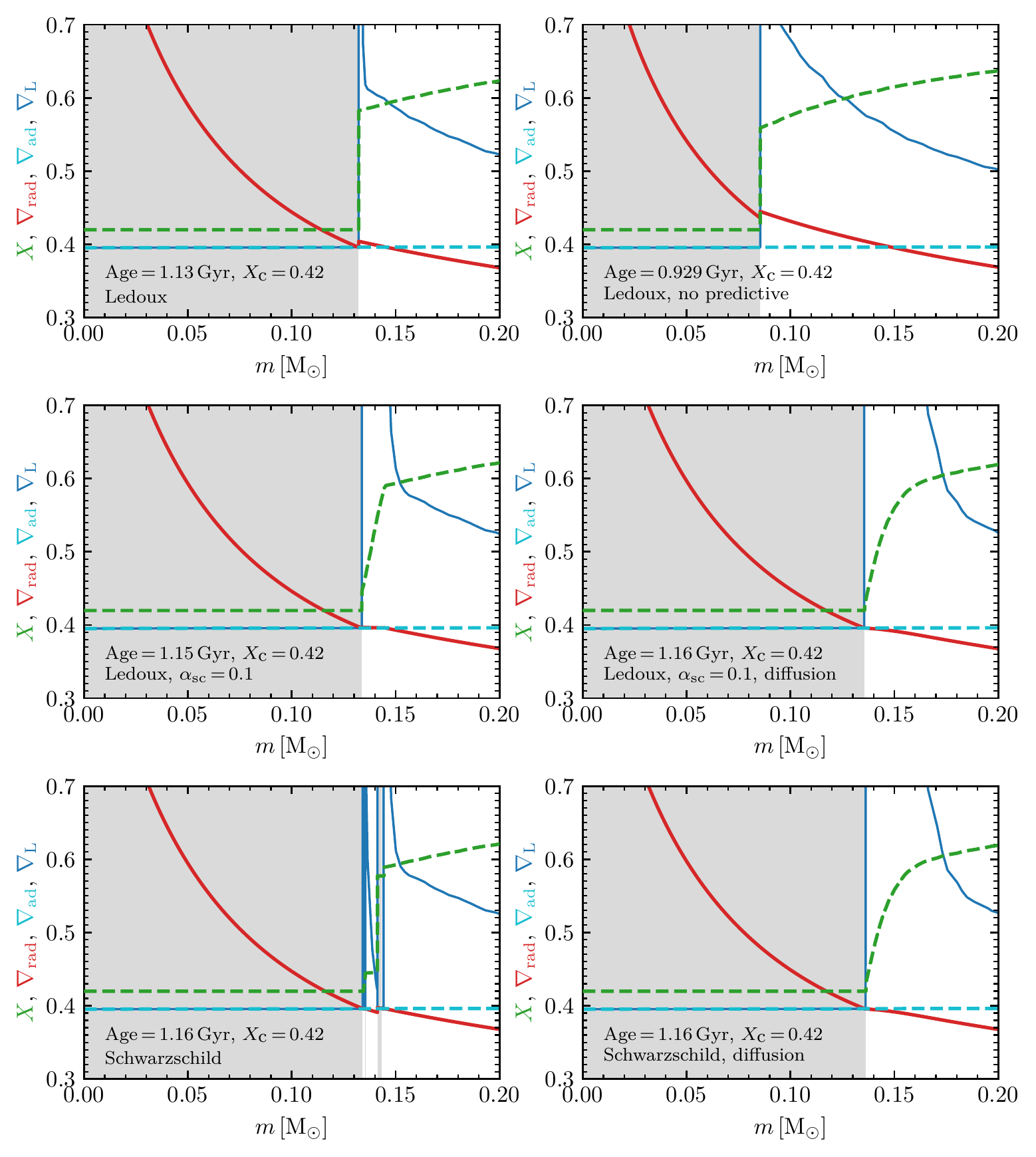}
\caption{Profiles of $\gradr$, $\grada$, $\gradL$, and $\Xhyd$ as a
  function of mass coordinate, in the inner part of the $1.5\,\Msun$
  stellar model at $\Xhyd_{\rm c} = 0.42$. The panels show the
  separate runs described in the text. Gray shading indicates regions
  undergoing convection.  Unless otherwise indicated, all models used
  predictive mixing.}
\label{fig:M15-core-grads}
\end{center}
\end{figure*}

We evolve a $1.5\,\Msun$ star from the zero age main sequence (ZAMS)
to the terminal age main sequence (TAMS) using the predictive mixing
scheme at the convective core boundary; \added{this is the same mass
  and evolutionary stage considered in Section 5.1.2 of
  \citet{Gabriel2014}}. Figure~\ref{fig:M15-core-bdy} plots the mass
coordinate of the convective-core boundary as a function of MS age,
showing results from separate runs using Schwarzschild and Ledoux
criteria, and additional runs with the incremental inclusion of
semi-convection (in just the Ledoux case) and then element diffusion
(in both cases). \added{The semi-convection is modeled using the
  \citet{Langer1985} scheme with an efficiency parameter $\alpha_{\rm
    sc}=0.1$ (see \mesatwo\ for a complete description of the
  semi-convection implementation in \MESA).} For comparison, the
figure also shows the outcome of using the Ledoux criterion but no
predictive mixing; in contrast to the other cases which broadly agree
with one another, the core growth is inhibited and the H-burning
lifetime correspondingly truncated.

Note that in Figure~13 of \mesatwo\ the results obtained with the
Ledoux criterion show a \emph{shrinking} convective core; this
behavior was due to a separate problem arising from over-smoothing of
the composition gradient \citep[see][]{Moore2016} and has since been
rectified in \MESA.  For completeness, we include this case in
Figure~\ref{fig:M15-core-bdy}.

Figure~\ref{fig:M15-core-grads} plots the profiles of $\gradr$,
$\grada$, $\gradL$ and $\Xhyd$, in the inner part of the $1.5\,\Msun$
star nearing the halfway point of its MS evolution (a core H mass
fraction $\Xhyd_{\rm c}=0.42$). In the upper row, the left panel
illustrates the run with the Ledoux criterion plus predictive mixing
(the dotted curve in Figure~\ref{fig:M15-core-bdy}), while the right
panel shows the run with the Ledoux criterion but without predictive
mixing (the black curve, ibid.). Clearly, without predictive mixing
$\gradr$ remains significantly larger than $\grada$ on the convective
side of the boundary, which as discussed previously is physically
inconsistent. When using predictive mixing, however, the profiles
satisfy $\gradr=\grada$ on the convective side, and closely match
those seen in the left panel in Figure~6 of \citet{Gabriel2014}.  The
small bump in \gradr\ just above the boundary is Schwarzschild
unstable but Ledoux stable.

The middle panels of Figure~\ref{fig:M15-core-grads} show the runs
with the Ledoux criterion and predictive mixing, and the incremental
addition of semi-convection (left) and then element diffusion
(right). \deleted{The semi-convection is modeled using the
  \citet{Langer1985} scheme with an efficiency parameter $\alpha_{\rm
    sc}=0.1$ (see \mesatwo\ for a complete description of the
  semi-convection implementation in \MESA).} Inside the core boundary,
the profiles are almost identical to those shown in the upper-left
panel but just outside the boundary, the semi-convection converts the
composition discontinuity into a steep gradient and flattens the bump
in \gradr\ to a neutral, $\gradr = \grada$ profile. Element diffusion
further softens the abundance profile, as shown in the middle-right
panel. Note that \replaced{contrary to the results of
  \citet{Michaud2007}, element diffusion does not move the convective
  boundary, now that the predictive mixing correctly determines its
  position.}{element diffusion has only a small effect on the location
  of the convective boundary; this is barely noticeable in
  Figure~\ref{fig:M15-core-grads}, but a slight extension of the
  boundary can be seen in Figure~\ref{fig:M15-core-bdy} toward the
  later part of the MS, for the two cases including diffusion.}

The lower panels of Figure~\ref{fig:M15-core-grads} show the runs
using the Schwarzschild criterion and predictive mixing, without
(left) and with (right) element diffusion. \replaced{In the left
  panel, the bump in \gradr\ effectively creates two convection
  regions --- the core itself, and an adjacent shell just outside the
  core. This leads to an abundance profile with discontinuities at the
  core/shell interface and also at the upper edge of the shell;
  between these discontinuities, the profile is flat due to the
  convective mixing within the shell.}{In the left panel, the
  abundance profile shows a chaotic staircase-like profile, due to
  mixing by transient convective shells that appear and disappear
  from one timestep to the next (two of these shells can be seen in
  the figure).} The shells do not appear in the Ledoux plots (middle
and upper panels) because the region outside the core is stabilized in
its entirety by the abundance gradient: $\gradr < \gradL$. This serves
as a good illustration of the earlier discussion
(Section~\ref{s.predictive-mix}) of how the Schwarzschild and Ledoux
criteria can sometimes lead to different outcomes. It is important to
note, however, that the location of the core boundary is the same in
all cases with predictive mixing; the differences only appear in the
inhomogeneous region beyond the boundary which arises from slow H
burning outside the core.

The lower-right panel of Figure~\ref{fig:M15-core-grads} shows that
adding element diffusion removes the abundance discontinuities,
replacing them with a smooth gradient. The resulting profiles appear
almost identical to the Ledoux case shown in the middle-right panel of
the figure (and compare also the curves with diffusion in
Figure~\ref{fig:M15-core-bdy}).

\subsection{Evolution of a Retreating Convective Core on the Main
  Sequence}
\label{s.core_retreat}

\begin{figure}
\begin{center}
\includegraphics{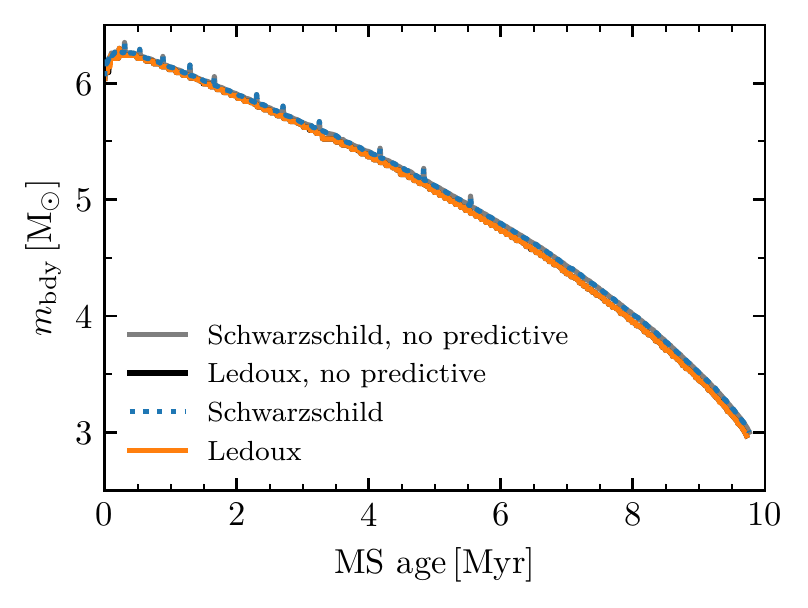}
\caption{The mass coordinate $m_{\rm bdy}$ of the convective-core
  boundary as a function of
  MS age, for the $16\,\Msun$ stellar model discussed in
  Section~\ref{s.core_retreat}. Different line styles/colors show the
  separate runs described in the text.}
  \label{fig:M16-core-bdy}
\end{center}
\end{figure}

\begin{figure*}
\begin{center}
\includegraphics{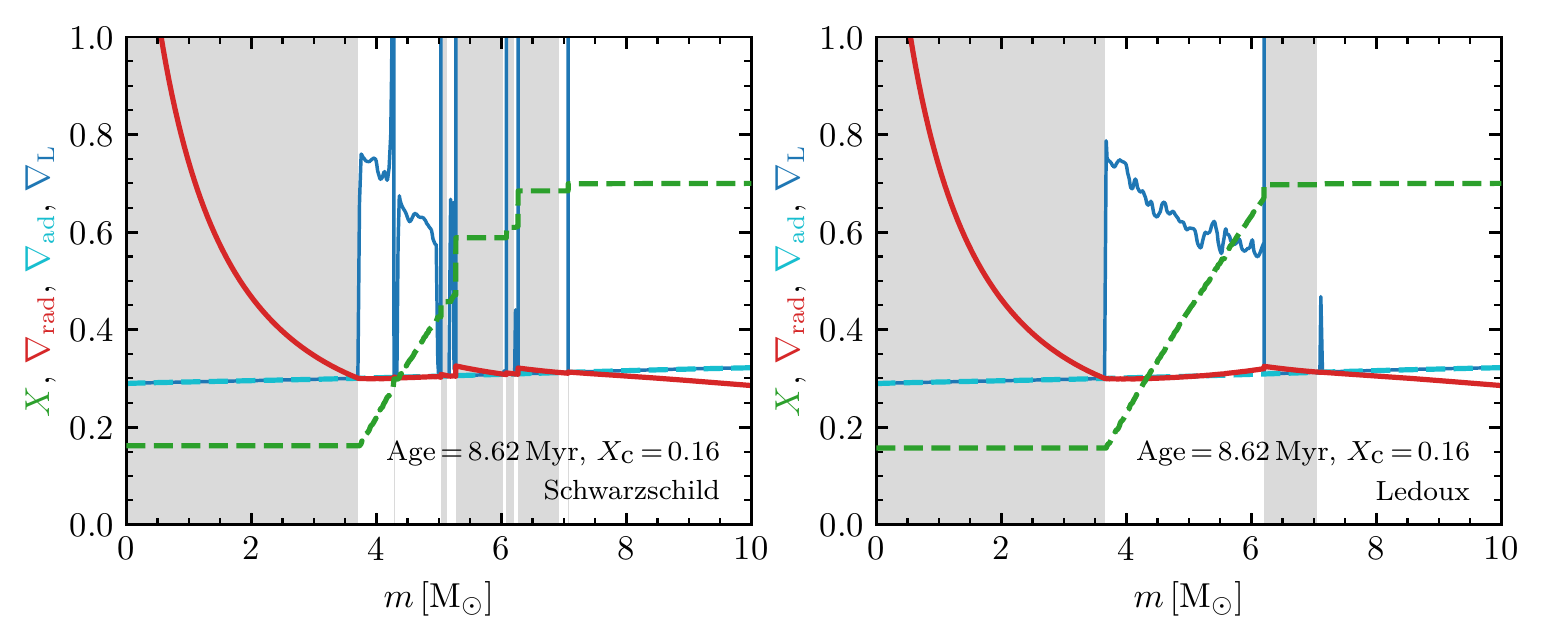}
\caption{Profiles of $\gradr$, $\grada$, $\gradL$, and $\Xhyd$ as a
  function of mass coordinate, in the inner part of the $16\,\Msun$
  stellar model at $\Xhyd_{\rm c}\approx 0.15$. The panels show the
  separate runs described in the text. Gray shading indicates regions
  undergoing convection.  Both models use predictive mixing at the
  convective core boundary.}
\label{fig:M16-core-grads}
\end{center}
\end{figure*}

We now evolve a $16\,\Msun$ star from ZAMS to TAMS using the new
predictive mixing scheme at the convective core boundary; \added{this
  is the same mass and evolutionary stage considered in Section~5.1.1
  of \citet{Gabriel2014}}.  Figure~\ref{fig:M16-core-bdy} plots the mass
of the convective core as a function of MS age, showing results from
separate runs using the Schwarzschild and Ledoux criteria, and with
and without predictive mixing. The agreement between these four cases
is very close. However, as was the case in the preceding section,
there are differences outside the convective core. These can be seen
in Figure~\ref{fig:M16-core-grads}, which plots the profiles of $\gradr$,
$\grada$, $\gradL$ and $\Xhyd$ near the end of the star's MS evolution
($\Xhyd_{\rm c}=0.15$), for the two runs with predictive mixing.

Even though both runs exhibit the same core structure, with $\gradr =
\grada$ at the convective side of the core boundary, the inhomogeneous
region left behind by the retreating core is very different. The H
abundance obtained with the Schwarzschild criterion shows the same
staircase-like profile seen in the lower-left panel of
Figure~\ref{fig:M15-core-grads}, again due to mixing by transient
convective shells. These shells are not present when the Ledoux
criterion is used, with the exception of a persistent solitary shell
at the top of the inhomogeneous region (corresponding to where the
core boundary was located at the ZAMS); the behavior of this shell is
discussed by \citet[][their Section 5.5.1; and compare also against
  their Figure~4]{Gabriel2014}. Between the shell and the core
boundary, the abundance profile from the Ledoux run remains relatively
smooth. The different abundance profiles in the two runs will have a
direct influence on the \BV\ frequency profile, and therefore on the
oscillation frequencies of the stellar model.

\subsection{Evolution of the Convective Core during Core He Burning}
\label{s.helium_core}

\begin{figure}
\begin{center}
\includegraphics{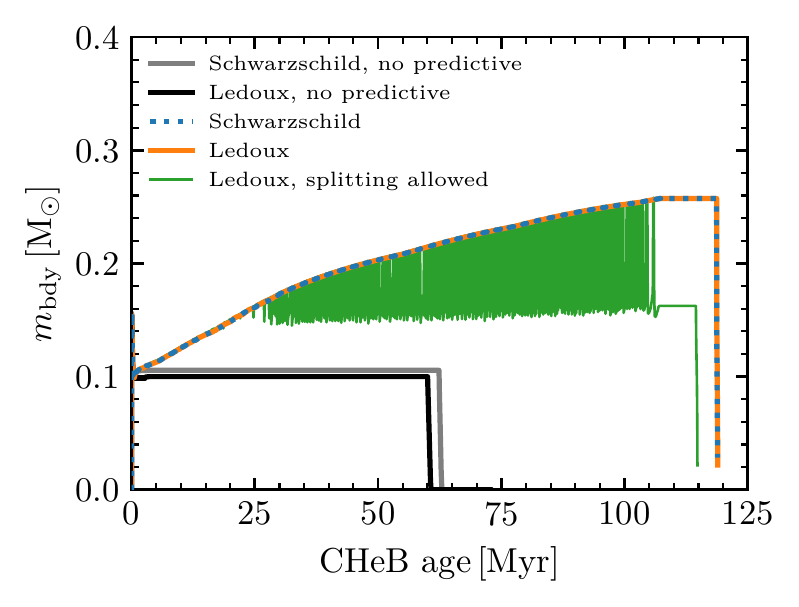} 
\caption{The mass coordinate $m_{\rm bdy}$ of the convective-core
  boundary plotted as a function of CHeB age, for the $1\,\Msun$
  stellar model discussed in Section~\ref{s.helium_core}. Different
  line styles/colors show the separate runs described in the text.}
\label{fig:M1-core-bdy}
\end{center}
\end{figure}

\begin{figure}
\begin{center}
\includegraphics{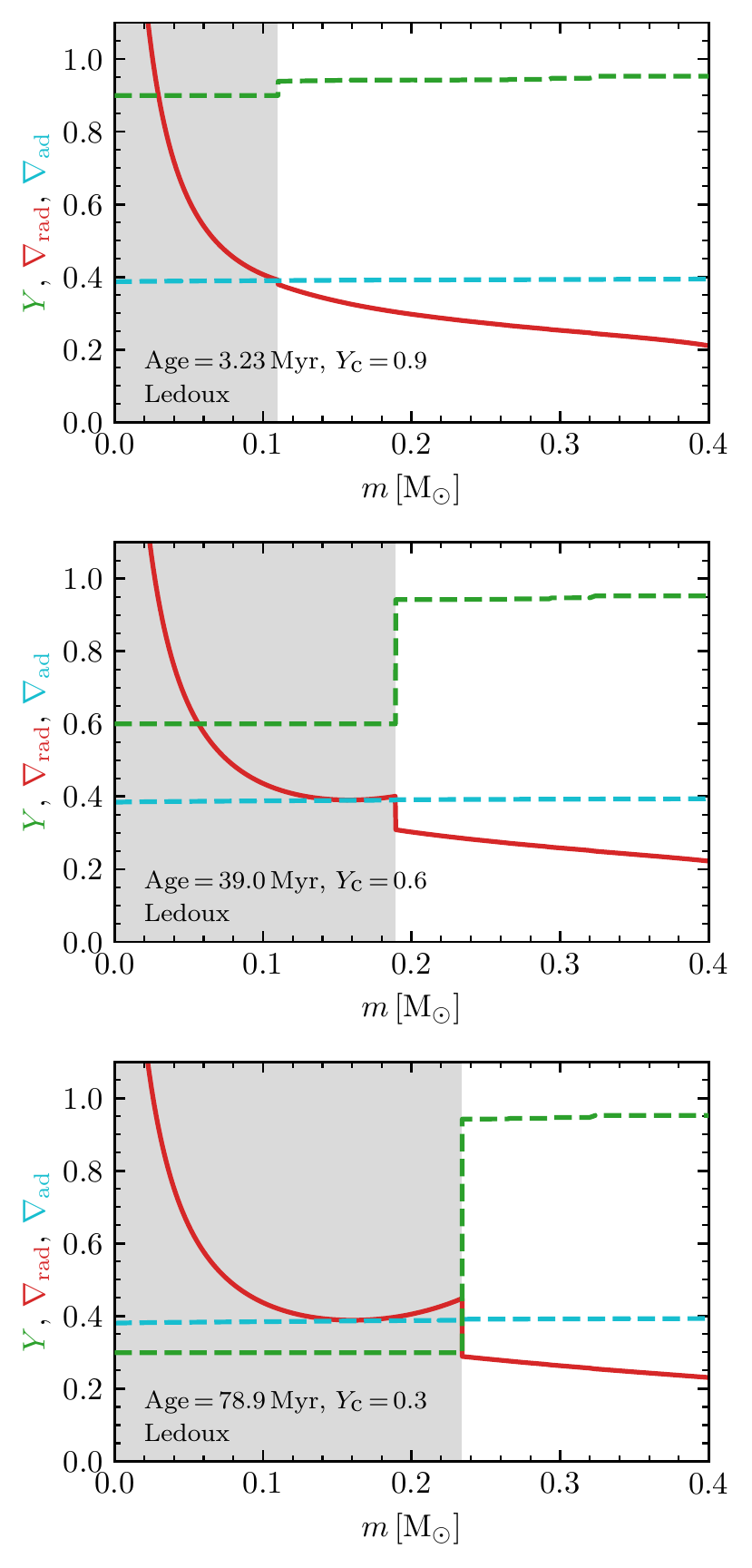}
\caption{Profiles of $\gradr$, $\grada$, and $\Yhel$ as a function of
  mass coordinate, for the $1\,\Msun$ stellar model. The panels
  correspond to different stages during CHeB: $\Yhel_{\rm c}\,=\,0.9$
  (upper), $\Yhel_{\rm c}\,=\,0.6$ (middle), and $\Yhel_{\rm
    c}\,=\,0.3$ (lower). Gray shading indicates regions undergoing
  convection.}
\label{fig:M1-core-grads}
\end{center}
\end{figure}

\begin{figure}
\begin{center}
\includegraphics{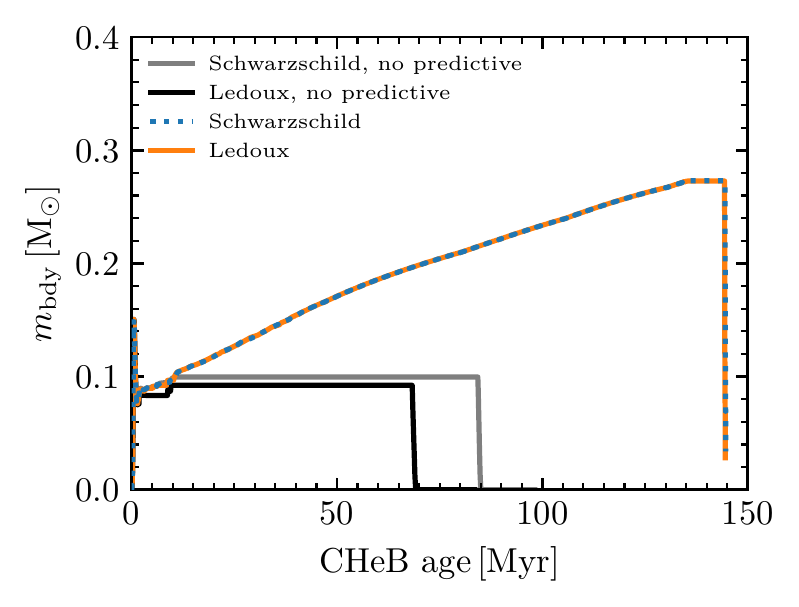} 
\caption{The mass coordinate $m_{\rm bdy}$ of the convective-core
  boundary plotted as a function of CHeB age, for the $3\,\Msun$
  stellar model discussed in Section~\ref{s.helium_core}. Different
  line styles/colors show the separate runs described in the text.}
\label{fig:M3-core-bdy}
\end{center}
\end{figure}

\begin{figure*}
\begin{center}
\includegraphics{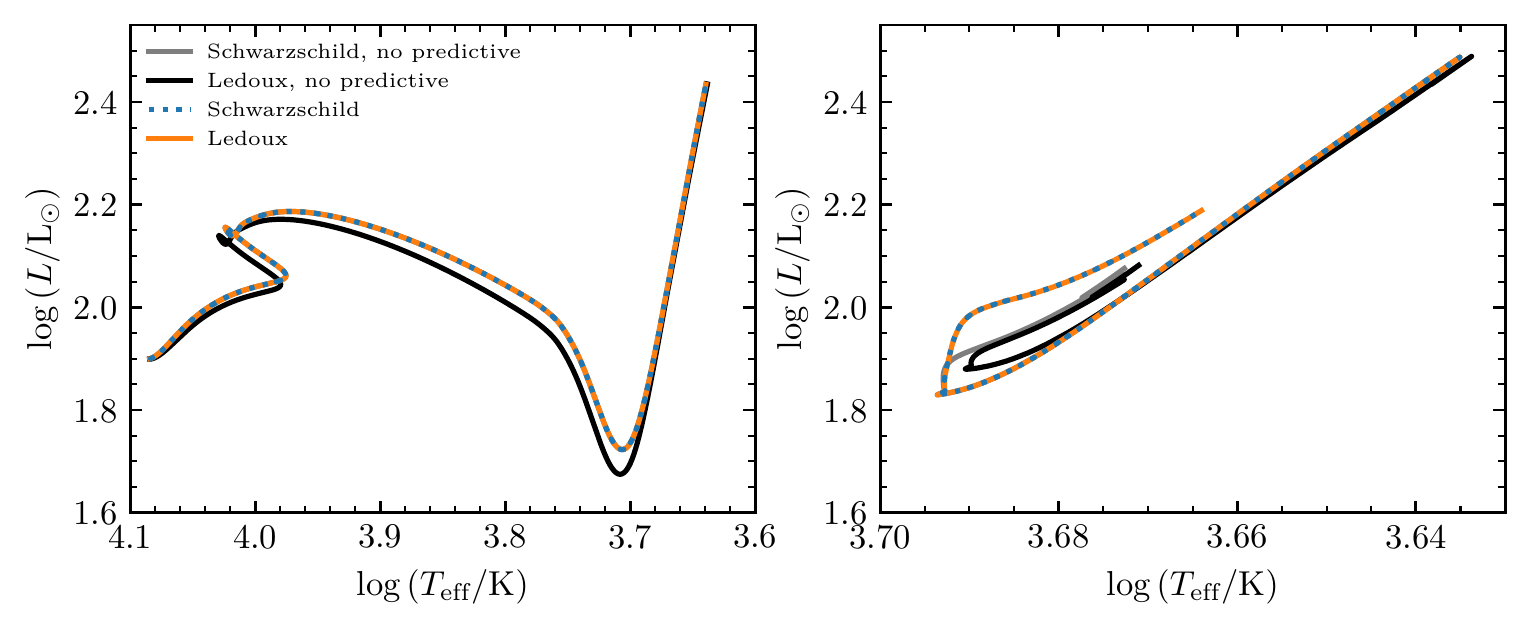}
\caption{Evolution of the $3\,\Msun$ stellar model in the HR diagram,
  from ZAMS to the beginning of CHeB (left), and throughout the CHeB
  phase (right). Separate tracks show the different cases considered in the
  text; in the left panel, the Schwarzschild track without predictive
  mixing lies beneath the tracks with predictive mixing.}
\label{fig:M3-hrd}
\end{center}
\end{figure*}

As reviewed by \citet{Salaris2017}, the modeling of mixing in low- and
intermediate-mass stars during core He burning (CHeB) is particularly
challenging. Correct treatment of convective boundaries is complicated
by the fact that the $\gradr$ profile within the core convection
region develops a local minimum at some point during CHeB evolution
(see the middle and lower panels of
Figure~\ref{fig:M1-core-grads}). This is a consequence of the complex
behavior of the physical quantities (opacity, temperature, density,
etc.)  involved in the expression for $\gradr$. With further outward
propagation of the convective boundary, the mixing of fresh He into
the core can lower the radiative gradient throughout the core to such
an extent that $\gradr=\grada$ at the local minimum of $\gradr$. When
this happens, the part of the convection region interior to the
minimum becomes decoupled from the part exterior to the minimum: the
convection region has split. This phenomenon was first discussed by
\citet{Eggleton1972}\added{, and a variety of ad-hoc approaches have
  been proposed to follow the subsequent evolution, mostly focused
  around the narrative that the exterior part undergoes partial mixing
  with the adjacent radiative region until it reaches convective
  neutrality \citep[see, e.g.,][and references
    therein]{Castellani:1985aa}.}

Another problem appears near the end of CHeB. At that point, even 
small amounts of He added to the core (which is almost totally
depleted in He) will enhance the rate of energy production and
thus the luminosity, resulting in an increase in $\gradr$. This
increase leads to a sudden growth in the core boundary and a
``breathing pulse''. The He is then quickly burned in the core and
the star re-adjusts itself. The existence of these breathing pulses
remains controversial and it is still unclear whether they are
numerical or physical
\citep{Caputo1989,Cassisi2003,farmer_2016_aa,Constantino2017}.  All of
these problems are clearly described and illustrated in
\citet{Salaris2017}.

\replaced{When predictive mixing is applied to CHeB in \MESA, the
  scheme includes additional checks that address the aforementioned
  issues.  When a convection region splits, the cells near the growing
  boundary should no longer be able to mix with the rest of the
  region. Therefore, \MESA\ attempts to prevent the scheme from
  precipitating a split.}{To manage these complexities, the predictive
  mixing scheme must be modified.  When a convection region splits, it
  is no longer meaningful to re-evaluate $y$ using opacities and other
  data calculated on the assumption of uniform composition throughout
  (Section~\ref{s.predictive-mix}), because the radiative region
  appearing at the split point prevents the free exchange of material
  between the adjacent convection regions. Although in principle we
  could resort to the partial mixing mentioned above, in practice it
  is not clear how this might be implemented within a diffusive mixing
  framework. \citet{Constantino2015} have developed an overshoot-like
  prescription which appears useful for mimicking the convective
  neutrality achieved by partial mixing (see their Section 2.3.3), but
  it involves a number of unconstrained parameters.  Therefore, on
  grounds of simplicity and pragmatism --- and recognizing that better
  approaches may become apparent in the future --- we modify the
  predictive scheme to prevent it from causing a convection region to
  split in the first place.} This involves a new control parameter,
\texttt{predictive\char`_superad\char`_thresh}; if during the
predictive mixing iterations the super-adiabaticity $\gradr/\grada-1$
drops below this threshold anywhere in the mixed region, then the code
backs off the mixing by one cell and updates the model convective
diffusivities and convective velocities in the usual manner.

Further functionality, controlled by a new parameter
\texttt{predictive\char`_avoid\char`_reversal}, also helps to prevent
splitting and breathing pulses. When this parameter is set to the name
of a \MESA\ isotope, then the code monitors how the predictive mixing
alters the abundance evolution of that isotope in the convection
region. If it would cause this evolution to reverse (i.e., switch from
decreasing to increasing, or vice-versa), then the code backs off the
mixing by one cell and updates the model as before. Thus, for
instance, setting this parameter to \texttt{`he4'} during CHeB ensures
that the predictive mixing scheme does not cause the core He abundance
to increase across a timestep.

To illustrate the preceding discussion, \replaced{we evolve a
  $3\,\Msun$ star through CHeB}{we evolve a $1\,\Msun$ star through
  CHeB; this is the same mass considered by
  \citet{Constantino2015}}. Figure~\ref{fig:M1-core-bdy} plots the mass
of the convective core as a function of CHeB age (defined as the time
elapsed since the central $\Yhel$ drops below 0.98), showing results
from separate runs with and without predictive mixing, and using the
Schwarzschild and Ledoux criteria. For the cases with predictive
mixing, we adopt a value of 0.005 for the
\texttt{predictive\char`_superad\char`_thresh} parameter, and set
\texttt{predictive\char`_avoid\char`_reversal} to \texttt{`he4'} to
prevent any reversal in the core He
abundance. Figure~\ref{fig:M1-core-bdy} also shows the results from an
additional Ledoux/predictive mixing run \replaced{with no reversal
  avoidance}{where we allow the core to split by \emph{not} setting
  the \texttt{predictive\char`_avoid\char`_reversal} and
  \texttt{predictive\char`_superad\char`_thresh} controls}.

Figure~\ref{fig:M1-core-bdy} shows that without predictive mixing the
core is prevented from growing, and the CHeB lifetime significantly
curtailed, irrespective of whether the Schwarzschild or Ledoux
criteria are used (see also Figure~15 of \mesatwo). With predictive
mixing but no splitting allowed, however, the core grows steadily
until He is exhausted, and no breathing pulses are seen. There is
almost no difference between the Schwarzschild and Ledoux
cases. \replaced{Without reversal avoidance, the core-mass evolution
  appears noisier; around a CHeB age $\approx\,45\,\Myr$, and then
  later around $\approx\,115\,\Myr$, the core undergoes repeated
  sequences of splittings and rejoinings which cause its boundary to
  oscillate between its original position and the $\gradr$
  minimum. Beyond $\approx\,125\,\Myr$ these oscillations develop into
  large breathing pulses, which have the ultimate effect of prolonging
  the star's CHeB lifetime.}{When the core is allowed to split, the
  evolution is much noisier. Starting at an age $\approx\,25\,\Myr$,
  the core undergoes episodes of splitting and rejoining that repeat
  on a short timescale. Toward the end of the evolution, as the core
  helium abundance becomes very small, the timescale between
  successive splittings becomes longer, until the core finally splits
  without rejoining. The overall CHeB lifetime of the model is
  shortened by $\approx\,6\,\Myr$ relative to the cases where
  splitting is avoided.}

Figure~\ref{fig:M1-core-grads} plots the profiles of $\gradr$,
$\grada$ and $\Yhel$ for the \replaced{$3\,\Msun$}{$1\,\Msun$} star at
three points during its CHeB evolution, corresponding to core helium
mass fractions $\Yhel_{\rm c} = 0.9$, 0.6 and 0.3. The profiles are
all from the run with the Ledoux criterion and predictive mixing.  In
the upper panel, a local minimum in $\gradr$ has yet to develop, and
the core boundary satisfies the $\gradr=\grada$ equality on its
convective side. In the middle and lower panels, the local minimum in
$\gradr$ can clearly be seen; in these cases, the predictive mixing
has extended the convection region as far as possible without pushing
the minimum $\gradr$ below the threshold set by the
\texttt{predictive\char`_superad\char`_thresh} control.  \MESA\ treats
the region between the $\gradr$ minimum and the convective boundary as
fully convective. On the convective side of this boundary $\gradr >
\grada$, which is physically inconsistent \replaced{but is the price
  to be paid for avoiding the core splitting To avoid this physically
  inconsistent result requires either allowing the convective region
  to split or the use of an ad hoc technique to force $\gradr=\grada$
  from the \gradr\ minimum to the edge of the convective region.}{but
  cannot be remedied with predictive mixing alone: any further
  extension of the boundary would cause the convection region to
  split. As discussed above, fixing this inconsistency requires some
  way of modeling the partial mixing expected to occur in the part of
  the convection region between the $\gradr$ minimum and the
  boundary.}

\added{The abundance profiles plotted in
  Figure~\ref{fig:M1-core-grads} show a sharp transition between the
  He-depleted core and the He-rich radiative region above. Although
  not shown, the carbon and oxygen abundance profiles exhibit
  corresponding jumps at the core boundary. Similar results are
  obtained by \citet{Constantino2015} with their ``maximal overshoot''
  scheme (cf. their Figure 2); and those authors also find a core mass
  evolution during CHeB that closely resembles the outcome from
  predictive mixing (cf. their Figure~8 and our
  Figure~\ref{fig:M1-core-bdy}). These similarities are not
  coincidental; although the predictive mixing and maximal overshoot
  schemes have different narratives and implementations, both have the
  effect of growing the core boundary during CHeB to the greatest
  extent permitted without causing the convection region to split. The
  larger cores that result from this growth appear to provide a better
  match to \emph{Kepler} asteroseismic period spacings, when compared
  with other mixing schemes that produce smaller cores
  \citep{Constantino2015}; and with certain assumptions about
  post-CHeB evolution, the larger cores can also provide a
  satisfactory fit to observational cluster counts
  \citep{Constantino2016}.}

To explore whether the predictive mixing performs equally well for a
\replaced{lower-mass star that has passed through the He flash, we
  also evolve a $1\,\Msun$ star through CHeB}{higher-mass star that
  has not passed through the He flash, we also evolve a $3\,\Msun$
  star through CHeB; this is the same mass and evolutionary stage
  considered in Figure~15 of \mesatwo}. Figure~\ref{fig:M3-core-bdy}
plots the mass of the convective core as a function of CHeB age,
showing results from separate runs with and without predictive mixing,
and using the Schwarzschild and Ledoux criteria. For the cases with
predictive mixing, we again adopt a value of 0.005 for the
\texttt{predictive\char`_superad\char`_thresh} parameter, and set
\texttt{predictive\char`_avoid\char`_reversal} to \texttt{`he4'} to
prevent any reversal in the core He abundance. As before, we find that
the predictive mixing allows the core to grow steadily; and that the
Schwarzschild and Ledoux criteria give essentially the same outcome.

\added{As a visual summary of how predictive mixing influences a
  star's evolution, Figure~\ref{fig:M3-hrd} plots evolutionary tracks
  of the $3\,\Msun$ model in the Hertzsprung-Russell (HR) diagram, for
  the same combinations of mixing and stability criteria considered in
  Figure~\ref{fig:M3-core-bdy}. The left panel focuses on the MS and
  red giant branch phases, and the right panel on the CHeB phase. In
  the left panel, the case with the Ledoux criterion but without
  predictive mixing stands out from the other three as having a
  slightly reduced luminosity. This behavior arises because the
  boundary of the hydrogen-burning convective core is incorrectly
  positioned during the early MS evolution, retarding the growth of
  the core (the same effect can be seen for the $1.5\,\Msun$ model in
  the upper panels of Figure~\ref{fig:M15-core-grads}). During the
  subsequent CHeB phase, all four tracks are similar until slightly
  after the luminosity minimum, when the helium-burning convective
  core starts to grow; this growth is retarded in both cases without
  predictive mixing, leading to reduced luminosities and the shorter
  CHeB lifetimes seen in Figure~\ref{fig:M3-core-bdy}. For the cases
  with predictive mixing, there is no difference between the
  Schwarzschild and Ledoux, either on the MS or after.}
  
\subsection{Evolution of the Bottom of the Surface Convective Region in a Low-Mass Star}
\label{s.envelope_conv}

\begin{figure}
\begin{center}
\includegraphics{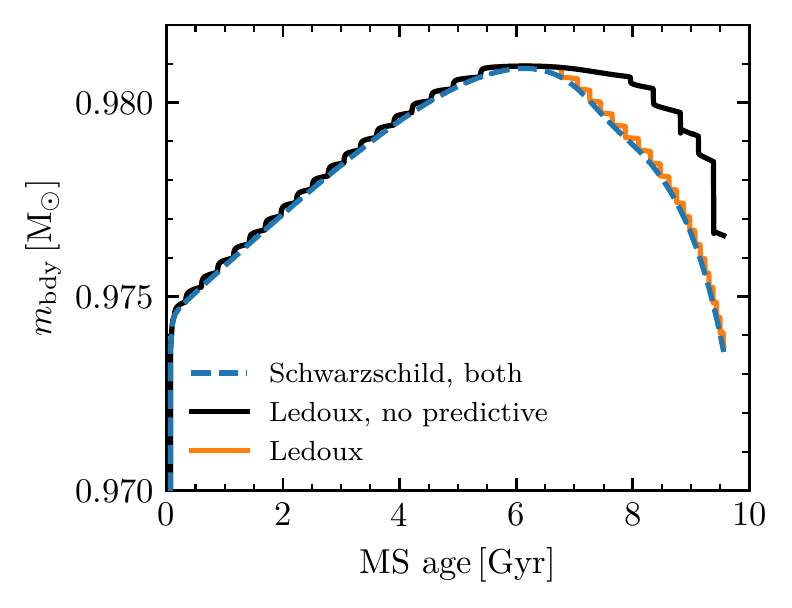}
\caption{The mass coordinate $m_{\rm bdy}$ of the lower boundary of
  the envelope convection region plotted as a function of MS
  age, for the $1\,\Msun$ stellar model discussed in
  Section~\ref{s.envelope_conv}. Different line styles/colors show the
  separate runs described in the text.}
\label{fig:M1-env-bdy}
\end{center}
\end{figure}

\begin{figure*}
\begin{center}
\includegraphics{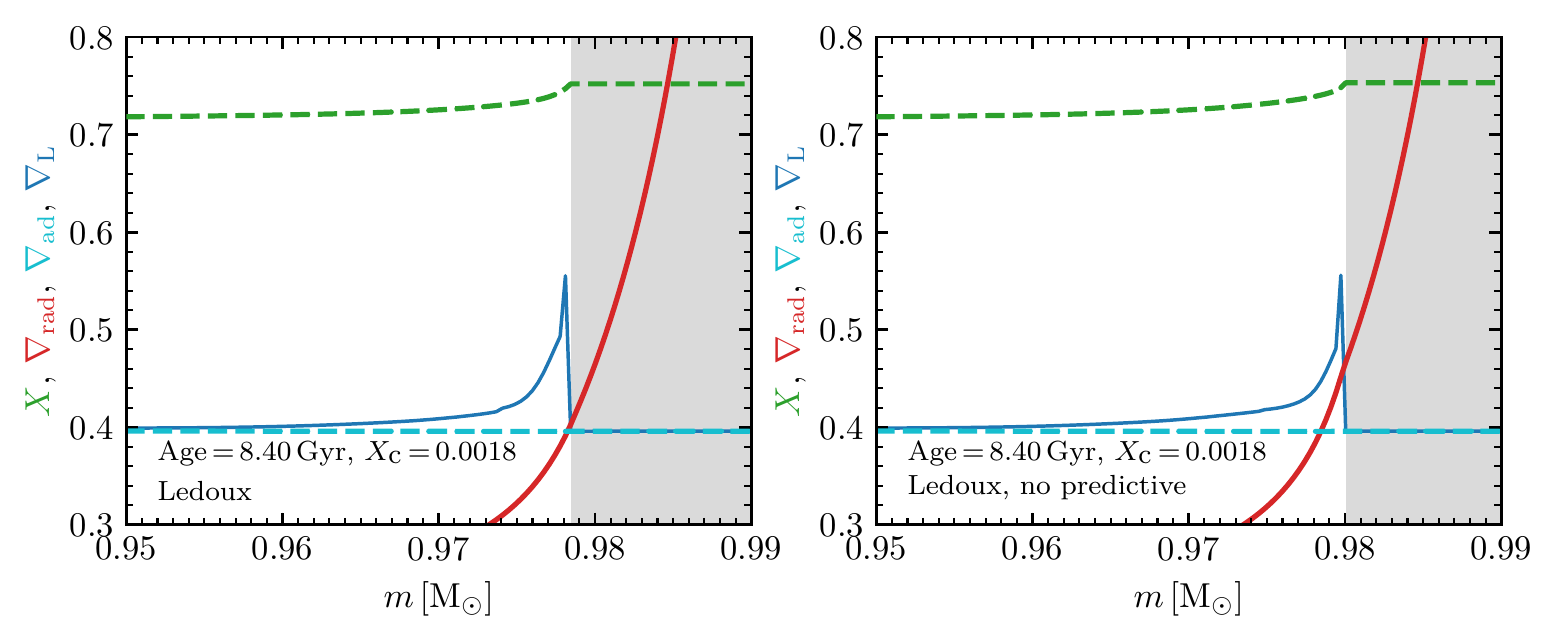}
\caption{Profiles of $\gradr$, $\grada$, $\gradL$, and $\Xhyd$ as a
  function of mass coordinate, in the outer envelope of the $1\,\Msun$
  stellar model at an age 8.40\,\Gyr. The panels show the separate runs
  described in the text. Gray shading indicates regions undergoing
  convection.}
\label{fig:M1-env-grads}
\end{center}
\end{figure*}

We now evolve a $1\,\Msun$ star from ZAMS to TAMS, using the
predictive mixing scheme to position the lower boundary of the
convective envelope. We include element diffusion in these
calculations; when it is excluded, the composition remains completely
uniform throughout the stellar envelope, and predictive mixing makes
no difference whatsoever to the
evolution. Figure~\ref{fig:M1-env-bdy} plots the mass coordinate of
the convective boundary as a function of MS age, showing results from
separate runs with and without predictive mixing, and using the
Schwarzschild and Ledoux criteria.

The four runs are in agreement until an age $\approx$\,6.5\,\Gyr; after
this point, the downward growth of the region boundary is slower in
the run that does not include predictive mixing with the Ledoux
criterion. Figure~\ref{fig:M1-env-grads} plots the profiles of
$\gradr$, $\grada$, $\gradL$ and $\Xhyd$, in the outer part of the
$1\,\Msun$ star at an age $8.40\,\Gyr$. The left panel illustrates the
run with the Ledoux criterion plus predictive mixing, while the right
panel shows the run with the Ledoux criterion but without predictive
mixing. The former shows that $\gradr = \grada$ on the convective
(upper) side of the convective boundary, while the latter has $\gradr
> \grada$ consistent with the boundary growth being retarded.

\subsection{Effect of Timesteps and Mesh Size}
\label{s.timestep_mesh}

\begin{figure}
\begin{center}
\includegraphics{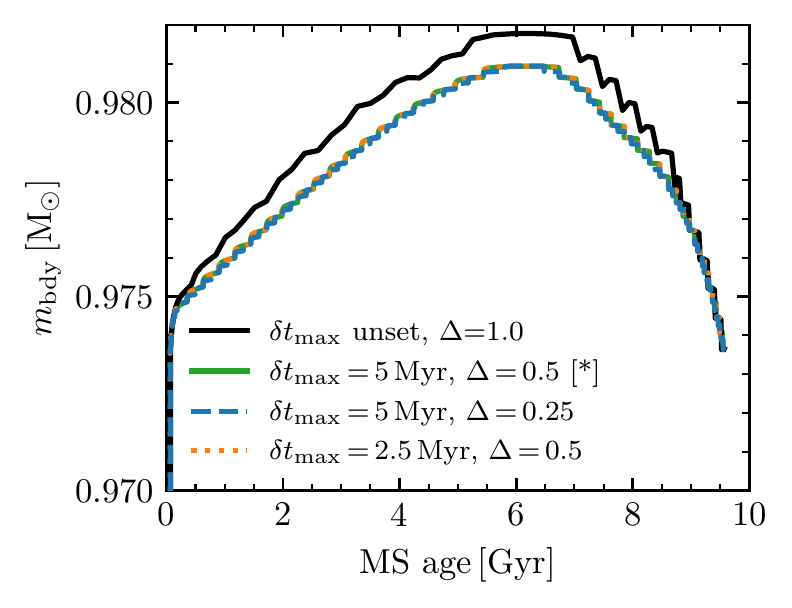}
\caption{The mass coordinate $m_{\rm bdy}$ of the lower boundary of
  the envelope convection region plotted as a function of MS age, for
  the $1\,\Msun$ stellar model. Different line styles/colors show the
  separate runs with alternative timestep ($\dt_{\rm max}$) and mesh
  resolution ($\Delta$) choices. \added{The choices adopted in
    Section~\ref{s.envelope_conv} are marked with an asterisk [*].}}
\label{fig:M1-env-bdy-convg}
\end{center}
\end{figure}

\begin{figure}
\begin{center}
\includegraphics{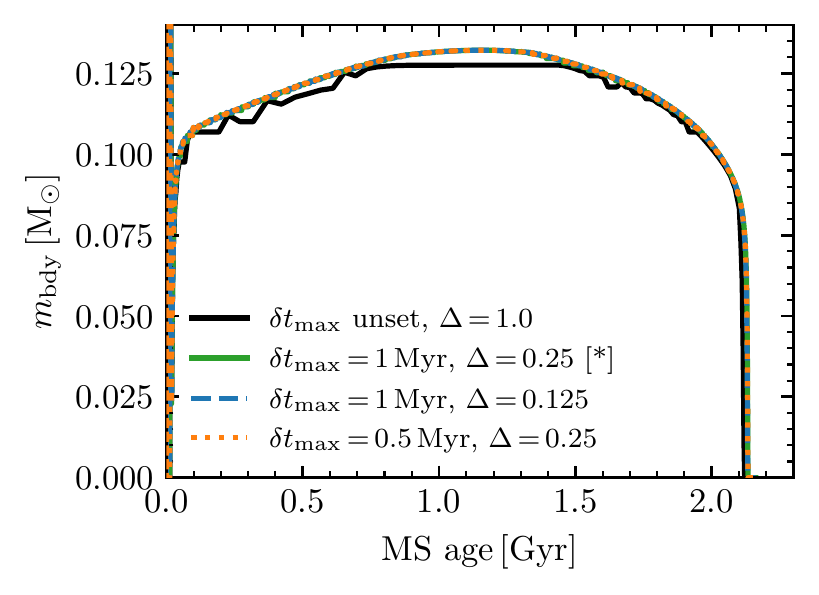}
\caption{The mass coordinate $m_{\rm bdy}$ of the convective-core
  boundary plotted as a function of MS age, for the $1.5\,\Msun$
  stellar model. Different line styles/colors show the separate runs
  with alternative timestep ($\dt_{\rm max}$) and mesh resolution
  ($\Delta$) choices. \added{The choices adopted in
    Section~\ref{s.growing_core} are marked with an asterisk [*].}}
\label{fig:M15-core-bdy-convg}
\end{center}
  \end{figure}

\begin{figure}
\begin{center}
\includegraphics{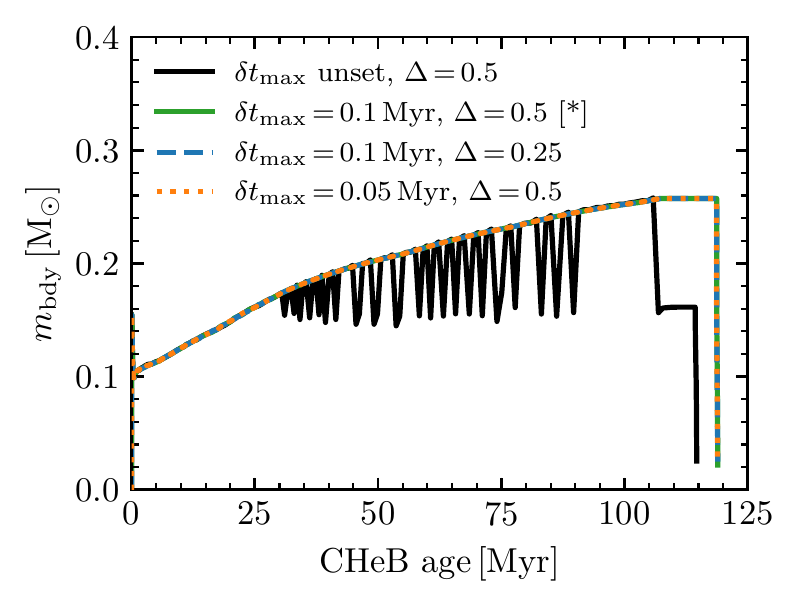}
\caption{The mass coordinate $m_{\rm bdy}$ of the convective-core
  boundary plotted as a function of CHeB age, for the $1\,\Msun$
  stellar model. Different line styles/colors show the separate runs
  with alternative timestep ($\dt_{\rm max}$) and mesh resolution
  ($\Delta$) choices. \added{The choices adopted in
    Section~\ref{s.helium_core} are marked with an asterisk [*].}}
\label{fig:M1-core-bdy-convg}
\end{center}
\end{figure}

We now demonstrate how limiting the maximum timestep $\dt_{\rm max}$
(set by the \texttt{max\char`_years\char`_for\char`_timestep} control)
and changing the mesh resolution parameter $\Delta$ (set by the
\texttt{mesh\char`_delta\char`_coeff} control; see Section B.4 of
\mesatwo\ for further details) influences the results presented in the
previous sections.

\replaced{Figure~\ref{fig:M1-env-bdy-convg} shows how the position of
  the convective envelope boundary in the $1\,\Msun$ MS model varies
  with $\dt_{\rm max}$ and $\Delta$. All runs use the Ledoux criterion
  with predictive mixing. The results presented for the tightest
  settings agree, suggesting convergence.  The result obtained with
  $\dt_{\rm max}$ unset and $\Delta$\,=\,1.0 (the default values) is
  not converged.}  {First we consider the effects of changing timestep
  and resolution on the position of the convective envelope boundary
  in the $1\,\Msun$ model considered in Section~\ref{s.envelope_conv},
  focusing specifically on the case with the Ledoux criterion and
  predictive mixing.  The results presented previously in
  Figure~\ref{fig:M1-env-bdy} are calculated using $\dt_{\rm max} =
  5\,\Myr$ and $\Delta = 0.5$. Figure~\ref{fig:M1-env-bdy-convg}
  demonstrates that halving either $\dt_{\rm max}$ or $\Delta$ has
  little effect on these results, confirming that the calculations are
  converged. Such settings need to be applied when a converged result is desired from 
MESA for this calculation.}
  
\replaced{Figures~\ref{fig:M15-core-bdy-convg} and
  \ref{fig:M1-core-bdy-convg} repeat the exercise for the position of
  the core convection boundary in a $1.5\,\Msun$ MS model and
  $1\,\Msun$ CHeB model respectively. The results presented for the
  tightest settings agree, suggesting convergence.  The results
  obtained with the default choices of $\dt_{\rm max}$ and $\Delta$
  are unconverged.}{Figures~\ref{fig:M15-core-bdy-convg} and
  \ref{fig:M1-core-bdy-convg} repeat this exercise for the position of
  the core convection boundary in the $1.5\,\Msun$ MS model and
  $1\,\Msun$ CHeB model, respectively. The results presented previously are clearly converged, and this exercise clarifies
the MESA settings that should be used for this calculation.}

\section{Element Diffusion}\label{s.diffusion}

Section~9 of \mesathree\  describes in detail the old implementation of
element diffusion in \MESA. Section~9.3.4 points out limitations to
those methods, namely: (1) electron degeneracy was not
properly accounted for in the diffusion equations,
and (2) strong Coulomb interaction
introduced theoretical uncertainties for the diffusion
coefficients. These two issues are especially important when modeling
diffusion in WDs. Here we describe
the impact of degeneracy and present new methods to incorporate its effects. We also discuss recent updates to
diffusion coefficients and potential approaches for further
improvements.

\subsection{Degeneracy and the Approach in \mesathree}
\label{s.degeneracy_errors}

The approach to diffusion presented in Section~9 of \mesathree\ assumes all particles
obey the ideal gas law. Electron degeneracy pressure
can significantly modify the EOS and violate this assumption.

For a plasma species $s$ (i.e.,~electrons and ions) with partial pressure $P_s$, mass density $\rho_s$, 
charge density $\rho_{es}$, number density $n_s$, and temperature $T$, 
the \citet{Burgers69} equations for diffusion are
\begin{align}
\begin{split}
\label{eq:Burgers1}
\frac{\dif P_s}{\dif r} + \rho_s g - \rho_{es} E 
= &\sum_{t \neq s} K_{st} (w_t - w_s) \\
&+ \sum_{t \neq s} K_{st} z_{st} \frac{m_t r_s - m_s r_t}{m_s + m_t},
\end{split}
\end{align}
\begin{align}
\begin{split}
\label{eq:BurgersEnergy}
\frac 5 2 n_s& k_{\rm B} \frac{\dif T}{\dif r} \\
=& - \frac 2 5 K_{ss}z_{ss}'' r_s 
 - \frac 5 2 \sum_{t \neq s} K_{st} z_{st} \frac{m_t}{m_s+m_t}(w_t - w_s)   \\
&  - \sum_{t \neq s} K_{st} \left[\frac{3 m_s^2 + m_t^2z_{st}'}{(m_s+m_t)^2} +
\frac 4 5 \frac{m_sm_t}{(m_s + m_t)^2}z_{st}'' \right] r_s \\
& + \sum_{t \neq s} K_{st} \frac{m_sm_t}{(m_s + m_t)^2} \left(3 + z_{st}' - \frac 4 5 z_{st}'' \right) r_t.
\end{split}
\end{align}
The resistance coefficients $K_{st}$, $z_{st}$, $z_{st}'$, and
$z_{st}''$ are defined in Equation~(86) of \mesathree.
With $S$ representing the total number of plasma species, we must solve for $2S+2$ unknowns: $S$
diffusion velocities ($w_s$), $S$ heat flow vectors ($r_s$),
the electric field ($E$), and the gravitational
acceleration ($g$). The Burgers equations above for each
species provide $2S$ equations, so we can close the system with two
additional constraints, which are no net
flow of mass or electric current due to diffusion,
\begin{equation}
\label{eq:mass_flux}
\sum_s \rho_s w_s = 0,
\end{equation}
\begin{equation}
\label{eq:charge_flux}
\sum_s \rho_{es} w_s = 0~.
\end{equation}
This gives a total of $2S+2$ equations.

When electrons are degenerate, Equation~(\ref{eq:Burgers1}) is difficult
to apply since $\dif P_{\rm e}/\dif r$ no longer takes a simple
analytic form. Moreover, the temperature term appearing on the left hand side
of Equation~(\ref{eq:BurgersEnergy}) clandestinely assumes an ideal gas law. \cite{Burgers69}
defines the temperature for each species as $T_s \equiv P_s / n_s
k_{\rm B}$ and assumes thermal equilibrium
between all species so that $T \equiv T_s$.
The quantities $P_s$ and $n_s$ are defined in terms of moments of a
Maxwellian distribution function, but the Fermi-Dirac distribution
function for electrons no longer reduces to a Maxwellian form when they are
degenerate, and hence $T_{\rm e} \neq P_{\rm e}/n_{\rm e}k_{\rm B}$.
If the electrons remain in thermal equilibrium with their surroundings
while failing to satisfy an ideal-gas relation for their temperature,
the Burgers treatment assigns an incorrect temperature to degenerate
electrons for the $\dif T/\dif r$ term in Equation~(\ref{eq:BurgersEnergy}).

Furthermore, the approach to diffusion described in \mesathree\ follows
\cite{Thoul94} in rearranging and rescaling all
equations into one matrix system with units convenient for
solving numerically,
\begin{equation}
\label{eq:Thoul3}
\frac P{K_0} \Bigg( \alpha_i \frac{\dif \ln P}{\dif r} + \nu_i
\frac{\dif \ln T}{\dif r} +
\sum_{\substack{j=1 \\ j \neq \rm e}}^S \gamma_{ij} \frac{\dif \ln
  C_j}{\dif r} \Bigg) = \sum_{j = 1}^{2S+2} \Delta_{ij} W_j~.
\end{equation}
The sum on the left hand side skips the electron index because
$C_{\rm e} \equiv 1$ by construction, and so we
save resources by not evaluating its gradient unnecessarily.
Here, indices ${i=1,2,\ldots S}$ encode the $S$ equations given by Equation~\eqref{eq:Burgers1},
indices ${i=S+1,S+2,\ldots,2S}$ encode the $S$ equations given by
Equation~\eqref{eq:BurgersEnergy}, and indices $i=2S+1,2S+2$ encode the $2$
constraints of no current or mass flux.
For definitions of the various coefficients and matrices in
Equation~\eqref{eq:Thoul3}, consult \mesathree\ and \cite{Thoul94}.
We repeat a few particularly relevant definitions here.
First, let $C_s = n_s /
n_{\rm e}$ denote the species concentration, where $n_{\rm e}$ is the electron number density.
Second, define the total concentration as $C = \sum_s
C_s$. Then the quantity $\alpha_i$ appearing in
Equation~(\ref{eq:Thoul3}) above is defined as 
\begin{equation}
\alpha_i = 
\begin{cases}
C_i / C & i = 1,2,\ldots S, \\
0  & i = S+1, \ldots 2S+2.
\end{cases}
\end{equation}
The term $\alpha_i \,\dif \ln P /\dif r$ in
Equation~(\ref{eq:Thoul3}) is meant to capture contributions of the
driving terms $\dif P_s/\dif r$ in Equation~(\ref{eq:Burgers1}). But this
correspondence only holds if the ratio of the partial
pressure  $P_s$ for species $s$ to the total pressure $P$ is
given by
\begin{equation}
\label{eq:ratio}
\frac{P_s}{P} = \frac{C_s}{C} = \frac{n_s/n_{\rm e}}{\sum_t n_t/n_{\rm
    e}} = \frac{n_s}{n},
\end{equation}
where $n$ is the total number density.
This holds as long as all pressures are ideal-gas. However, once electron
degeneracy modifies the equation of state, $P$ does not
scale linearly with $n$, and so
Equation~(\ref{eq:ratio}) fails for all species in the
plasma. This means the $\alpha_i$ term no longer accurately
represents the information in the Burgers equations for the diffusion
velocity of any species.

Moreover, the prefactor $P/K_0$ in Equation~(\ref{eq:Thoul3})
also assumes ideal gas for each species. The quantity
${K_0 = 1.144 \times 10^{-40} (T/10^7 \, {\rm K})^{-3/2} n_{\rm e}^2}$
simply scales out some of the
information common to all diffusion coefficients in the units used for
Equation~(\ref{eq:Thoul3}). \cite{Thoul94} assume an ideal gas to
simplify the prefactor in Equation~\eqref{eq:Thoul3} to
\begin{equation}
\frac{P}{K_0} = 2.00 \frac{(T/10^7 \, {\rm K})^{5/2}}{(\rho/100 \,
  \grampercc)} \left( \sum_s C_s \right) \left( \sum_s A_s C_s \right),
\end{equation}
where $A_s$ is the mass of species $s$ in atomic mass units.
This scaling was propagated into the \mesa\ diffusion routine
described in \mesathree. Since ideal gas pressure can be significantly
smaller than total pressure when electrons are degenerate, this
prefactor for Equation~(\ref{eq:Thoul3}) is systematically too small
for degenerate plasmas. This can result in diffusion velocities that
are many orders of magnitude smaller than obtained by a proper solution.

\begin{figure}
\begin{center}
\includegraphics[width=\apjcolwidth]{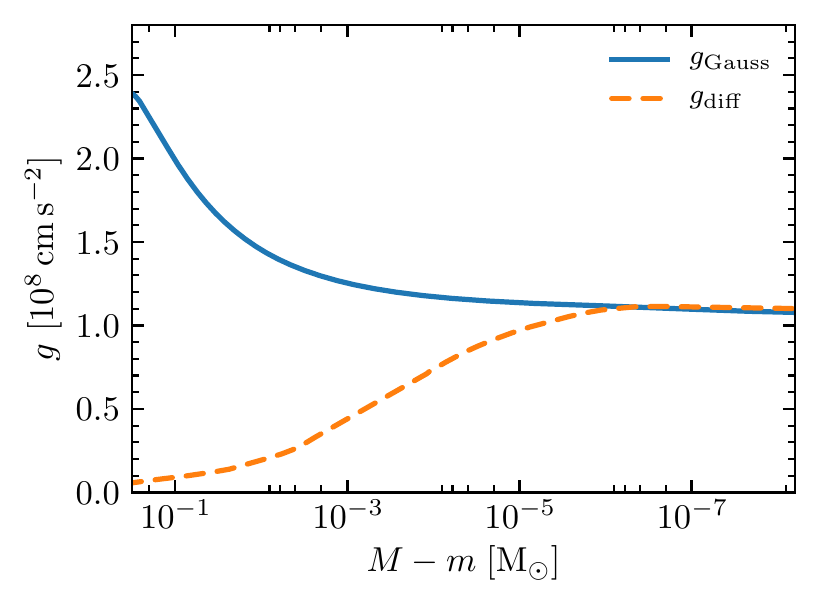}
\caption{
The gravitational acceleration reported by the diffusion routine
described in \mesathree\ compared with ${g_{\rm Gauss} = Gm/r^2}$
for a $0.6 \, \Msun$ \MESA\ WD model. 
}
\label{fig:gfield}
\end{center}
\end{figure}

We can verify that there are problems in the degenerate regime by
looking at the local gravitational acceleration $g_{\rm diff}$, which is solved for
simultaneously with the diffusion velocities in the diffusion routine described
by \mesathree. \MESA\ also reports the gravitational acceleration
independent of the diffusion routine, $g_{\rm Gauss} = Gm /r^2$.
For a \MESA\ WD model, layers below the surface quickly become degenerate,
and the difference between $g_{\rm diff}$ and $g_{\rm Gauss}$
is significant
(Figure~\ref{fig:gfield}). This reflects the fact that the solutions
given by the diffusion routine scale with a pressure that is far too
small in the interior.

\subsection{New Methods}
\label{s.diffusion_methods}

We now describe new methods that have been introduced to avoid the limitations discussed in Section~\ref{s.degeneracy_errors}.

\subsubsection{Recasting the Burgers Equations}

The problems with Equation~\eqref{eq:Thoul3}
demonstrated in Figure~\ref{fig:gfield} can be
circumvented by solving the Burgers equations directly as presented in
Equations~(\ref{eq:Burgers1}) and~(\ref{eq:BurgersEnergy}). When avoiding the rescaling of the Burgers equations
that was originally adopted from \cite{Thoul94}, no limitations on the form
of total pressure are present.

To that end, we recast the diffusion solver into the form given in
Appendix~\ref{app.diffusion}. This form closely follows the general
approach presented by \cite{Thoul94} for arranging the full set of
equations into a single matrix equation, but enters the Burgers equations into that matrix structure
without rescaling any quantities. We therefore avoid making
any additional ideal-gas assumptions  beyond those already
present in the Burgers equations.

\subsubsection{Resolving the Degeneracy Problem}\label{s.skirting}

Electron degeneracy makes it difficult to evaluate the term
$\dif P_s/\dif r$
in Equation~\eqref{eq:Burgers1} in the case of electrons,
but it is possible to form a closed set of diffusion equations that
makes no explicit reference to this equation for the electrons.
Even in many applications involving WDs, each ion species can be treated as approximately ideal, and hence
Equation~\eqref{eq:Burgers1} remains useful for ions.
We are then left with
just two problematic equations out of the system of $2S+2$ equations: Equations~(\ref{eq:Burgers1})
and~(\ref{eq:BurgersEnergy}) for the electrons.

For the $S-1$ species of ions in the system,
we can write $S-1$ Equations~\eqref{eq:Burgers1} in the form
\begin{align}
\begin{split}
\label{eq:BurgersMomentum}
&n_s k_{\rm B} T \frac{ \dif \ln T}{\dif r}
+ n_s k_{\rm B} T \frac{\dif \ln n_s}{\dif r} + n_s A_s m_{\rm p} g - n_s \bar{Z}_s
\qe E \\
&= \sum_{t \neq s} K_{st} (w_t - w_s)
+ \sum_{t \neq s} K_{st} z_{st} \frac{A_t r_s - A_s r_t}{A_s + A_t},
\end{split}
\end{align}
where $\bar{Z}_s$ is the average charge of species $s$ obtained using \citet{Paquette86WD}.
Taking this together with $S$ Equations~\eqref{eq:BurgersEnergy} and the two
constraints on current and mass flux, we have a total of $2S + 1$
equations. If we drop $g$ as an unknown and treat it as a fixed input to the diffusion
routine in \mesa\ using ${g = Gm/r^2}$, we are left with $2S + 1$
unknowns. This gives a closed system of diffusion
equations with no explicit reference to the problematic
Equation~\eqref{eq:Burgers1} for electrons.   This is the form of diffusion equations
described in Appendix~\ref{app.diffusion}.

The thermal diffusion terms (those including $\dif T/\dif r$ 
in Equation~\ref{eq:BurgersEnergy}) still contain ideal-gas assumptions as described in Section~\ref{s.degeneracy_errors}.
Fortunately, in WD cores where strong electron degeneracy occurs,
electron conduction leads to efficient thermal transport,
resulting in small temperature gradients. With $\dif T/\dif r \ll T/H$, where
$H = P/\rho g$ is the local scale height,
the heat flow vectors (representing kinetic energy carried along a
temperature gradient by diffusing particles) become negligible:
$r_s \ll w_t$ for all $w_t$. 
Thus for WD interiors the system of diffusion equations can be simplified
by dropping the $S$ heat flow terms, removing the need for
the $S$ Equations~\eqref{eq:BurgersEnergy}.
Indeed, according to \cite{Iben85} and \cite{Paquette86WD},
thermal diffusion leads only to small corrections to the
diffusion velocities for degenerate WD interiors.

Therefore, following \cite{Iben85}, we provide options for
neglecting thermal diffusion in electron degenerate regions, setting $r_s = 0$ and dropping
Equation~(\ref{eq:BurgersEnergy}) for each species.
Equation~\eqref{eq:BurgersMomentum} then simplifies to the following $S-1$
equations that no longer depend on $r_s$ for the ions:
\begin{align}
\begin{split}
\label{eq:Iben}
\frac{1}{n_s} & \sum_{t} K_{st} ( w_t -  w_s) + \bar{Z}_s \qe E \\
&= A_s m_{\rm p} g + k_{\rm B} T \frac{\dif \ln T}{\dif r} + k_{\rm B} T
\frac{\dif \ln n_s}{\dif r},
\end{split}
\end{align}
which matches Equation~(10) from \cite{Iben85}.
Together with the $2$ constraints, this leaves a 
simplified set of $S+1$ equations for $S+1$ unknowns: $S$ diffusion
velocities $w_s$ and the electric field $E$.

\begin{figure}
\begin{center}
\includegraphics[width=\apjcolwidth]{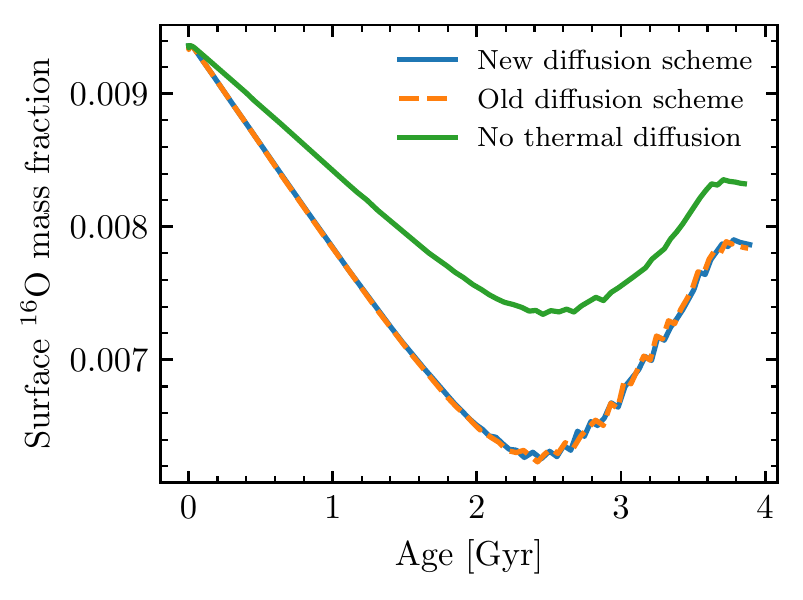}
\caption{
Surface $^{16}{\rm O}$ mass fraction of a $1.25 \, \Msun$ star over
its MS lifetime. It first decreases as diffusion
causes sedimentation. Then it increases after the small surface convection zone
begins to grow, catching the receding $^{16}{\rm O}$ and mixing it
back toward the surface.
}
\label{fig:surfO}
\end{center}
\end{figure}

Thermal diffusion terms tend to enhance
gravitational settling velocities \citep{Iben92}. This can 
be seen in Figure~\ref{fig:surfO} for a $1.25 \, \Msun$ star on the
MS, where the solvers that include
thermal diffusion speed the sedimentation of $^{16} \rm O$ away from
the surface relative to the solver that neglects thermal
diffusion. \mesa\ also provides
options for smoothly transitioning between diffusion velocities
obtained with and without thermal diffusion (averaging between the two
solutions in a blending region as a function of electron degeneracy
parameter). By default, this transition region occurs when the electron
chemical potential is near $\mue \sim k_{\rm B} T$, but
it is left to the user to decide on an appropriate
range of electron degeneracy over which thermal diffusion should be
shut off, if at all.
The effect of blending between solvers with and
without thermal diffusion is to suppress
the thermal enhancements to diffusion velocities, smoothly
pushing the enhancements to zero as electrons reach a degeneracy threshold.
The implementation for the simplified set of diffusion
Equations~\eqref{eq:Iben} and the smooth turn-off of thermal diffusion terms as a
function of degeneracy are described in Appendix~\ref{app.diffusion}.

In order to confirm that we recover the correct behavior on the MS,
we compare results obtained with different diffusion
routines for a $1.25 \, \Msun$ star in Figure~\ref{fig:surfO}. Here
the results based on \cite{Thoul94} are valid, since no significant
departures from ideal-gas behavior are present near the surface.
The results obtained with the new scheme are in agreement.

\subsubsection{Diffusive Equilibrium}

\mesatwo\ and \mesathree\ show abundance profiles for WDs that have reached diffusive equilibrium in their outer layers.
Figure~23 of \mesatwo\ compares the diffusive tails of H and
He to an analytic expression from \cite{Althaus03} and finds good
agreement. However, \cite{Althaus03} note that their analytic
expression for diffusive equilibrium follows \cite{Arcoragi80} in
assuming an ideal gas, and the equilibrium abundance profiles from their
evolutionary models deviate from the analytic expression due to the inclusion of
electron degeneracy. Similarly, the He layer of the WD model shown in Figure~43 of \mesathree\   is
partially degenerate, and hence the driving forces for diffusion
should be modified in this region.

\begin{figure}
\begin{center}
\includegraphics[width=\apjcolwidth]{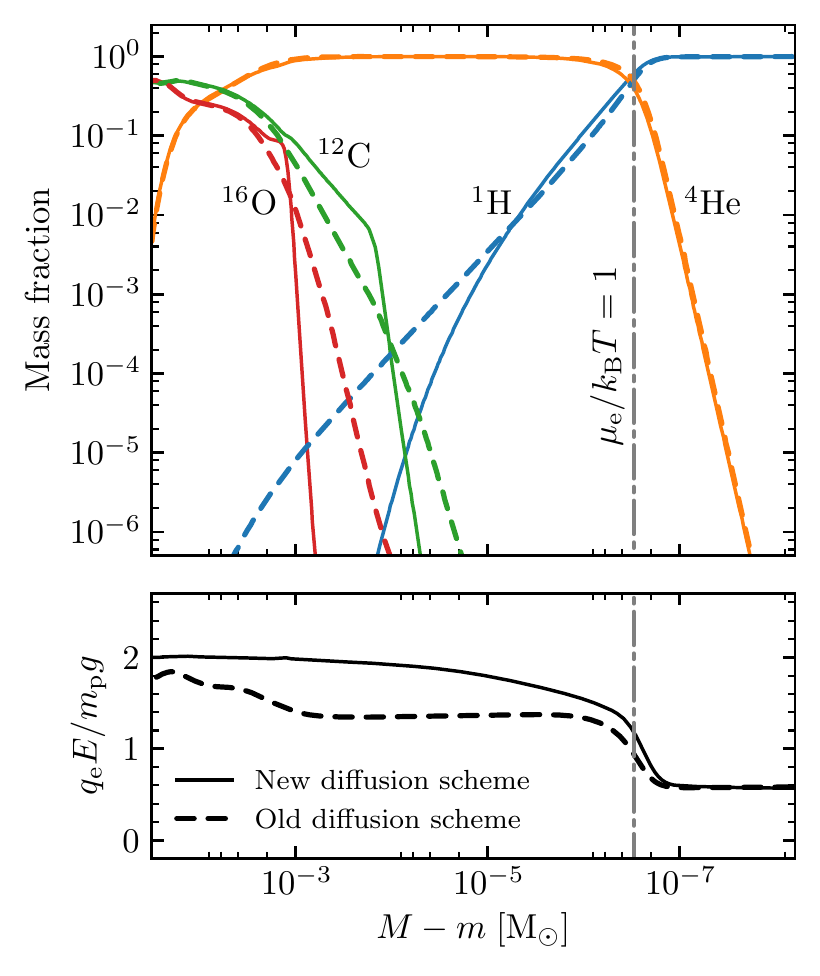}
\caption{
Abundance profiles in ${0.6 \, \Msun}$ \mesa\ WD models at
${T_{\rm eff} = 5,000 \, \rm K}$ after evolving for
$4 \, \rm Gyr$ to approach diffusive equilibrium in the outer layers.
The old equations assume an ideal gas; the new equations
include the effects of electron degeneracy.
}
\label{fig:diff_equilibrium}
\end{center}
\end{figure}

For a fully-ionized isothermal ideal gas the electric field that serves as one of
the driving forces for diffusion in Equation~\eqref{eq:BurgersMomentum} takes the
form %\citep{ChangBildsten}
$\qe E = [A/(Z+1)] m_{\rm p} g$.
In contrast, in the limit of strong electron degeneracy, the electric
field approaches $\qe E = (A/ Z) m_{\rm p} g$.
When He is the background material, the electric-to-gravitational force ratio $\qe E/m_{\rm p} g$ increases from
$4/3$ to $2$. In this limit, any
trace isotopes with $A/Z = 2$ see no net sedimentation force ($Z\qe E - A
m_{\rm p} g = 0$), while H with $A/Z = 1$ sees a significant
upward sedimentation force ($Z\qe E - A m_{\rm p} g > 0$). This extra
buoyant force on H in a degenerate He background pushes the diffusive tail further toward the surface relative to the ideal-gas case,
as shown in Figure~\ref{fig:diff_equilibrium}.
With the proper handling of electron degeneracy described in
Section~\ref{s.diffusion_methods}, our \mesa\ models now
agree with the time-dependent diffusion models shown in 
Figure~18 of \cite{Althaus03}.

\subsubsection{Radiative Levitation}\label{s.levitation}

Radiative levitation is included as an optional extra
term. The Burgers equations are
modified with a  extra forcing term by taking
${\rho_s g \to \rho_s(g - g_{{\rm rad},s})}$, as shown in
Equation~(99) of \mesathree. Our implementation continues to follow
\cite{Hu11} but no longer employs their matrix
structure for the Burgers equations;
details of how the $g_{{\rm rad},s}$ terms are handled with
the updated diffusion schemes can be found in
Appendix~\ref{app.diffusion}.

\subsection{Updated Diffusion Coefficients}
\label{s.coefficients}

The \cite{Paquette86coeff} diffusion coefficients have served
as the standard for stellar diffusion problems. The scattering
cross-sections for these coefficients are calculated using a screened
Coulomb potential
\begin{equation}
\label{eq:screened_coulomb}
V_{12}(r) = \frac {\bar{Z}_1 \bar{Z}_2 \qe^2} r \exp(-\lambda/r)~,
\end{equation}
with the screening length chosen as
$\lambda = \text{max}(\lambda_{\rm D},\bar a_{\rm i})$, where
$\lambda_{\rm D}$ is the Debye length,
$\bar a_{\rm i} = (3/4 \pi n_{\rm i})^{1/3}$ is the average interionic
distance, and $n_{\rm i}$ is the ion density.  This choice is a crude
but effective way to handle the strongly coupled regime; as shown in
\mesathree, this yields reasonable agreement with diffusion
coefficients calculated from molecular dynamics.

\cite{Stanton16} provide updated calculations of
collision integrals for screened Coulomb interactions and suggest improvements to the treatment of screening length. They
provide fitting functions and tables that can be used with any choice
of screening length. In \mesa\ we follow their
suggested screening prescription.
The electron screening length is given by a Thomas-Fermi approximation
that accounts for non-relativistic degeneracy:
\begin{equation}
\lambda_{\rm e} = \left( \frac{4 \pi \qe^2 n_{\rm e}}{\sqrt{(k_{\rm B} T)^2
+ \left(\frac 2 3 E_{\rm F}\right)^2}} \right)^{-1/2},
\end{equation}
where $E_{\rm F} = \hbar^2(3 \pi^2 n_{\rm e})^{2/3}/2m_{\rm e}$ is the electron Fermi energy.
The direct inclusion of degeneracy increases $\lambda_{\rm e}$.
The ion screening lengths are the Debye lengths for each species,
\begin{equation}
\lambda_i = \left( \frac{4 \pi \bar{Z}_i^2 \qe^2 n_i}{k_{\rm B} T} \right)^{-1/2}.
\end{equation}
To prevent ions from screening below the inter-ionic spacing, \cite{Stanton16} introduce an
approximate ion-sphere for each species
$a_i \equiv ( 3 \bar{Z}_i/4 \pi n_{\rm e})^{1/3}$,
and define an ion-sphere coupling parameter
\begin{equation}
\Gamma_{i} \equiv \frac{(\bar{Z}_i \qe)^2}{a_i k_{\rm B} T}.
\end{equation}
Their net effective screening length is then
\begin{equation}
\lambda_{\rm eff} \equiv \left[\frac{1}{\lambda_{\rm e}^2} +
  \sum_{i}\frac{1}{\lambda_i^2} \left( \frac 1 {1 + 3 \Gamma_i} \right) \right]^{-1/2}.
\end{equation}

This construction enforces
a minimum on the screening length at
approximately the ion-sphere radius $a_i$ for each species, 
similar to the strict
minimum at $\bar a_{\rm i}$ set by \cite{Paquette86coeff}.
\cite{Stanton16} point out that this adjustment to the ion screening
length is physically motivated by the ion pair distribution functions
in a strongly coupled plasma, where the occupation probability within the
ion-sphere radius is negligible, and hence no ions are present to
provide screening beneath that cutoff.
The proper handling of degeneracy in the electron screening length
makes it unnecessary to impose any particular minimum there, so
there is no longer any ad hoc appeal to a universal minimum
screening length.

For repulsive Coulomb potentials of the form in
Equation~\eqref{eq:screened_coulomb}, \cite{Stanton16} provide fits
and tables of collision integrals and coefficients that we now use to
calculate the resistance coefficients $K_{st}$ for inclusion in the
Burgers equations in \mesa. They do not provide fits
for attractive potentials, and \cite{Paquette86coeff} note that
interactions with these potentials behave significantly differently
from those with repulsive potentials when screened. Hence, \MESA\ 
continues to use the \cite{Paquette86coeff} coefficients for
electron-ion terms, and adopts \cite{Stanton16} for all ion-ion coefficients.
In any case,
it is evident from Equation~(94) in \mesathree\ that the resistance
coefficients approximately follow ${K_{st} \propto \mu_{st}^{1/2}}$,
where $\mu_{st}$ is
the reduced mass of particles $s$ and $t$; so, electron-ion resistance
coefficients are generally negligible compared to the ion-ion terms.

The calculations of \cite{Paquette86coeff}
overestimate the electron-ion resistance
coefficients in the case where electrons are degenerate. This is
because diffusion and resistance coefficients are generally calculated
assuming that the velocity distributions of all particles are
Maxwellian, and the coefficients roughly scale as $K_{st} \propto
v_s^{-2} v_t^{-2}$. When the electrons become degenerate, their
characteristic kinetic energies are of order $E_{\rm F} \gg k_{\rm B}
T$, and so their velocity distribution skews toward larger
velocities. This results in smaller resistance coefficients $K_{st}$,
overestimating the impact of electron-ion
drag. However, the overestimate results in
coefficients that remain negligible compared to ion-ion terms, and no
attempt is made to correct it in \mesa.

For repulsive potentials,
the coefficients from \cite{Stanton16} generally agree with
those of \cite{Paquette86coeff} to within a few percent.   In
strongly coupled WD interiors the \cite{Stanton16} coefficients lead to $\sim 10 \%$ shorter diffusion timescales
due to a screening length that is allowed to be
somewhat smaller than the minimum value imposed by
\cite{Paquette86coeff}:
$\lambda_{\rm eff} < \bar a_{\rm i}$.  Future prospects for further improvements to diffusion coefficients
include the recent progress on effective potential methods from
\cite{Daligault16} and \cite{Shaffer17}.

\subsection{Diffusion-Induced Flashes on He WDs}
\label{s.hewdflash}

Diffusion-induced H shell flashes on low-mass ($M \la 0.4 \, \Msun$)
He WDs are known to alter their cooling times \citep{Althaus00,Althaus01}
and seismic properties \citep{Althaus13}.
\cite{Istrate16seis,Istrate16mod} use
\mesa\ to model this process, generating tables of cooling
timescales and comparing \mesa\ models with those of
\cite{Althaus13}.

\begin{figure}
\begin{center}
\includegraphics[width=\apjcolwidth]{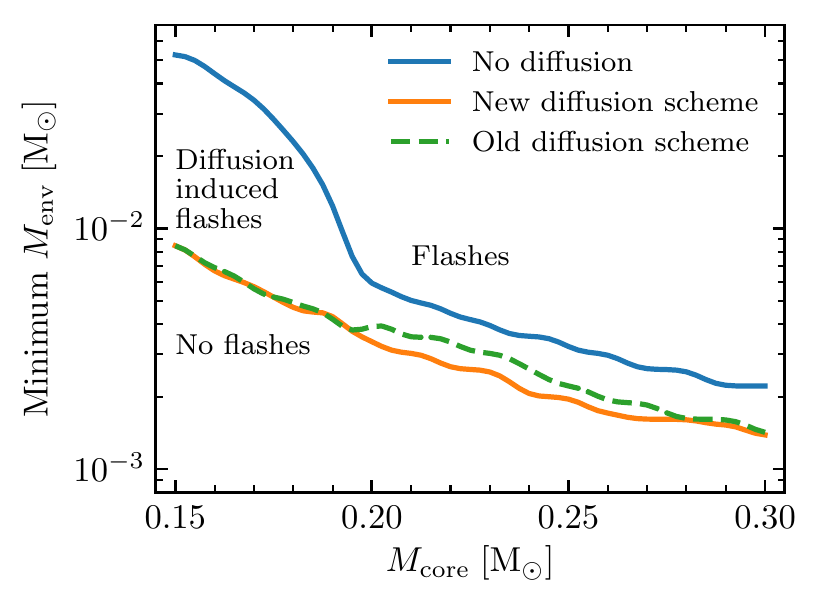}
\caption{
Minimum envelope mass $M_{\rm env}$ for which a H shell flash occurs on
a He WD core mass $M_{\rm core}$ for $\Zmet = 0.02$ \mesa\ models
with and without diffusion. The regime for the phenomenon of diffusion-induced
flashes lies between the boundaries for models with and without diffusion.
}
\label{fig:HeWDflash}
\end{center}
\end{figure}

Figure~\ref{fig:HeWDflash} shows an exploration 
of the H shell flash domain for a large grid of $\Zmet = 0.02$
\mesa\ models over a range of He-core and H-envelope masses. Here the envelope mass is
defined as the total mass of H-rich material ($\Xhyd > 0.01$) at
the surface at the beginning of the WD cooling track. Lines
show the minimum envelope masses for which H shell flashes
occur given various diffusion prescriptions.

For a given core mass, there is a range of
envelope masses that exhibit shell flashes only
if diffusion is included, but this range depends on the diffusion
prescription. The two lower lines for models including diffusion in
Figure~\ref{fig:HeWDflash} differ only in the handling of
electron degeneracy in the diffusion scheme.
This illustrates the importance of properly handling
degeneracy as described in Section~\ref{s.diffusion_methods},
since the diffusion-induced
flashes are typically ignited by CNO burning in the diffusive tail of
H that reaches into the partially degenerate He layers.
WDs in this mass range often experience cycles of many H
flashes, depleting H incrementally until insufficient H
remains to ignite another flash. The disagreement between
diffusion prescriptions on the minimum envelope mass for flashes is
therefore significant, as this will determine
the total number of flashes and final H mass that sets the
ultimate cooling timescale for an object.

To explore the full range of parameters presented in
Figure~\ref{fig:HeWDflash},
our WD models were built by artificially stripping the H
envelope down to a specific mass coordinate above the He core of
a $1.0 \, \Msun$ model ascending the RGB.  For a discussion of
\mesa\ models including proto-WD formation and the
resulting H envelope masses, see \cite{Istrate16mod}.

\subsection{Heating from $^{22} \rm Ne$ Settling}\label{s.heating}

In the strongly degenerate limit, 
$\qe E/m_{\rm p} g \approx 2$ for C/O WD cores.
For an isotope where $A/Z \neq 2$,
the electric and gravitational fields result in a net
force that drives diffusion. For $^{22}{\rm Ne}$ in
cooling WD interiors, this force is 
$F = Z \qe E - A m_{\rm p} g \approx - 2 m_{\rm p} g$, causing $^{22}{\rm Ne}$ to sediment toward the
center and deposit energy as it moves deeper into the
gravitational potential
\citep{BildstenHall,DeloyeBildsten,Berro08,Berro10}. This
heating can prolong the WD cooling timescale, especially at late
times when the WD is very dim and radiates away the energy slowly.
This effect may be especially important for explaining WD luminosity
functions in old and metal-rich open clusters such as NGC~6791, where
abundant $^{22} \rm Ne$ is available in WD interiors to provide
heating. 

\mesa\ now offers an option to include this heating term in the energy equation (see Section~\ref{sec:grav-settl})  when
diffusion is enabled.
The specific rate at which energy is deposited is
\begin{equation}
\begin{aligned}
\epsilon_{\rm 22} 
= \frac{ |F| v_{22}} {(A m_{\rm p})/X_{22}}
\label{eq:ne22_heat}
= \left( 22 m_{\rm p} g - 10 \qe E  \right) \frac{X_{22} v_{22}}{22 m_{\rm p}}.
\end{aligned}
\end{equation}
The $^{22}{\rm Ne}$ diffusion velocity ($v_{22}$)
and electric field \replaced{is}{are} calculated in the diffusion routine and
then used to evaluate the above heating term.
Note that the updates to diffusion described in Section~\ref{s.diffusion_methods}
are essential for correctly calculating both the diffusion velocity
and magnitude of the driving force in the degenerate interior of the
WD.

\begin{figure}
\centering
\includegraphics{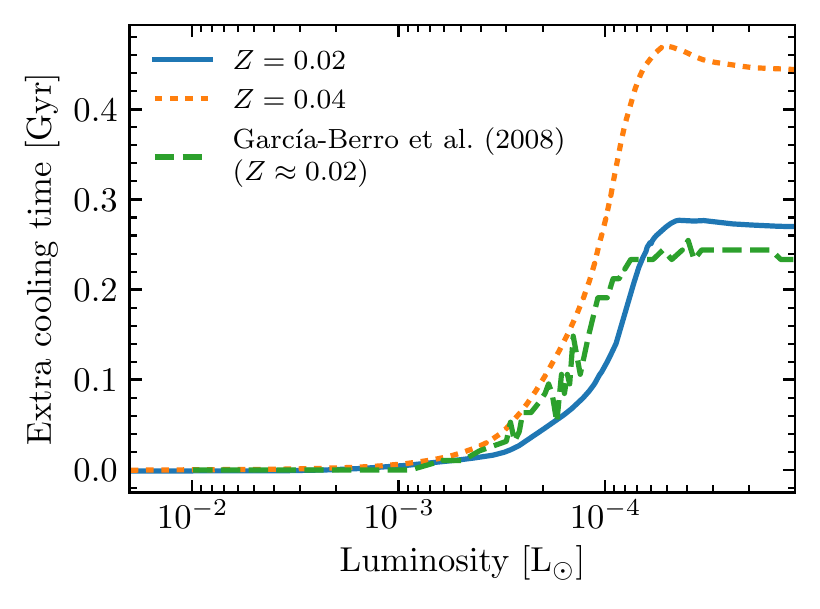}
\caption{Extra cooling time required to reach a given luminosity for
  $0.6 \, \Msun$ WD models including heating from $^{22}\rm Ne$
  settling, relative to models neglecting this heating. 
  \added{For comparison, we also show a result from \cite{Berro08} for a $0.6
  \, \Msun$ WD with an Oxygen-dominated core composition.}
  Figure~\ref{fig:wd_cool} shows the same quantity including other physical processes such as crystallization for the same
  ${\Zmet = 0.02}$ WD model shown here.
}
\label{fig:Ne22}
\end{figure}

Figure~\ref{fig:Ne22} shows the delay in WD cooling from introducing
$\epsilon_{22}$ into $0.6 \, \Msun$ models. These models turn off
diffusion for $\Gamma > 175$, so $\epsilon_{22}$ is only active in
material for which crystallization has not yet occurred. The
time-delays shown in Figure~\ref{fig:Ne22} are in good agreement with
those shown by \cite{DeloyeBildsten} and \cite{Berro08} for comparable cases.

%%% Local Variables:
%%% mode: latex
%%% TeX-master: "paper"
%%% End:

\section{Implicit Hydrodynamics}\label{s.hydro}

In \mesathree\ we describe implicit shock-capturing hydrodynamics
capabilities based on the use of an artificial viscosity.
We now add an option for using an approximate Riemann
solver, the HLLC (Harten-Lax-van Leer-Contact) solver introduced by
\citet{Toro1994}. (See also \citealt{Batten1997} for an early implicit
implementation of HLLC.)  The HLLC method
provides improved shock capturing and energy conservation by avoiding
the need for artificial viscosity.  However, the methods presented in
\mesathree\ are still included in \MESA\ so that users
may continue to apply them.

\subsection{Implementation of HLLC}
\label{s.hllc}

Accurate shock-capturing methods evaluate the flux of hydrodynamical
conserved quantities by extrapolating the solution on each interface
between zones over the course of the timestep.  The different methods
for projecting the solution into the future are known as different
``Riemann solvers''.  HLLC is designed to accurately capture
the evolution of contact discontinuities. When implemented on a
Lagrangian grid, HLLC is able to evolve purely advective flows without
any contact smearing \citep{Cheng2007, Duffell2011, Cheng2012,
  Cheng2014}.

\mesaone\ and \mesathree\ discussed the evolution of a velocity
variable $v$, defined at cell faces.  When using HLLC,
\MESA\ instead evolves a cell-centered velocity $u$.

\begin{figure}[htb!]
\includegraphics[width=\columnwidth]{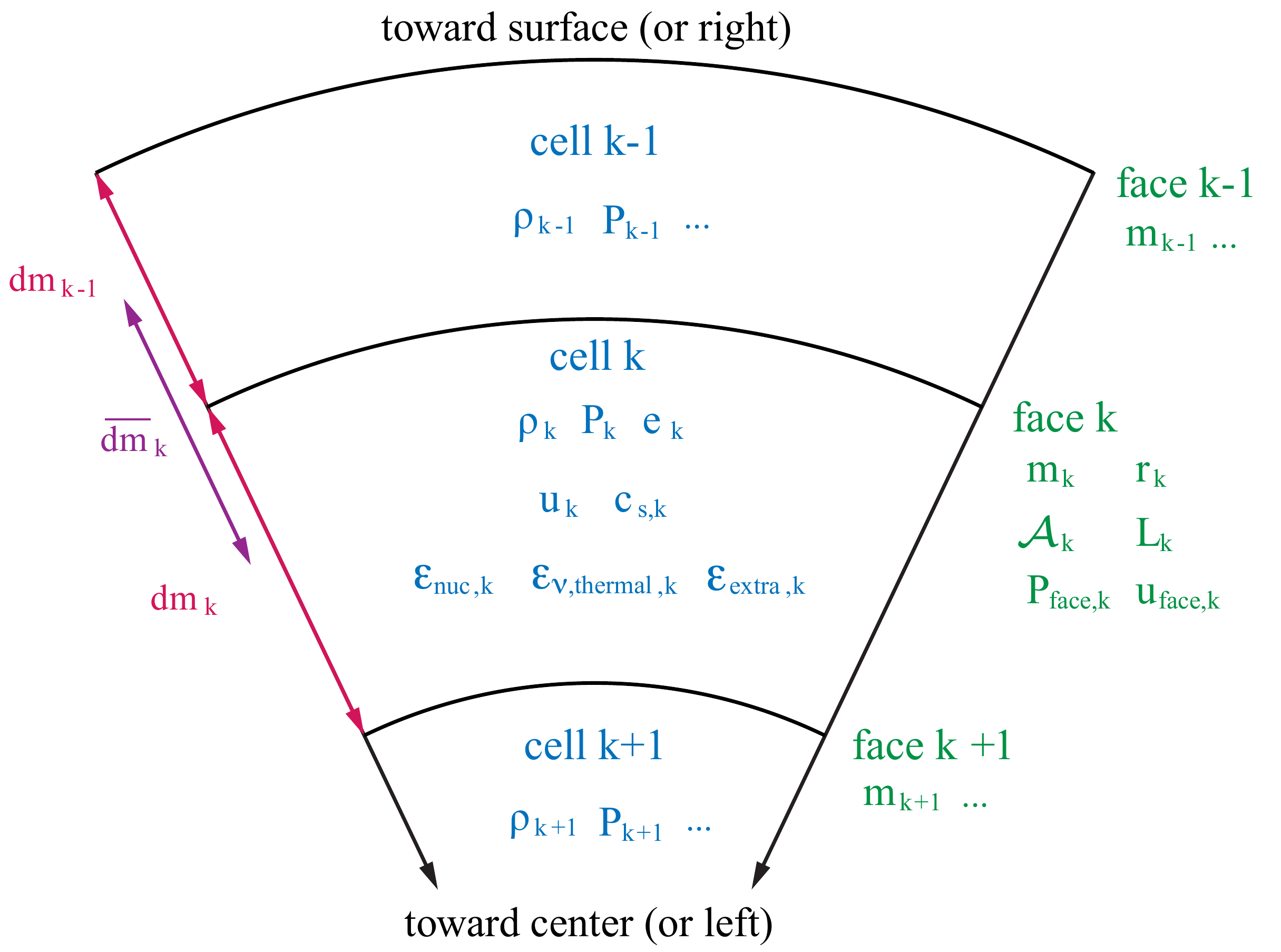}
\caption{Cell and face variables relevant for
  hydrodynamics in \MESA\ when using HLLC.}
\label{f.structure}
\centering
\end{figure}

We solve a Riemann problem at the cell face with
index $k$.  The cell to the left (toward the center) is the cell with
index $k$; the cell to the right (toward the surface) is the cell with
index $k-1$.  The cell face radius is $r_k$.  The mass contained
within an individual cell is $dm_k$.  The mass enclosed from the
center of the star to the cell face is $m_k$.  For convenience, we
define the face area as $\area_k = 4 \pi r_k^2$.  Thermodynamic variables
(e.g.,~$P_k$, $\rho_k$) are defined at cell centers by mass.
Figure~\ref{f.structure} shows the layout of cells.

The calculation begins by making estimates for the density and velocity
at the left and right of the face.
Explicit codes sometimes use multipoint polynomial interpolation based
on values in neighboring cells to improve the reconstruction of the
values at the face.  However for an implicit code such as \mesa, that
would introduce dependencies in the partial derivatives for the
Jacobian that would violate the necessary block tridiagonal
structure (see Appendix~B in \mesatwo).
To avoid this problem, we use the cell center density and velocity
alone to estimate the values at the edges of that cell.  The variables
for the left and right values are named relative to the edge rather
than the cell, that is
\begin{equation}
\begin{aligned}
\rhoL &= \rho_k \quad & \rhoR &= \rho_{k-1} \\
\uL &= u_k      \quad &  \uR &= u_{k-1} ~.
\end{aligned}
\end{equation}
This choice limits the solution to be first-order accurate in
space.

Using an approach similar to \cite{Kappeli2014},
we reconstruct the pressure at the faces
assuming hydrostatic equilibrium.  The pressure derivative implied by
hydrostatic equilibrium at the face is
\begin{equation}
  \left(\frac{\dif P}{\dif m}\right)_{\mathrm{hse}} = -\frac{Gm_k}{\area_k r_k^2}~,
  \label{e.hse-face}
\end{equation}
and we reconstruct the pressure to the left and right of the face
\begin{align}
\begin{split}
  \PL &= P_k + \frac{dm_{k}}{2}\left(\frac{\dif P}{\dif m}\right)_{\mathrm{hse}}~, \\
  \PR &= P_{k-1} - \frac{dm_{k-1}}{2}\left(\frac{\dif P}{\dif m}\right)_{\mathrm{hse}} ~.
\end{split}
\end{align}
This choice improves the timescale over which 
hydrostatic equilibrium can be maintained when using HLLC, and
facilitates the process of switching from a
hydrostatic model to one in which a velocity variable is evolved.

The 1D Lagrangian context makes the implementation of HLLC
straightforward. In a Lagrangian code like \MESA\ there is
no mass flux across cell faces.  In hydrodynamics, there is no mass
flux across a contact discontinuity. HLLC
includes the contact wave, so we simply associate the contact
wave with the cell face.\footnote{In Section~\ref{s.RTI}, where we
  consider the effects of mass diffusion, we will need to slightly
  revise this association.}  As given by \citet{Toro2009}, the HLLC
estimate of the contact wave speed is
\begin{equation}
  \Ss =
  \frac{\uR\rhoR(\Sr - \uR) + \uL \rhoL (\uL - \Sl) + (\PL - \PR)}{\rhoR (\Sr - \uR) + \rhoL (\uL - \Sl)},
\label{eq:wave-speed}
\end{equation}
and hence $\uface$=$\Ss$.  Likewise, the pressure at the cell face is the pressure at the contact wave, $\Pface$=$\Ps$, where
\begin{equation}
  \begin{split}
  \Ps = \frac{1}{2}
   &\left[\left(\rhoR(\uR-\Sr)(\uR-\Ss) + \PR\right) \right. \\
  +&\left.\left(\rhoL(\uL-\Sl)(\uL-\Ss) + \PL\right) \right] ~,
\end{split}
\end{equation}
and $\Sl$ and $\Sr$ are the fastest wavespeeds moving to the left
and right, respectively.  To evaluate these, we assume the simple and most conservative bounds on the signal velocities, 
\begin{equation}
\begin{split}
  \Sl & = \min\left(\uL - \csL, \uR - \csR\right)~, \\
  \Sr & = \max\left({\uL + \csL, \uR + \csR}\right) ~,
\end{split}
\end{equation}
where \csL\ and \csR\ are the sound speeds on the left and right sides
of the cell boundary, respectively.
Having obtained values for \uface\ and \Pface, we now formulate
the versions of the equations used when HLLC is enabled.

In the Lagrangian picture, the cell boundaries move with the fluid
velocity, such that the \replaced{fluences}{net fluxes} for mass, momentum, and energy
from cell $k$ to cell $k-1$ are extremely simple 
\citep{Cheng2014}
and given by
\begin{equation}
\begin{split}
  F_{\rho,k} &= 0~,\\
  F_{p,k} &= \area_k P_{\mathrm{face},k}~,  \\
  F_{e,k} &= \area_k P_{\mathrm{face},k} u_{\mathrm{face}, k} + L_k~.
  \label{e.energy-flux1}
\end{split}
\end{equation}
The $ L_k$ term in the energy \replaced{fluence}{flux} does not come from
the solution of the Riemann problem, but from the fact that \MESA\
also evolves a luminosity variable that reflects the radiative or
convective transport of energy.

The finite volume form of the mass conservation equation remains the same
as that given in \mesaone,
\begin{equation}
  \ln r_k = \frac{1}{3} \ln \left [ r_{k+1}^3 + \frac{3}{4\pi}\frac{dm_k}{\rho_k} \right ]~.
\end{equation}
However, the equation for the radius (Equation 29 in \mesathree) has changed.  The new equation for the radius is
\begin{equation}
  r_k = r_{{\rm start},k} + u_{{\rm face},k} \dt ~,
\end{equation}
where $\dt$ is the timestep.
For numerical precision, we re-write this as
\begin{equation}
  \underbrace{ \exp \left( \ln r_k - \ln r_{{\rm start},k} \right ) - 1}_{\mathtt{expm}} = \frac{u_{{\rm face},k} \dt}{r_{{\rm start},k}} ~,
\end{equation}
where this recasting allows use of \texttt{crlibm} \citep[][see also \mesathree]{de-dinechin_2007_aa}
function \texttt{expm} to evaluate the function $\exp(x)-1$ to machine
precision (as indicated by the underbrace).

The local radial momentum equation for cell $k$ is
\begin{equation}
  \begin{split}
    \frac{u_k - u_{{\rm start},k}}{\dt} =
    &-\frac{1}{2}\left(\frac{G m_k}{r_k^2} + \frac{G m_{k+1}}{r_{k+1}^2}\right) \\
    &+ \frac{P_k}{dm_k} \left (\area_k - \area_{k+1}\right) \\
    &+ \frac{1}{dm_k} \left(F_{p,k+1} - F_{p,k}\right)~.
  \end{split}
  \label{e.momentum}
\end{equation}
On the right hand side, the first term is gravitational, the second is
a geometric source term that arises from putting the equation in
conservation-law form, and the final term is the momentum \replaced{fluence}{flux} in
and out of the cell found by HLLC.

The local total energy
conservation equation for cell $k$ is
\begin{equation}
  \begin{split}
    e_k - e_{{\rm start},k}
    +\frac{1}{2}\left(u_k^2 - u_{\mathrm{start},k}^2\right)
    -G\mC\left(\frac{1}{r_{\rm C}} - \frac{1}{r_{\rm C,start}}\right) = \\
    \dt \left[\frac{1}{dm_k} \left(F_{e,k+1} - F_{e,k}\right)
      + \epsilon_{{\rm nuc},k}
      - \epsilon_{\nu,k}
      + \epsilon_{{\rm extra},k}\right].
  \end{split}
  \label{e.energy}
\end{equation}
(See Section~\ref{s.implicit_hydro} for a discussion of how this
energy accounting is related to that typically used in stellar
evolution calculations.)
We define the cell center quantities $\mC$ and $r_{\rm C}$ to be
$r$ and $m$ at the center of mass of the cell.  The terms on
the left split the local total energy into internal, kinetic, and
potential components.  The right side gives the energy in and
out of the cell and the energy sources and sinks in the cell.  Energy loss
from neutrinos due to nuclear reactions is already subtracted from the nuclear
burning term, $\epsilon_{\rm nuc}$, so only the neutrino energy
loss rate from thermal processes, $\epsilon_\nu$, is explicitly
accounted for in Equation~\eqref{e.energy}. Other processes are
accounted for via $\epsilon_{\rm extra}$.

As in \mesaone, the temperature differences of interior cells $T_k$ are set by
energy transport across boundaries,
\begin{equation}
T_{k-1} - T_k = {\overline{dm}_k} \left [ \nabla_{T,k} \left(\frac{\dif P}{\dif m}\right)_{\rm hse} 
\frac{\overline{T}_k}{\overline{P}_k}
\right ]
\enskip ,
\label{e.temperature}
\end{equation}
where $\nabla_{T,k}$ is provided by $\MESA$ module $\mlt$ (see
Section~5.1 in \mesaone) and the overbars indicate quantities at
the cell faces (see Figure~\ref{f.structure}).  This equation
  relates temperatures of neighboring cells; the
  actual temperature in each cell is then fixed by a surface boundary condition.

\MESA's HLLC includes the effects of rotation in
the shellular approximation (see \mesatwo, Section~6.1) and can also
include a post-Newtonian correction to the gravitational force. (For an
example application to neutron star envelopes, see \mesathree,
Section~5.3).  These capabilities require modifications to the momentum
equation.  In both cases, they can be treated as a rescaling of the
local gravitational constant $G \to f G$.  In the case of rotation,
the rescaling factor is $f_P$ (\mesatwo, Equation~23).  In the
post-Newtonian case,
it is $(1-2 G m/(r c^2))^{-1/2}$.  Therefore, when either of these is
used with the hydrodynamics capabilities described in this
section, the rescaling is applied to the $G$ in pressure
reconstruction (Equation~\ref{e.hse-face}) and separately to each $G$
(for cell $k$ and $k+1$) in the momentum equation
(Equation~\ref{e.momentum}).

\subsection{Mesh Refinement}
\label{s.amr}
During a typical stellar evolution run,
\MESA\ controls its meshing using ``mesh functions'' that limit the
maximum allowed change of various quantities between adjacent cells
(see Section 6.5 in \mesaone\ and Section B.4 in \mesatwo). 
With HLLC, the criteria to split or merge cells are written
solely in terms of the radial coordinate in order to simplify the
adjustments to the mesh in response to large changes in density
before and after a shock.  Cells split when they decompress
enough that their radial extent becomes too large, and they merge with
a neighbor when they compress enough that their radial extent becomes too small.

The refinement criteria can use either linear ($x$=$r$)
or logarithmic ($x$=$\ln r$) radius.  The user selects a target number
of cells, $\Ncells$.  \MESA\ translates this into a target cell size,
$dx_{\mathrm{target}}$=$(x_{\mathrm{surface}} - x_{\mathrm{center}}) /
\Ncells$.  A cell is considered too large if
$dx_k/dx_{\mathrm{target}} > f_{\mathrm{long}}$ and a cell is
considered too small if
$dx_{\mathrm{target}}/dx_k > f_{\mathrm{short}}$.\footnote{By default
  $\Ncells = 1000$, $f_{\mathrm{short}} = 4.0$ and
  $f_{\mathrm{long}} = 1.5$, but these parameters are configurable at
  run-time.}  The refinement then proceeds iteratively.  At each
iteration, \MESA\ selects the smallest and largest cells.  If the
largest cell is too large, it is split.  If the smallest cell is too
small, it is merged unless the magnitude of the difference between its
velocity and that of either neighbor is a significant
fraction of the local sound speed: this prevents merging in the immediate
vicinity of the shock where there are sharp jumps in velocity.  The refinement proceeds up to some maximum number of
iterations, though in practice the procedure typically stops before then because
no more cells satisfy the criteria to be split or merged.

A cell merges with its smaller neighbor, unless they have a different most-abundant
chemical species, in which case the cell merges with the other neighbor instead.  When a cell is split,
differences in quantities such as density and
chemical abundances between the two child cells are determined by
interpolation from the neighboring cells.

\subsection{Time Resolution}\label{s:hydro_time}

Since the hydrodynamics equations are being solved implicitly, \MESA\
is not subject to the Courant-Friedrichs-Lewy (CFL) timestep condition
for numerical stability.  The size of the \MESA\ timestep is instead
limited by the restrictions on the allowed changes in the structure of
the star.  The usual timestep controls continue to apply.

While numerical stability does not require the restrictive CFL
timestep condition, the choice of timestep does affect the accuracy of
the solution.  A CFL-like limit is often also applied because it can
be a convenient additional way to restrict timesteps along with the
other options.  Such a restriction allows for well-converged
solutions. The timestep can be limited
by the requirement that
\begin{equation}
  \label{e.CFL-like}
  \min \left\{\frac{dr_k}{|u_k| + \cs{,k}}\right\} < f_t \times \dt ~,
\end{equation}
where $f_t$ is a user parameter.
Unlike in an explicit code where a similar minimum must be
evaluated over all cells, in \MESA\ the minimum is taken only over
cells for which
\begin{equation}
  \frac{\max\{|u_k - u_{k+1}|, |u_k - u_{k-1}|\}}{\cs{,\rm max}} < f_u~,
\end{equation}
where $ \cs{,\rm max}$ is the maximum evaluated over nearby cells, and
$f_u$ is a user parameter.\footnote{The value of $f_t$ is similar to
  the values of a CFL parameter in an explicit code, while $f_u$ in
  the examples is typically a small value like $10^{-2}$. The
  description of these limits is
  schematic and the reader is referred to the source code
  for the precise implementation details.}
This means that only regions near the shock front limit the timestep.
The option for additional limitations on where this condition is
evaluated (e.g.,~in mass) are provided.

\subsection{Hydrodynamic Test Problems}

In order to test the HLLC implementation, comparisons 
are now made to problems with known solutions.

\begin{figure}[!htb]
\includegraphics[width=\columnwidth]{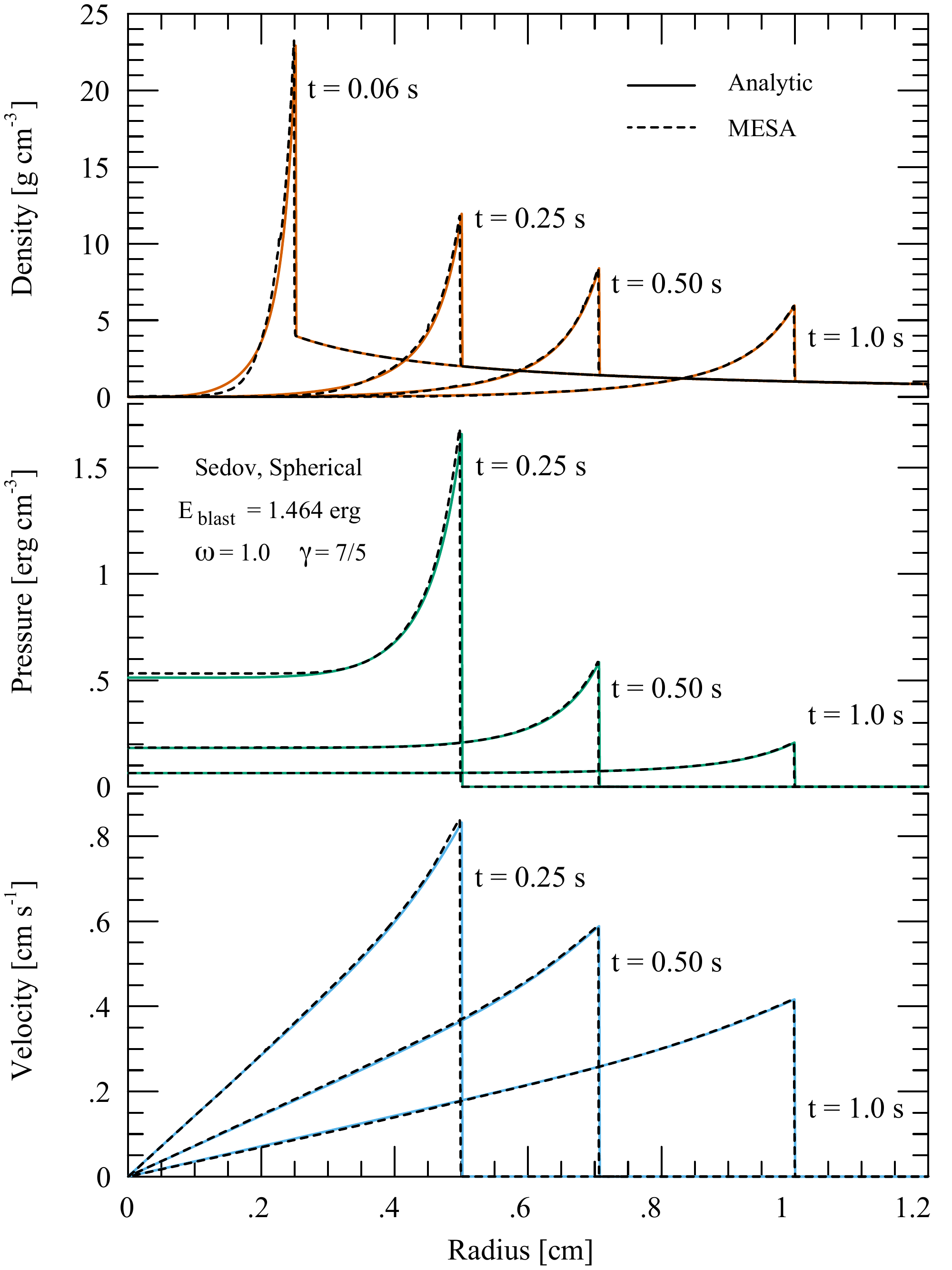}
\caption{
Sedov blast wave density, pressure and velocity profiles at the
labeled times. The analytic (black) and \MESA\ (colored) curves 
show solutions for a shock propagating down a $\rho=\rho_0 r^{-1}$
density profile with an adiabatic index $\gamma$=7/5.
Deviations from the analytic solutions are
$\lesssim$2\%.
}
\label{f.sedov3panel}
\centering
\end{figure}

\begin{figure}[!htb]
\includegraphics[width=\columnwidth]{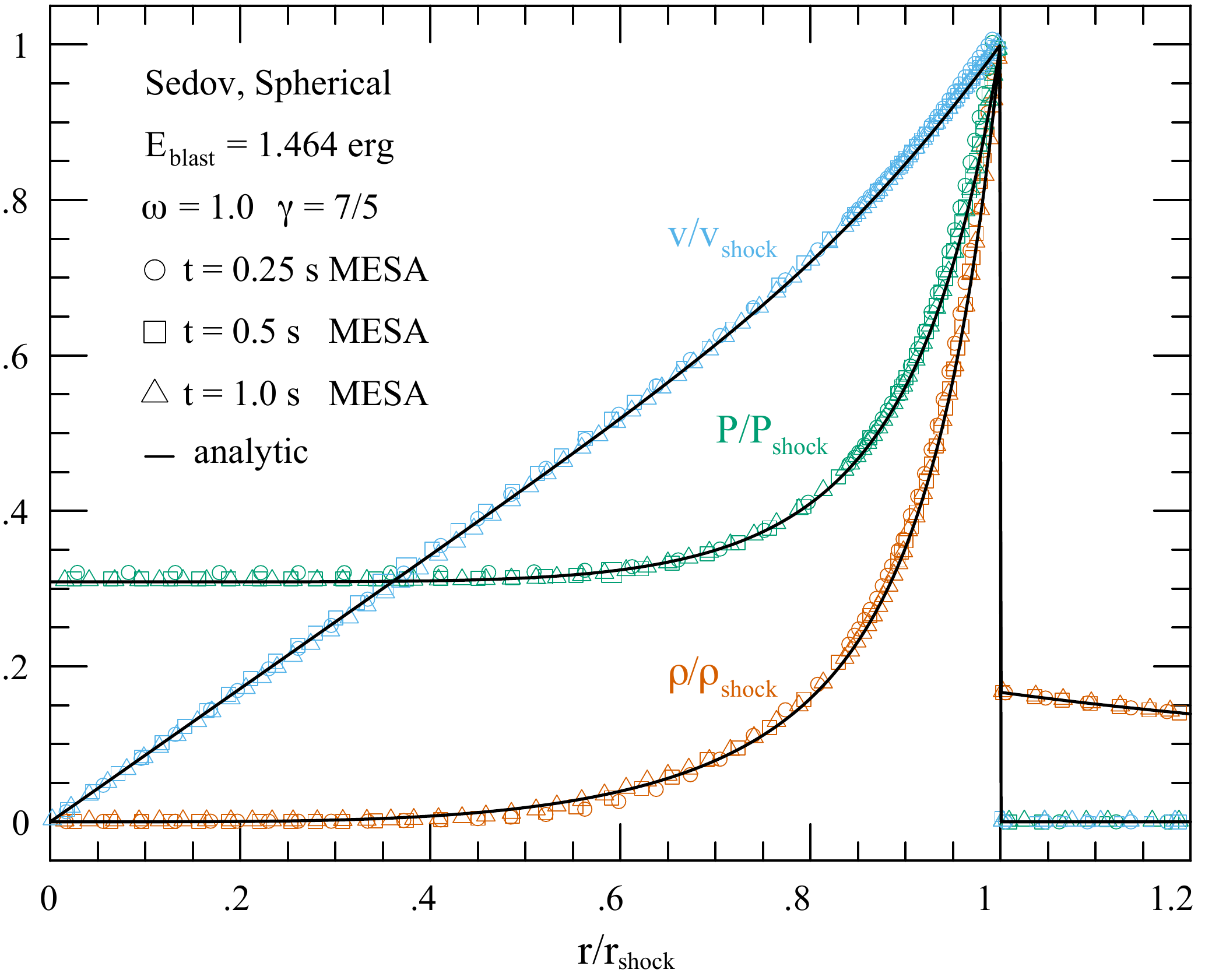}
\caption{
Sedov blast wave self-similarity of the analytic (black curves) and
\MESA\ (colored symbols) solutions. Scaled velocity $v/v_{\rm shock}$,
pressure $P/P_{\rm shock}$, and density $\rho/\rho_{\rm shock}$
profiles for a shock propagating down a $\rho=\rho_0 r^{-1}$ density
profile at the three different times are overlaid. Symbols for each
epoch mark cell locations.  Deviations from the analytic
self-similar solutions are $\lesssim$2\%. 
}
\label{f.sedov_selfsim}
\centering
\end{figure}

\begin{figure}[!htb]
\includegraphics[width=\columnwidth]{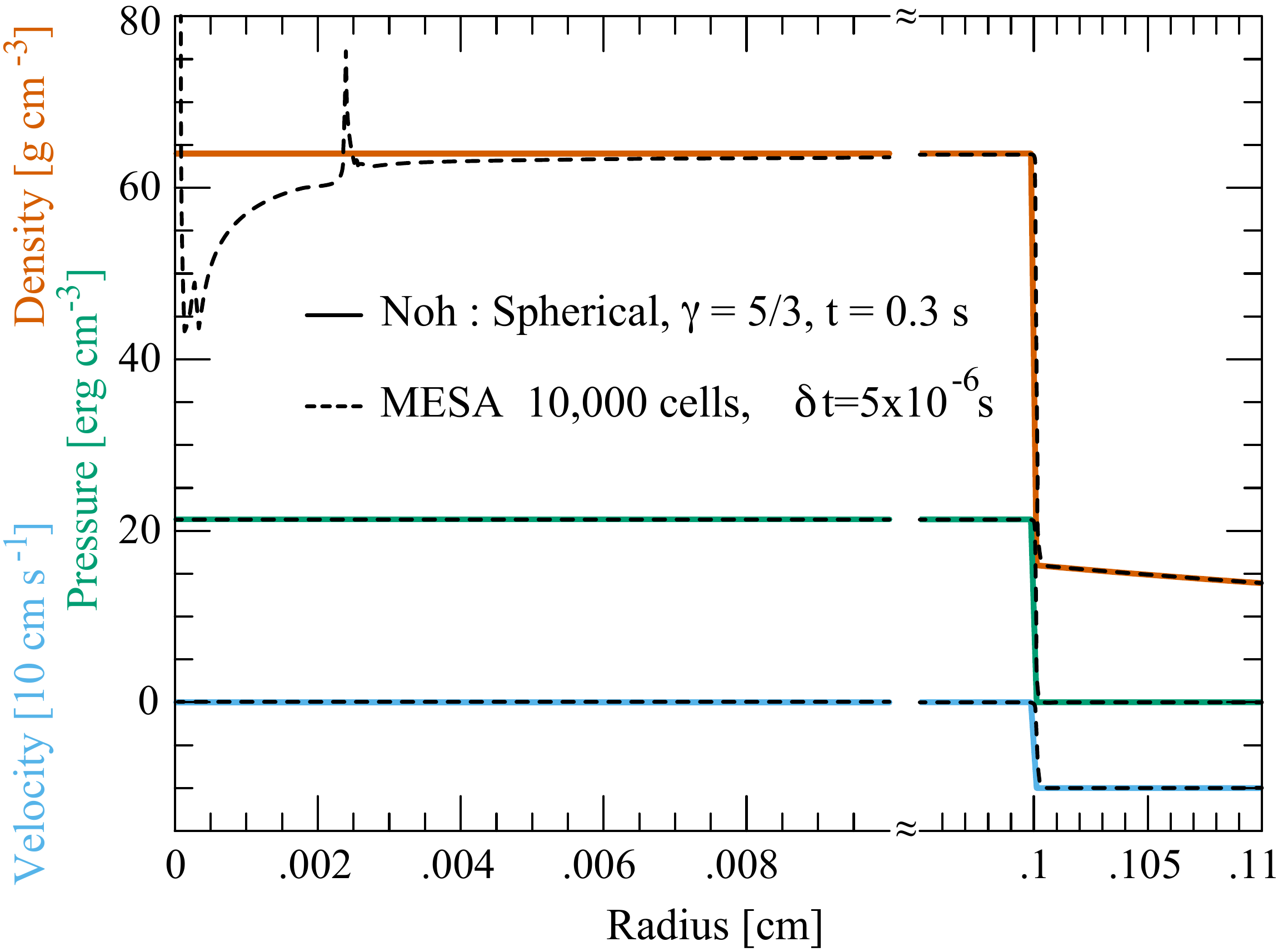}
\caption{
Analytic (colored solid) and \MESA\ (black dashed) solutions for the density, pressure, and 
velocity at $t$=0.3\,s in the Noh problem. Disagreements near the
center are due to wall-heating as discussed in the text.}
\label{f.noh}
\centering
\end{figure}

\subsubsection{Sedov Blast Wave}
\label{s.tests.sedov}

In the
Sedov blast wave problem, an energy
$\eblast$ is deposited at the origin at time zero in a domain with a
non-uniform density profile $\rho=\rho_0 r^{-\omega}$, where $\rho_0$
and $\omega$ are constants. We assume an ideal-gas EOS
with a constant adiabatic index $\gamma$, that is
$P = (\gamma - 1) \rho e$.

Generation of numerical Sedov solutions 
is discussed in \citet{kamm_2007_aa}.
A constant initial density profile, $\omega$=0, is frequently used in
verification tests \citep[e.g.,][]{gehmeyr_1994_aa,fryxell_2000_aa}.
Although a power-law
initial density profile is more challenging for verification
studies, we explore such a profile because
density gradients are prevalent in astrophysics.

To model a shock propagating down a linear density gradient in spherical
geometry we set $\omega$=1, $\gamma$=7/5, $\rho_0$=1\,g\,cm$^{-3}$, and
$P_0$=0\,erg\,cm$^{-3}$ in the analytic solution;
while we set $P_0$=10$^{-6}$~erg~cm$^{-3}$ in \MESA\ 
as a stable numerical approximation to zero pressure.
The initial blast energy, $\eblast$=1.464 erg, is
determined by choosing that $r_{\rm shock}$=1 cm at $t$=1 s and then
calculating the Sedov energy integral.
Figure \ref{f.sedov3panel} shows the evolution of the density,
pressure, and velocity. As the shock propagates outward from the origin
these quantities monotonically decrease as mass is swept up by the
shock.  
The spherical Sedov problem admits a similarity solution.
Figure \ref{f.sedov_selfsim} demonstrates that \MESA\ maintains
the analytic self-similar profiles at different times.

\subsubsection{Noh Problem}
\label{s.tests.Noh}

\citet{noh_1987_aa} describes a standard verification problem
that tests the ability to transform kinetic energy into internal
energy, and the ability to follow supersonic flows.
A sphere of cold gas with an ideal-gas EOS and constant adiabatic index $\gamma$, that is
$P = (\gamma - 1) \rho e$, is initialized with a uniform,
radially inward speed of 1\,cm\,s$^{-1}$.
A shock forms at the origin and
propagates outward as the gas stagnates. For an initial gas density
$\rho_0$=1\,g\,cm$^{-3}$, the analytic solution in spherical geometry 
for $\gamma$=5/3 predicts a density 64\,g\,cm$^{-3}$ in the stagnated
gas.

Figure \ref{f.noh} shows the analytic and \MESA\
profiles for the density, pressure, and material speed at $t$=0.3 s.
Most implementations, including \MESA's, produce anomalous
``wall-heating'' near the origin \citep[although see][]{gehmeyr_1997_aa}.
As the shock forms at the origin the momentum equation tries to
establish the correct pressure level. However, numerical dissipation
generates additional entropy. The density near the origin drops
below the correct value to compensate for the excess internal energy
\citep[e.g.,][]{noh_1987_aa, rider_2000_aa}. Thus, the density profile
is altered near the origin while the pressure profile remains at the
correct constant level in the post-shock region.  

Figure \ref{f.noh} shows the analytic solution and \MESA \ solution
for a fixed timestep of $\delta t$=5$\times$10$^{-6}$\,s and 10,000 
cells.  Deviations from the analytic solutions are $\lesssim$1\%,
except for the density near the origin and the shock front.  A
convergence exercise with different fixed timesteps and spatial resolutions 
suggests spatial resolution is relatively more important in the \MESA\ solutions 
than temporal resolution for the Noh problem.

\begin{figure}[htb!]
  \includegraphics[width=\apjcolwidth]{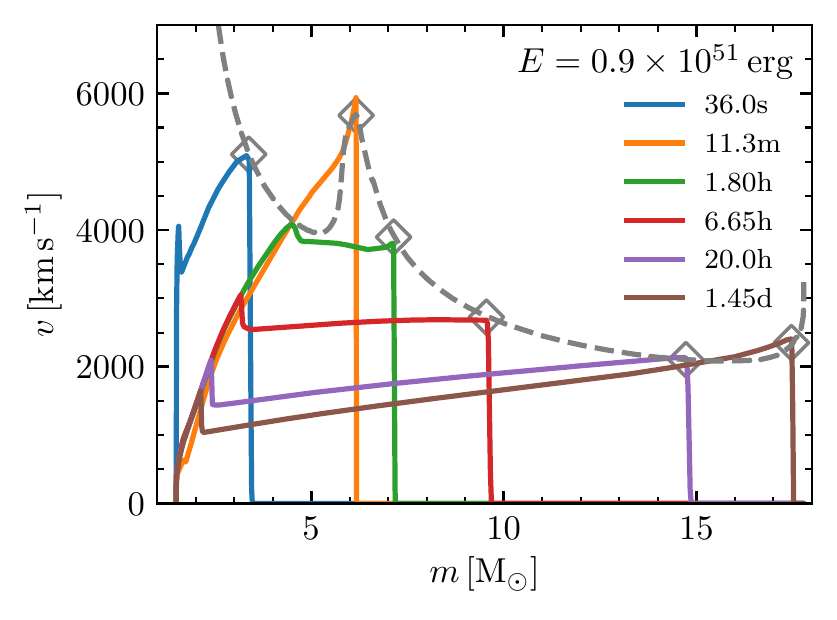}
  \includegraphics[width=\apjcolwidth]{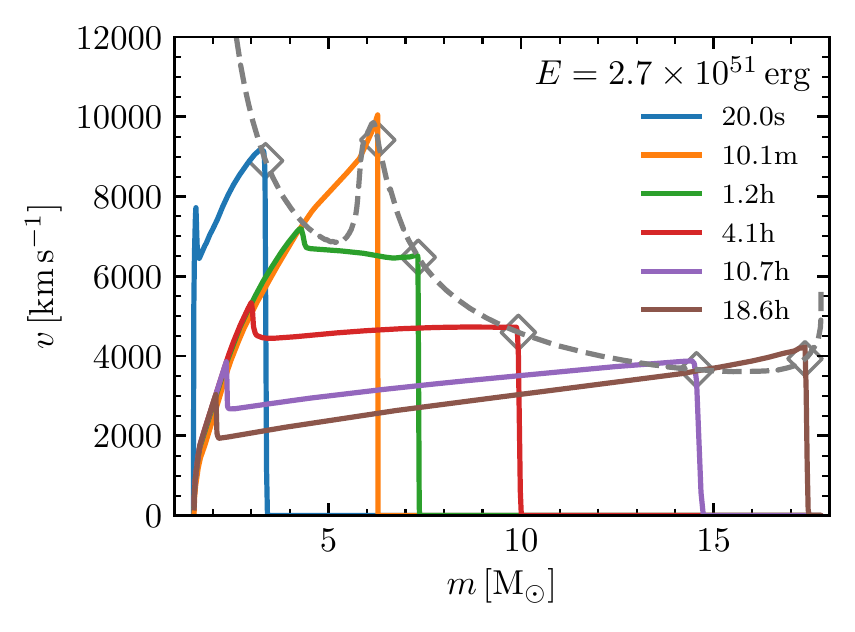}
\caption{Comparison of \MESA\ with the analytic results of
  \citet{Matzner1999} and \citet{Tan2001}.
  This started as a 19 \Msun\ model at ZAMS;
  at explosion it is 17.79 \Msun\ with $M_{\rm center} = 1.5\,\Msun$.
  The upper panel shows an explosion with
  $E =0.9 \times 10^{51}$\,\erg; the lower panel shows $E = 2.7 \times 10^{51}\,\erg$.
  The \replaced{grey}{gray} dashed curve shows the analytic
  prediction for the post-shock fluid velocity given the density
  profile of the initial model.  The solid curves show velocity
  profiles from the \MESA\ calculation at specific times.
  The unfilled diamonds indicate where on the dashed curve the two
  should be compared.
  \label{f.matznermckee}}
\centering
\end{figure}

\subsubsection{Supernova Shock}
\label{s.tests.mmk}

The problem of a supernova (SN) shock moving through a stellar envelope has
been extensively studied.  For a radiation-dominated strong shock,
a simple analytic expression for the
shock velocity is provided by \citet{Matzner1999},
\begin{equation}
  v_{\rm sh} = \alpha \left(\frac{E}{m-M_{\rm center}}\right)^{1/2} \left[\frac{m - M_{\rm center}}{\rho r^3}\right]^{0.19}~,
\label{e.matznermckee}
\end{equation}
where we adopt $\alpha=0.736$ as suggested by \citet{Tan2001}.  The explosion has an energy $E$.  The
mass that enters into this expression is the mass entrained by the shock and so differs from
the Lagrangian mass coordinate ($m$) by the mass of the remnant 
($M_{\rm center}$).  Since the material in the shocked
envelope has an adiabatic index of $4/3$,
the \citet{Matzner1999} prediction for the post-shock
velocity is $v_{\rm MM} = 6 v_{\rm sh} / 7$.

\MESA\ defines the shock location to be the outermost point
where the fluid Mach number exceeds~1, as measured in the rest frame
of the star.  Since the primary application of these capabilities are
blast waves propagating into approximately static stellar envelopes,
this shock detection criterion suffices.
Figure~\ref{f.matznermckee} compares the velocity in a \MESA\ model
(the 19~\Msun\ model of SN1999em; see Section~\ref{s.sneiip}) with
$v_{\rm MM}$.  We show explosions with two different energies,
$E = 0.9 \times 10^{51}\,\erg$ and $E = 2.7 \times 10^{51}\,\erg$.
Both cases have $M_{\rm center} = 1.5\,\Msun$.
Typical differences are at the few percent level.

\subsubsection{Weak Shock Propagation}
\label{s.tests.weak}

We now explore weak shocks with Mach numbers $\mathcal{M} = 1.2 - 2.2$
propagating outward in the hot stellar envelope of a classical
nova progenitor. The model is a $0.8\, \Msun$ WD. The H/He envelope
extends from $r=7.1 \times 10^8$\,cm to $r=7.85 \times
10^8$\,cm with densities $\rho = 10 - 100$\,g\,cm$^{-3}$ and temperatures
$T \approx 10^7$\,K.

\begin{figure}
\centering
\includegraphics{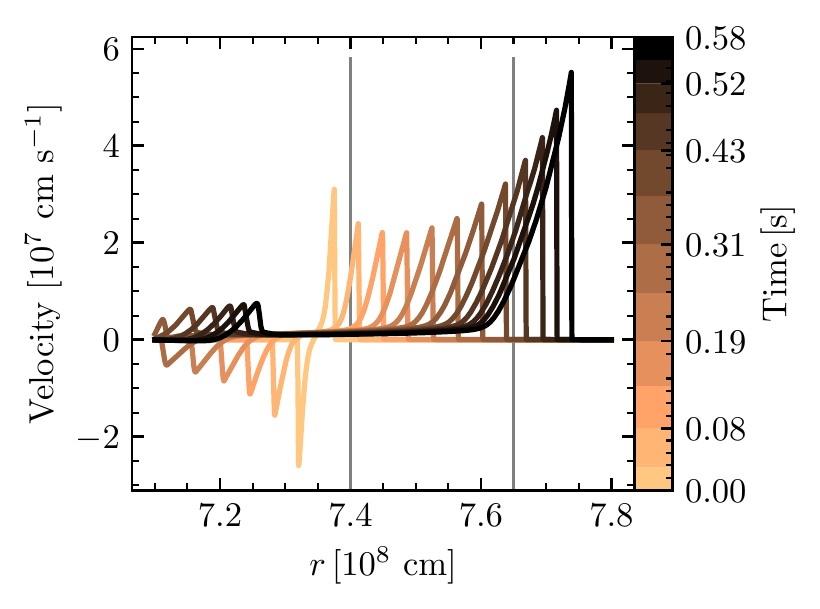}
\caption{Velocity at different times in the envelope after
  $6\times10^{42}$ ergs has been injected. One shock front travels
  upward and grows as it enters the outer atmosphere, and another
  pulse travels downward and reflects off of the inner boundary. The
  thin \replaced{grey}{gray} lines denote the region of study.}
\label{Vprofiles}
\end{figure}

After excising the core
we run the model with HLLC enabled
for 100\,s, corresponding to $\approx$\,50 sound crossing times in the
outer envelope, to allow the envelope to settle. 
Afterwards, the envelope has a total energy of $-9.16 \times
10^{45}$\,\erg,
with $-9.38 \times 10^{45}$\,\erg\ in potential energy, $2.22 \times
10^{44}$\,\erg\ in thermal energy, and a negligible kinetic energy  $1.2 \times 10^{29}$\,\erg. 
We turn off convective energy transport to study the properties of
weak shocks. To create weak shocks, 
we inject $0.5\% - 5\%$ of the total
thermal energy  into a single cell with mass $dm = 1.6 \times 10^{25}$\,g at
$r=7.3475\times 10^{8}$\,cm over $10^{-4}$\,s.
Figure~\ref{Vprofiles} shows the resulting upward and downward
propagating shocks.
We restrict our region of study to the region where the upward and
downward shocks are well-separated, in the radius range of $r=7.4-7.65 \times
10^8$\,cm (denoted with thin \replaced{grey}{gray} lines in Figure~\ref{Vprofiles}).
We do not study the properties of the downward shock and its
artificial reflection from the ``floor'' of our model.  

\begin{figure}
\centering
\includegraphics{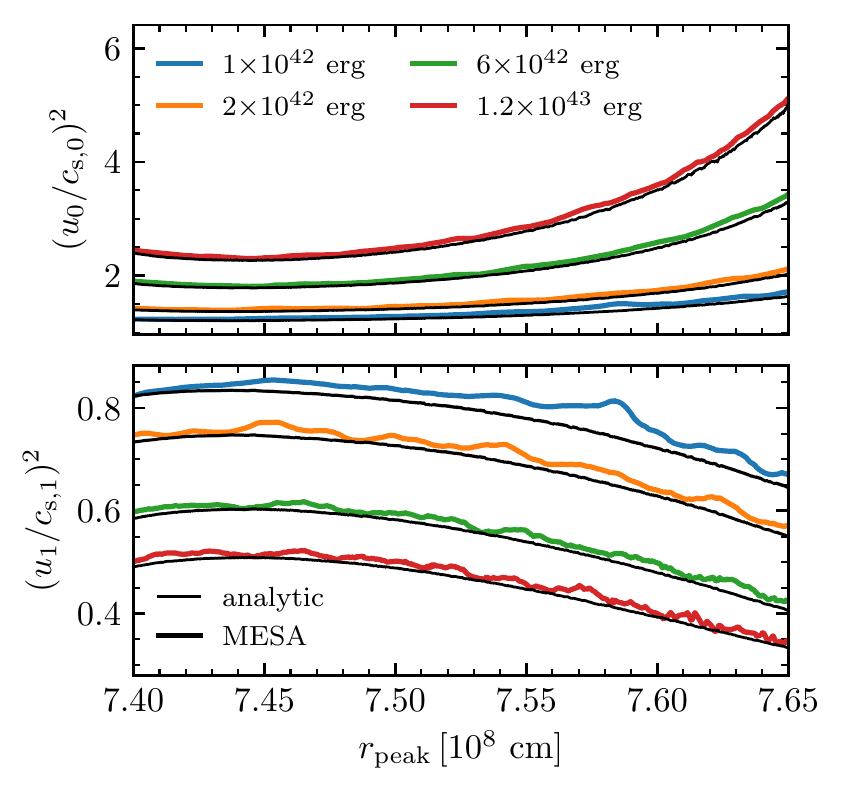}
\caption{Comparison of the \MESA\ calculation (colored lines) with
  analytic expressions (thin black lines) for $(u_0/c_{\mathrm{s},0})^2$
  (upper) and $(u_1/c_{\mathrm{s},1})^2$ (lower) for different energies
  injected.}
\label{UnshockedShockedZeldovich}
\label{ShockedZeldovich}
\end{figure}

 We define $u_0 = \dif r_\mathrm{peak}/\dif t$ as the shock velocity,
 where $r_\mathrm{peak}$ is defined as the
 radial location with maximum fluid velocity. We compare the
 properties of the shock to analytic expectations for cases where
 $\gamma$ is identical in the pre- and post-shock material. Pre-shocked
 quantities carry a 0 subscript, and shocked quantities carry a 1, and
 we use the sound speeds, $c_{\rm s}$, and pressures, $P$, on either side
 of the discontinuity. Following \cite{Zeldovich67}, in the rest frame of the shock front, the
 pre-shock gas travels into the shock front at velocity 
\begin{equation}
\left (\frac{u_0}{c_{\mathrm{s},0}}\right )^2 = \frac{(\gamma - 1)+(\gamma + 1)P_1/P_0}{2\gamma}. 
\label{Zeldovich1.83}
\end{equation}
The post-shock velocity $u_1$ has magnitude $|u_1| = |u_0 -
u_\mathrm{peak}|$, where $u_\mathrm{peak}$ denotes the fluid velocity
$u$ at $r_\mathrm{peak}$.  The analytic expression is
\begin{equation}
\left (\frac{u_1}{c_{\mathrm{s},1}}\right )^2 = \frac{(\gamma - 1)+(\gamma + 1)P_0/P_1}{2\gamma}. 
\label{Zeldovich1.84}
\end{equation}
Local shocked quantities are evaluated at the cell with the maximum
Lagrangian fluid velocity, while pre-shocked quantities are evaluated
at the cell in the initial \MESA\ profile (before the shock has
propagated) with the same mass coordinate as the shock front when it
reaches $r_\mathrm{peak}$. The thin black lines in the upper and
lower panels of Figure \ref{UnshockedShockedZeldovich} are the
right-hand side of Equations~\eqref{Zeldovich1.83} and
\eqref{Zeldovich1.84}, respectively, for shocks produced by different
amounts of injected energy. Colored lines show the left-hand side of each equation
as calculated from the \MESA\ model.

\begin{figure}
\centering
\includegraphics{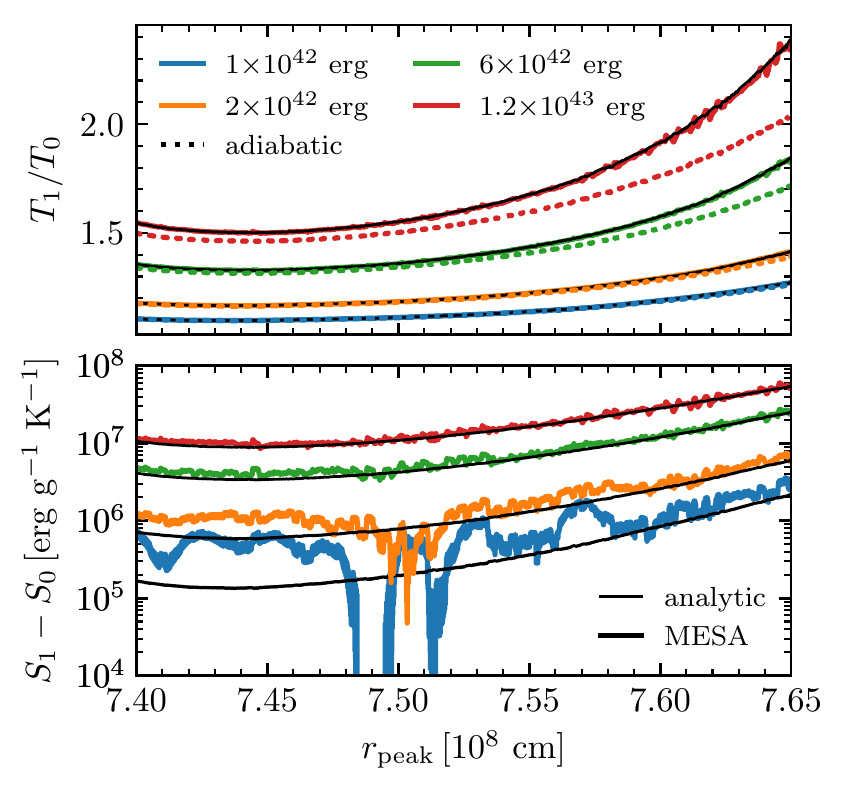}
\caption{Comparison of the \MESA\ calculation (colored lines) to the expectations of
  shock theory (thin black lines) for the temperature increase (upper) and entropy
  increase (lower) for different energies injected.
  Colored dotted lines in the upper plot indicate the temperature
  change for a purely adiabatic compression.}
\label{TemperatureZeldovich}
\end{figure}

We now compare the temperatures $T_0$ and $T_1$ of the pre- and
post-shock gas. We expect
\begin{equation}
\frac{T_1}{T_0} =  \frac{P_1}{P_0}  \left [ \frac{(\gamma - 1)(P_1/P_0)+(\gamma + 1)}{(\gamma - 1)(P_1/P_0)+(\gamma - 1)}\right ].
\label{T_Theory}
\end{equation}
The thin black lines in the upper panel of Figure \ref{TemperatureZeldovich} show the right hand side of Equation~\eqref{T_Theory} 
and the solid colored lines correspond to quantities calculated by \MESA.
The colored dotted lines in the top plot show the temperature change for an adiabatic compression, 
\begin{equation}
\left( \frac{T_1}{T_0}\right)_{S=\mathrm{const}}  =  \left( \frac{P_1}{P_0} \right)^{(1-1/\gamma)}, 
\label{T_Theory}
\end{equation}
making it clear that for the weakest shocks, the temperature jump is that expected from an adiabatic compression. 
However, for stronger shocks, the temperature is higher due to the entropy increase associated with the shock. 
For a gas with specific heat capacity $c_{V}$, this entropy jump is 
\begin{equation}
S_1 - S_0
= c_{V} \mathrm{ln}\left \{  \frac{P_1}{P_0}  \left [ \frac{(\gamma -
      1)(P_1/P_0)+(\gamma + 1)}{(\gamma - 1)(P_1/P_0)+(\gamma -
      1)}\right ]^{\gamma} \right \},
\label{Zeldovich1.85}
\end{equation}
shown by the thin black lines in the bottom plot of Figure
\ref{TemperatureZeldovich}. The colored lines correspond to quantities
calculated by \MESA. The agreement is excellent for large entropy
jumps, but becomes noisy at lower injection energies
because extracting small changes from the background is then challenging.
For the weakest shock, the entropy changes are orders of
magnitude smaller than the background entropy $1.2 - 1.6 \times 10^{9}$
erg\,g$^{-1}$\,K$^{-1}$ in the region of interest.

%%% Local Variables:
%%% mode: latex
%%% TeX-master: "paper"
%%% End:

\section{Rayleigh-Taylor Instabilities}\label{s.RTI}

The outward moving shock in a core collapse SN explosion 
encounters multiple composition boundaries.
Across these boundaries the density gradient is steep, especially at the
H/He boundary. Post-shock, these
regions become unstable to the Rayleigh-Taylor instability (RTI).
Early analytics and 2D simulations
\citep{Chevalier1976,Chevalier1978,Weaver1980,Benz1990,Herant1991}
and modern 3D calculations \citep{Hammer2010,Wongwathanarat2015,Utrobin2017} 
show significant impact on the density, velocity, and composition
structure of the ejecta.

It has been known for decades that the resulting compositional mixing can
significantly alter the photometry of the SN. This effect has been 
roughly included in 1D modeling of Type IIP light curves  resulting
from explosions deep within a red supergiant
\citep{Eastman1993,Utrobin2007,Dessart2010,Dessart2011}.
The mass densities and energy densities are also smeared out
by the mixing from the RTI (see \citealt{Bersten2011} for an early
discussion raising this concern).
In their recent modeling of the Type IIP SN 1999em, 
\citet{Utrobin2017} capture the impact of the RTI using a 3D model
pre-breakout and connect to observable SN properties with a 1D
post-breakout radiation calculation.

To approximate the 3D effects of the RTI, we implement a scheme by \citet{Duffell2016}
that modifies the 1D spherical hydrodynamics equations.
This scheme has been recently applied to the specific case of core
collapse SN by \duffellinprep\ and is now implemented in \MESA\ for
use along with the HLLC scheme. In this section, we describe the \MESA\
implementation and compare to 3D calculations of \citet{Wongwathanarat2015}. 
The use of the resulting RTI-mixed ejecta for SN lightcurves
and velocities will be 
discussed in Section~\ref{s.sneiip}. 

\subsection{Implementation of Duffell RTI}\label{s.RTI_implement}

The \citet{Duffell2016} scheme evolves an additional
scalar quantity \alphaRTI\ representing the relative strength of
turbulent fluctuations.\footnote{The quantity \alphaRTI\ is denoted by
  $\kappa$ in \citet{Duffell2016} and \texttt{alpha\_RTI} within \MESA.} The evolution equation for
\alphaRTI\ is an advection-diffusion equation with source terms.
In Eulerian form, this is
\begin{equation}
\begin{split}
 \ddt{} \left(\rho \alphaRTI{}\right) 
  & + \frac{1}{r^2} \ddr{} \left[r^2 \left(\rho \alphaRTI{} u - \etaRTI{}\ddr{} \left(\rho \alphaRTI{} \right)\right)\right] \\
  & = \Sap + \Sam ~,
  \label{e.alpharti-continuous}
\end{split}
\end{equation}
where
\begin{equation}
\begin{split}
  \Sap &= \left(\ARTI + \BRTI\alphaRTI{} \right) \sqrt{\max\left(0, -\ddr{P} \ddr{\rho}\right)} ~,\\
  \Sam &= -\DRTI\alphaRTI{} \rho \cs{} r^{-1} ~,\\
  \etaRTI{} &= \CRTI\alphaRTI{} \cs{} r ~.
\end{split}
\label{eq:etartidef}
\end{equation}
The source and sink terms $\Sap$ and $\Sam$ represent growth and
decay of the turbulence, respectively.  These terms along with a
diffusion coefficient $\etaRTI{}$ are determined via scaling
arguments.  The dimensionless coefficients in front of these
quantities (growth coefficients $\ARTI$, $\BRTI$, diffusion
coefficient $\CRTI$, and decay coefficient $\DRTI$) are
determined by calibrating a suite of 1D models against 
3D hydrodynamics simulations. The original model of
\citet{Duffell2016} calibrates against 2D simulations; see
\duffellinprep, for the re-calibration of these constants to
3D simulations.  The values of the constants found by that 3D calibration are $\ARTI
= 10^{-3}$, $\BRTI = 2.5$, $\CRTI = 0.2$, and $\DRTI = 2.0$.  
In \MESA, these constants are adjustable so that the user may explore the
effect of varying them. For example, we show later the 
effect of $\DRTI = 2.0,\, 3.0, \,\rm{and}\, 4.0$ on the mass fractions  
in massive star SN models at shock breakout. 

Additionally, a diffusive term (with diffusivity
$\etaRTI{}$) appears in each of the mass, momentum, and energy
equations. For the sake of exploration in \MESA, we allow each diffusivity to
be scaled by an independent factor. With the diffusive term, the
mass \replaced{fluence}{flux} becomes (cf. Equation \ref{e.energy-flux1})
\begin{equation}
  F_\rho = \area \rho (u - \dot{r}) - \area \etaRTI{}\ddr{\rho}~,
\end{equation}
and the choice $u = \dot{r}$ (i.e.,~$\uface = \Ss$) no longer causes
this quantity to vanish. If no correction were applied, \MESA\ 
would no longer preserve the mass coordinates of zone faces.
In order to preserve the Lagrangian nature of the equations, we allow 
for an additional velocity between the cell face and the fluid.  
The advective flux introduced by the relative motion of the face will 
then exactly cancel this diffusive flux, restoring the Lagrangian 
nature of the scheme.  Assuming $\dot{r} = u + \delta u$, 
the no mass flux condition can be rewritten as
\begin{equation}
  \delta u = \etaRTI{}\frac{1}{\rho}\ddr{\rho} = \etaRTI{}\ddm{\rho}\left(\frac{1}{\rho}\frac{dm}{dr}\right) ~.
\end{equation}
The term in parentheses is equal to $\area$.  In the finite volume form evolved by \MESA, evaluating this condition at the cell face gives
\begin{equation}
  \delta u = \area\,\etaRTI{} \frac{\left(\rhoL - \rhoR
    \right)}{\overline{dm}}~.
\end{equation}
Therefore, we modify the HLLC equation $\uface = \Ss$ to
\begin{equation}
  \uface = \Ss + \delta u~,
\end{equation}
and proceed as in Section~\ref{s.hllc} (see
Equation~\ref{eq:wave-speed} and surrounding discussion).
Usually $|\delta u| \ll \Ss$, so in practice this is a small modification and the
HLLC scheme still works well.

For a scalar quantity $f$, the \replaced{fluence}{flux} is the sum of the diffusive
\replaced{fluence}{flux} plus the advective \replaced{fluence}{flux} $(\area \rho f \delta u)$ created by the
velocity shift $\delta u$, that is
\begin{equation}
  F_{f} = \underbrace{\area\,\etaRTI{} \ddr{\rho} f}_{\mathrm{advective}}  \underbrace{-\area\,\etaRTI{} \ddr{(\rho f)}}_{\mathrm{diffusive}}  = -\area\,\etaRTI{} \rho \ddr{f}~~.
\end{equation}
Rewriting the spatial derivative as a mass derivative gives
\begin{equation}
  F_{f} = -\etaRTI{} (\area\,\rho)^2 \ddm{f} ~.
\end{equation}
To evaluate the \replaced{fluences}{fluxes} for a cell $k$, we define
\begin{equation}
  \label{eq:sigma-rti}
  \sigRTI{,k} = \etaRTI{,k} (\area_k \rho_k)^2 \frac{1}{\overline{dm}_k}~,
\end{equation}
where
\begin{equation}
    \label{eq:eta-rti}
    \etaRTI{,k} = \CRTI \overline{\alphaRTI{}}_{,k}\overline{\cs{}}_{,k} r_k ~.
\end{equation}
The \replaced{fluences}{fluxes} across faces are
\begin{equation}
\begin{split}
  F_{\rho,k} &= 0~,\\
  F_{p,k} &= \area_k P_{\mathrm{face},k} - \sigRTI{,k} (u_{k-1} - u_k)~,\\
  F_{e,k} &= \area_k P_{\mathrm{face},k} u_{\mathrm{face}, k} + L_k - \sigRTI{,k}(e_{k-1} - e_k)~.
  \label{e.energy-flux}
\end{split}
\end{equation}

The finite volume version of Equation~\eqref{e.alpharti-continuous}
evolved by \MESA\ is
\begin{equation}
  \begin{split}
    &\alphaRTI{} - \alphaRTI{,\mathrm{start}} = \\
\dt \Bigg[ & \frac{1}{dm_k} \left(F_{\alpha,k+1} - F_{\alpha,k}\right) \\
    +& (\ARTI + \BRTI \alphaRTI{}) \left[\frac{1}{\rho_k}\sqrt{\max\left(0, -\ddr{P} \ddr{\rho}\right)}\right]_{\mathrm{start}} \\
    -& \DRTI \alphaRTI{,k} \left(\frac{2 \cs{,k}}{r_k + r_{k+1}}\right)_{\mathrm{start}}\Bigg]~,
  \end{split}
\end{equation}
where
\begin{equation}
  F_{\alpha,k} = -\sigRTI{,k}(\alphaRTI{,k-1} - \alphaRTI{,k})~.
\end{equation}
We evaluate the product of the $P$ and $\rho$ spatial derivatives as
\begin{equation}
  \ddr{P} \ddr{\rho} =
  \left(\frac{\overline{P}_k - \overline{P}_{k-1}}{dr_k}\right)
  \left(\frac{\overline{\rho}_k - \overline{\rho}_{k-1}}{dr_k}\right)~,
\end{equation}
where
\begin{equation}
  dr_k = \frac{dm_k}{4 \pi r_{\rm C}^2 \rho_k}~,
\end{equation}
which is numerically preferable to a subtraction of radial coordinates.
At sharp jumps in density and pressure, these source terms can diverge, and therefore options to 
smooth $\ddr{P} \ddr{\rho}$ are available, though they are off by default.  In practice, 
smoothing does not appear to be necessary in cases we have explored,
as HLLC typically 
smears out these sharp jumps over several cells in the model at the
shock, and RTI mixing then smooths out the jumps more post-shock.

\subsection{Comparing a Munich 3D Model to \MESA\ with Duffell RTI}

We now develop a \MESA\  analog to a specific 3D simulation of \citet{Wongwathanarat2015}.  This provides a 
comparison of the predictions from the \MESA\ implementation of the
RTI mixing described in the previous sub-section (which we refer to as Duffell RTI) 
with those obtained in a 3D simulation.  The
\citet{Wongwathanarat2015} progenitor model we use, L15-1-cw, 
has a mass of $15\,\Msun$  based on \citet{Limongi2000}. We refer to
this as the Munich L15 model.  Note, as made clear in
\citet{Wongwathanarat2015},
most prior studies simulating RTI in SN envelopes
disregard early-time asymmetries, relying on explosions that 
are initiated assuming spherical symmetry.  Since those explosion
asymmetries appear to have significant consequences, it is important
to start from a 3D model like L15-1-cw when
evaluating the use of \MESA\ for SNe.

To compare with the Munich L15 model, 
we construct a \MESA\ starting model with similar parameters.
Future studies of a variety of 3D models
will be necessary to assess the impact on our 1D results
of a variety of 3D asymmetries in the initial explosion.
Just as Duffell RTI allows 1D simulations to
capture many of the effects of the 3D RTI, it may be possible
to extend 1D codes in the future to include relevant
effects of explosion asymmetries in a self-consistent manner rather
than by expediencies such as we describe below for initializing the
$^{56}$Ni abundance.

\begin{figure*}[htb!]
  \centering
  \includegraphics[width=0.495\textwidth]{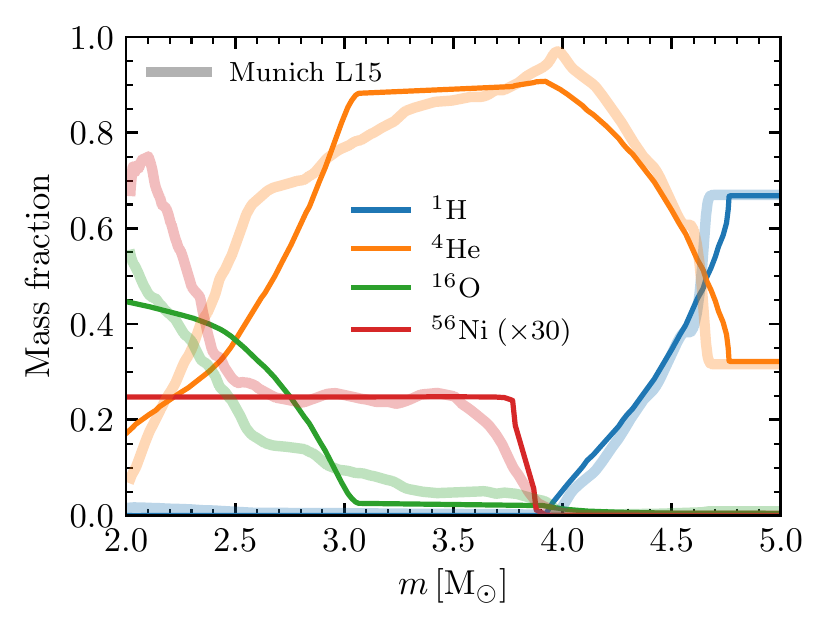}
  \includegraphics[width=0.495\textwidth]{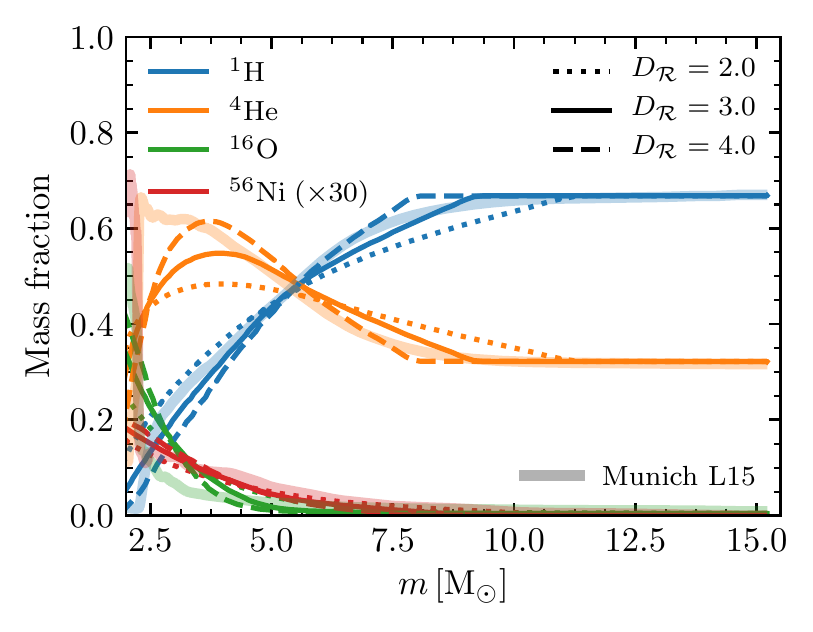}
  \caption{Comparison of abundances in \MESA\ models (thin
    lines) with 3D shell averages from the Munich L15 model (thick
    lines).  This is a comparison of analogous models at
    similar times, so the goal is to illustrate qualitative agreement.
    Left panel: for $\DRTI=3$, the time when the forward shock is at 4.8 \Msun\ and
    the reverse shock at about 4.0 \Msun. The \MESA\ $^{56}$Ni curve
    is the result of artificially inserting $^{56}$Ni in the model at
    this time. Right panel: the time the
    forward shock is at 14.7 \Msun, so near breakout, and for $\DRTI=2, 3$ and 4. 
    }
  \label{f.l15_abundances}
\end{figure*}

\begin{figure}[htb]
  \centering
\includegraphics[width=0.5\textwidth]{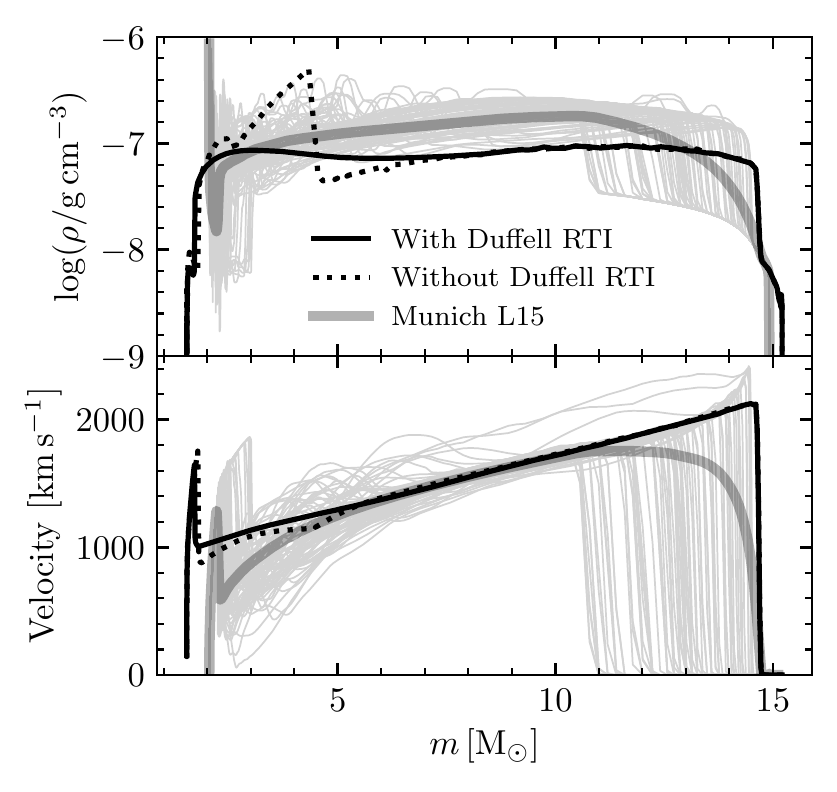}
\caption{Comparison of density (upper panel) and velocity (lower panel)
  profiles.  The solid black line shows the \MESA\ model using the
  Duffell RTI capabilities documented here with $\DRTI=3$; the dotted
  black line shows the same model run without the effects of the RTI.
  The thick gray lines show the 1D shell averages of the Munich model,
  while the fainter gray lines show the Munich model densities and
  velocities at different angles. }
\label{f.L15-rho-and-v}
\end{figure}

We now compare the 3D shell averages of \citet{Wongwathanarat2015} to \MESA\ with Duffell RTI 
enabled. The left panel of Figure~\ref{f.l15_abundances} shows 
the resulting abundances when the shock is at $4.8\,\Msun$, with the
thin lines from \MESA\ with $\DRTI=3$ 
and the  thick lines the 3D averages from the Munich L15
model. For H, He, and O, the \MESA\ lines agree with the Munich
model. If nothing is done to take into account the initial asymmetry
of the explosion, the radial extent of the 
$^{56}$Ni in the Munich model far exceeds what can be achieved in
\MESA\ by Duffell RTI mixing. Hence, at this moment in the model evolution, we
use the Munich L15 results 
to fix the extent of the \MESA\ distribution of $^{56}$Ni.  Later
mixing in the \MESA\ run is done by Duffell RTI. 
The right panel of Figure~\ref{f.l15_abundances} shows the comparison
with the Munich model just before shock breakout. For this case, we show 
three simulations with $\DRTI=2, 3 $ and 4. 

In Figure~\ref{f.L15-rho-and-v} we show the 
\MESA\ profiles of density (upper panel) and velocity (lower panel) at the moment 
when the shock is at $14.7\,\Msun$. The solid lines are with Duffell RTI enabled, while the dotted lines
are with it turned off. As shown by \citet{Wongwathanarat2015}, \citet{Utrobin2017}, and \duffellinprep\
the operation of the RTI removes the unphysical density feature produced
in 1D simulations without it.  Such features can be seen in 
Figures 2 of \citet{Eastman1994} and \citet{Dessart2011} and in the dotted black line in the 
upper panel of Figure~\ref{f.L15-rho-and-v}. Duffell RTI also alters the velocity structure of the material near the H/He boundary, as 
we discuss more in Section \ref{s.rtilight}. The thick gray lines in both plots 
show the 1D shell averages of the 3D Munich L15 model.  The fainter gray lines show the density and velocity  profiles for a variety of angles in the Munich model. 
The asymmetries of the shock in the Munich model lead to its location varying between mass
coordinates $10.5\,\Msun$ and $14.5\,\Msun$.  This variation with angle
leads to 1D shell averages that do not show a sharp shock feature, but
instead have more rounded shapes.   Since the 1D \MESA\ results have
the shock at a single mass coordinate, they are similar to Munich
profiles at a particular angle.  This difference must be considered
when comparing results from \MESA\  to shell averages from the Munich model.
It also shows that the time of shock breakout, which is well-defined in the 1D
model, varies with angle in the 3D model.

%%% Local Variables:
%%% mode: latex
%%% TeX-master: "paper"
%%% End:

\section{Light Curves and Velocity Evolution of Core Collapse Supernovae}
\label{s.sneiip}

We now present \MESA\ modeling of the ejecta evolution triggered by
core collapse in massive stars (roughly $M>8\,\Msun$).  The new \MESA\
capabilities enable self-consistent calculations of photometric
evolution of core collapse supernovae (SNe) using the \STELLA\ code
\citep{Blinnikov1998,Blinnikov2004,Baklanov2005, Blinnikov2006}.  A
public version of \STELLA\ is now included with the \MESA\ distribution, and
the interface from \MESA\ to \STELLA\ has been customized for ease of
use.\footnote{When using these capabilities
  one should cite this instrument paper and
  the following papers describing \STELLA\
  \citep{Blinnikov2004,Baklanov2005, Blinnikov2006}.}

Our main emphasis in this section is on the commonly observed
Type IIP ``plateau'' SNe that originate from energy deposited
deep in the core of a $M\approx 8-20\,\Msun$ red supergiant
\citep{Smartt2009}. We also exhibit how these new capabilities enable
simulations of core collapse events that occur after the star has lost
the majority of its outer hydrogen envelope, the Type IIb and Ib
SNe.

The new capabilities we present are provided by a powerful combination
of \MESA\ and \STELLA. Post core collapse evolution proceeds in two
distinct phases. First we use \MESA\ to evolve models from a few
seconds after the central explosion triggered by core collapse to a
time just before the outgoing shock reaches the stellar surface.
These calculations make use of HLLC
(Section~\ref{s.hydro}) and Duffell RTI (Section~\ref{s.RTI}).
Subsequently, we use \stella\ to evolve models through shock breakout
and beyond the end of the plateau, generating light curves and
velocities of the material at the photosphere and above.

Simulations using 3D models from the core collapse event to shock
breakout are computationally expensive but now feasible
\citep{Wongwathanarat2015, Utrobin2017}, and it will be a significant
contribution to have more of them available in the future.
To explore the subsequent $\ga$\,100 days of photometric and spectroscopic
evolution, 1D approximations are common.
The new capabilities with \MESA\ and
\STELLA\ also use a 1D approximation for both the pre-
and post-breakout evolution. This provides
a less computationally costly alternative for initial
exploration of the parameter space for potential progenitors prior to
or instead of doing a more realistic but more computationally costly
3D simulation.  The pair \MESA\ and \STELLA\ can produce useful results
in a few hours running on a modern multicore desktop workstation (see
Section~\ref{s.knobs}), while the 3D pre-breakout evolution and post-breakout spectral analysis
can take weeks running on a supercomputer.  \MESA\ and \STELLA\
are not a replacement for the more computationally expensive codes but
will be useful in conjunction with them.

Throughout this section, we present models that we characterize as
``similar to'' observed SNe.  We list the properties and
parameters of these models in Table~\ref{tab:models}. As we discuss in
Section~\ref{s.specificsne}, where we describe the procedure by which
these models were generated, they are not ``best-fit'' models.
Rather, they simply serve as illustrative cases of these new
capabilities.

\subsection{From Core Collapse to Near Breakout with \MESA}
\label{s.corecoll}

Models of massive stars can be evolved in \MESA\ up to the onset of
the rapid infall of the iron core (see \mesaone, \mesatwo,
\mesathree).  However, \MESA\ cannot model the core collapse event itself.
Hence, to transition from the onset of core infall to the
explosion phase, we rely on a a variety of approximate procedures
(\mesathree).

For the current efforts, our approach is as follows. We remove the
center section of the model at the location where the entropy per
baryon is 4\,\kB, excising the portion of the model that will have
collapsed to form a proto-neutron star.  This corresponds
approximately to the iron core,
typically at about $1.5\,\Msun$.  We allow the model to continue
infall until its inner boundary (IB) reaches $200$ km, near the location of
the stalled shock (thanks to H.-T. Janka, private communication, for
suggesting this scheme).  After the first few seconds, we account for
further fallback by removing negative velocity material at the
IB. We are not seeking a numerical model of
realistic fallback since that depends on 3D details of the explosion
that are beyond what \MESA\ can simulate.

\begin{figure*}[htb]
  \centering
  \includegraphics[width=0.495\textwidth]{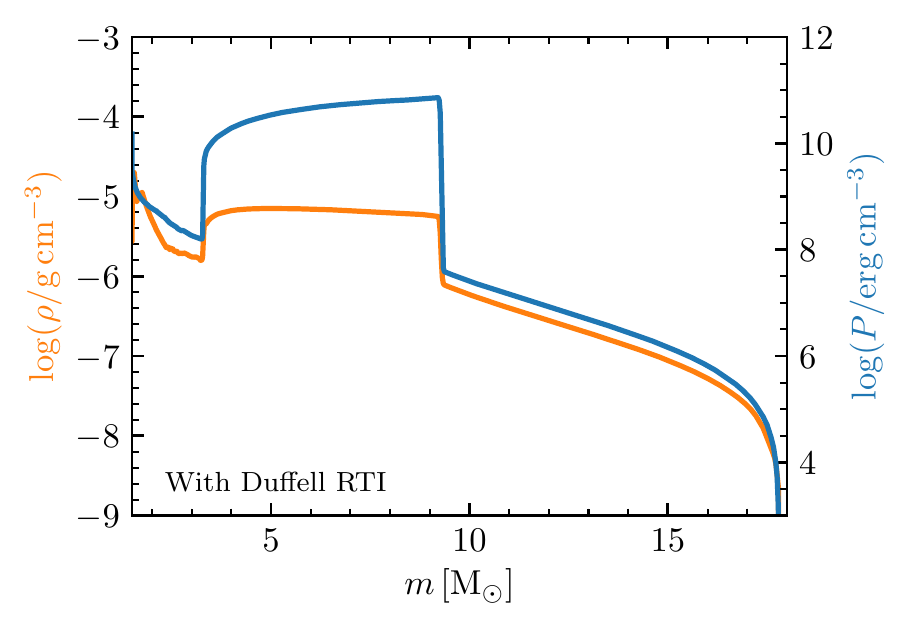}
  \includegraphics[width=0.495\textwidth]{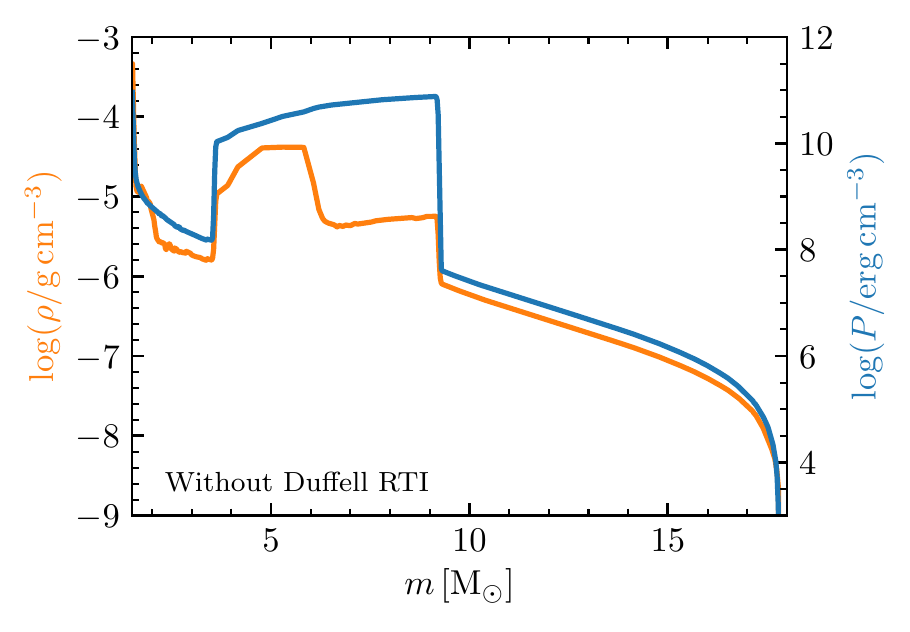}
  \caption{Density (left y-axis, orange curves) and pressure (right
    y-axis, blue curves) for \MESA\ models with Duffell RTI (left
    panel) and without any RTI effects (right panel) at a time when
    the forward shock is approximately halfway through the star.  The
    \href{http://mesa.sourceforge.net/mesa4.html}{online animated Figure} shows the time evolution of these and other
    quantities for each case.}
\label{f.RTI-movies}
\end{figure*}

The stellar explosion is induced by injecting energy in a thin layer
of approximately $0.01\,\Msun$ at the IB for 5\,ms,
at a rate sufficient to raise the total energy of the
model to a user-specified value.  In the subsequent evolution, nuclear
reactions are allowed to change abundances but not to
generate energy.  This choice is suitable because we
are not seeking accurate nucleosynthetic yields.
The explosion energy spent to
photodisintegrate the core to a mix of protons, neutrons, and alpha
particles is soon after roughly repaid by energy released as those
particles recombine to form products such as $^{56}$Ni.  Getting an
accurate accounting of the energy balance of that complex process is
beyond the scope of this paper and is not attempted in the following
examples.  Our choice to exclude nuclear energy generation
can be seen as a simplifying assumption that the cost of
photodisintegration is balanced by the return from later
recombination.  For users wishing to refine this, any excess change in
energy from nuclear reactions can be included in the
specification of the post-explosion total energy of the model.

The conservation of total energy throughout the run is estimated by
summing the per-step errors from post-explosion to near breakout.
At each timestep, we compare the actual
change in total energy between initial and final models for the step,
to the change expected from surface luminosity and neutrino losses
over the duration of the step.  The runs for the models reported below
typically show relative cumulative errors in conservation of total
energy of less than 1\%, with most of that error happening in the
first few minutes post explosion when the shock is most extreme.  For
later stages, the cumulative relative error is orders of
magnitude smaller.

The post-explosion evolution of the \MESA\ model is determined by the
shock traversal through the star and the resulting Duffell RTI.
Figure~\ref{f.RTI-movies} illustrates the difference between
models with and without the effects of the RTI by showing density and
pressure profiles.  They are shown when the forward shock is
about halfway through the star and when the reverse shock originating
at the H/He boundary has reached $\approx$\,$4\,\Msun$ on its way to the
center.  The reverse shock is primarily responsible for the large RTI
effects evident in the plots.  The online animated figure shows the
time evolution of these and many other quantities of interest from
seconds after explosion to near shock breakout.

\subsection{From Near Breakout through the Plateau: Introducing \STELLA}
\label{s.postbreakout}

To follow the evolution of the model through shock breakout
and beyond we use a multigroup (i.e.,~frequency-dependent) radiation hydrodynamics
code.\footnote{\MESA\ can be run through shock breakout and beyond,
  but we do not view gray opacity lightcurves as
  sufficient for quantitative comparisons to observed SNe.}  When the shock is
near breakout, we hand the \MESA\ model off to \STELLA\ in an
appropriate form which involves interpolating to the desired grid and
optionally adding circumstellar material (CSM) according to user specifications.
With that done, \MESA\ is finished, and \STELLA\ takes over (see
Section~\ref{s.handoff} for a discussion of how we select when to
hand off).

\STELLA\ \citep{Blinnikov1998,Blinnikov2004,Baklanov2005,
  Blinnikov2006} is able to model SN evolution at early times,
before the expansion is homologous.  It can also handle shock breakout
and interaction with circumstellar material outside the
conventional stellar photosphere.  \STELLA\ is an implicitly
differenced hydrodynamics code that incorporates multigroup radiative
transfer. The time-dependent equations are solved implicitly for the
angular moments of intensity averaged over fixed frequency bands.
\STELLA\ takes about the same amount of time for near-breakout
to post-plateau evolution as \MESA\ takes to simulate from
explosion to near-breakout: about an hour on current workstations.

\STELLA\ solves the radiative transfer equations in the intensity
momentum approximation in each frequency bin. We use from 40 to 200
frequency groups, enough to produce bolometric luminosities and
broad-band colors, but not sufficient to produce spectra.  Better
broad-band light curves can be produced with the larger number of
frequency groups, but 40 is sufficient for a bolometric lightcurve and
gives faster runtimes since each group must be represented by a
variable and an equation at each zone.  The opacity is computed based
on over 153,000 spectral lines from \citet{Kurucz1995} and
\citet{Verner1996}. The expansion opacity formalism from
\citet{Eastman1993} is used for line opacities taking
high velocity gradients into account.  The opacity also includes
photoionization, free-free absorption, and electron scattering.  LTE
is assumed in the plasma, which allows the
use of the Boltzmann-Saha distribution for ionization and level
populations. \STELLA\ does not include a nuclear reaction network except for
the radioactive decay chain initiated from $^{56}$Ni.  For
calculating the overall opacity, the code uses 16 species: H, He, C,
N, O, Ne, Na, Mg, Al, Si, S, Ar, Ca, a sum of stable Fe and
radioactive $^{56}$Co, and stable Ni and radioactive $^{56}$Ni. Energy
from nickel and cobalt radioactive decay is deposited as positrons
and gamma-rays and is treated in a one-group transport approximation
according to \citet{Swartz1995}.

\STELLA\ solves the conservation equations for mass, momentum, and
total energy on a Lagrangian comoving grid. It employs artificial
viscosity based on the standard von Neumann artificial viscous
pressure used for stabilizing solutions \citep{vonNeumann1950} and a
cold artificial viscosity used to smear shocks
\citep{Blinnikov1998, Moriya2013}.  The coupled equations of radiation
hydrodynamics are solved through an implicit high-order
predictor-corrector procedure based on the methods of \citet{Gear1971}
and \citet{Brayton1972}; see \citet{Blinnikov1996} and
\citet{Stabrowski1997} for details.

\begin{figure}[htb]
  \centering
\includegraphics[width=0.5\textwidth]{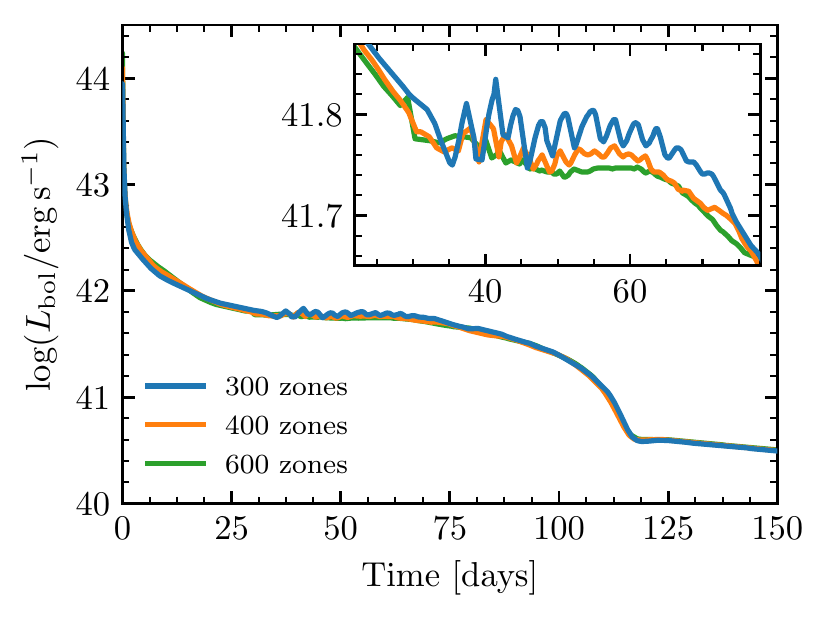}
\caption{Effect of number of \STELLA\ zones on a model (12A) similar to SN
  2012A.  The inset zooms in on the region of the plateau that
  exhibits numerical oscillations.  A small
  increase in the number of zones significantly reduces this artifact
  and doubling the number of zones almost completely removes it.}
\label{f.stella-zones}
\end{figure}

We explore the sensitivity of bolometric light curves
($L_{\rm bol}$) reported by \STELLA\ to the number of
frequency bins, spatial zoning, and error tolerances.
The result of our sensitivity study is that 40 frequency bins,
300 spatial zones, and an error tolerance $0.001$ for the Gear-Brayton
method typically give a converged model. In our experience using \MESA\ and
\STELLA\ for Type IIP SNe, we have not found cases that
require different values for number of frequency bins and error
tolerance. Some cases may need a larger number of zones in order to
minimize numerical artifacts producing spurious oscillations in the
light curve. This problem can often be
fixed by a relatively small increase in the number of zones; this is
shown for a case similar to SN 2012A in Figure~\ref{f.stella-zones}.

\begin{figure}[htb!]
  \centering
  \includegraphics[width=0.5\textwidth]{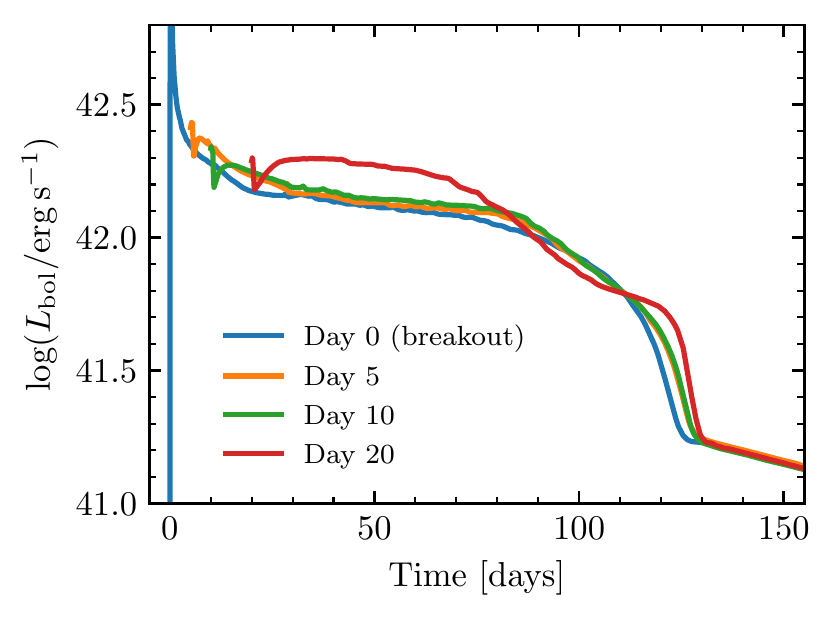}
  \caption{Bolometric lightcurves for Type IIP SNe obtained where the
    transition between \MESA\ and \STELLA\ has been done at different
    times relative to shock breakout.  Performing the handoff just
    before breakout (blue curve, day 0) is the recommended choice.}
  \label{f.mesa_to_stella_time}
\end{figure}

\subsection{Handing Off from \MESA\ to \STELLA} 
\label{s.handoff}

A time must be chosen to hand off the \MESA\ model to \STELLA.  This
choice is driven by a compromise between two considerations.  First,
RTI modeling ceases once \STELLA\ is running even though the effects
of RTI may not be complete at that time.  Therefore, one wants the
model to remain in \MESA\ as long as possible.
But second, \STELLA\ 
more accurately handles shock breakout and the outermost layers,
especially if any matter is placed above the photosphere or if
significant radiation is free-streaming from just
below the photosphere prior to shock breakout.
Moreover, the
sophisticated multigroup radiation transfer of \STELLA\ will do a
much better job than (gray) \MESA\ can at later times post breakout.
Hence for longer-term lightcurve evolution, this motivates the default
choice to perform this handoff just before breakout.

In order to illustrate the effects of this choice,
Figure~\ref{f.mesa_to_stella_time} shows bolometric light curves for
cases where the handoffs are done at different times.  Note that this
plot shows \MESA\ being forced to run post-breakout even though that
is not recommended.  The deviation of the light curves for later
handoff are primarily the result of \STELLA\ doing a better job
because of its multigroup radiation transfer rather than any
late-stage RTI effects being captured by \MESA\ that are missed by
\STELLA.  That is because, for this case, the H envelope is of normal
thickness and the reverse shock from the H/He boundary has time to
reach the center, completing essentially all of the RTI effects before
breakout.

In the runs presented in the remainder of this section, we choose to
do the \MESA-to-\STELLA\ handoff shortly before breakout, as
determined by the outgoing shock front reaching a location
$0.11\,\Msun$ below the surface of the model (this location is a
user-defined parameter).  Again, we note that in some cases the
reverse shock is still far from the center at this moment, and not all of
the RTI mixing has completed.  In particular, this is true for
models with a partially stripped envelope (see
Section~\ref{s.stripped}).  For now, this remains a caveat for the
user; a solution would be to have the post-breakout radiation
hydrodynamics code include a treatment of the effects of RTI.
\added{When presenting the results, we define $t=0$ as the time of
  shock breakout---which we identify using the peak of the bolometric
  luminosity---and not the (earlier) time of the \MESA-to-\STELLA\
  handoff.}

Because of \STELLA's treatment of radiation hydrodynamics,
we have not had to take the progress of the model toward
homologous expansion into consideration in selecting a time to hand off
from \MESA.  However this is a consideration for doing a handoff to
radiative transfer codes that assume homology.
More accurate spectral and lightcurve
modeling with full radiative transfer, such as \texttt{EDDINGTON}
\citep{Eastman1993}, \texttt{SEDONA} \citep{Kasen2006}, and \texttt{CMFGEN}
\citep{Dessart2010}, assume homologous expansion in their current
applications to SNe, and this should be considered when deciding the time to
hand off from another simulation. Indeed, \citet{Eastman1994} and
\citet{Dessart2011} discuss this challenge, especially for the
inner-most material that has not reached a homologous stage and can
still have a reverse shock running through it. Approximations
made in mapping to a thereafter homologous code can impact the
late-time photospheric velocity evolution and the nebular line width
predictions associated with the innermost ejecta.

\begin{figure}[!htb]
  \centering
\includegraphics[width=0.5\textwidth]{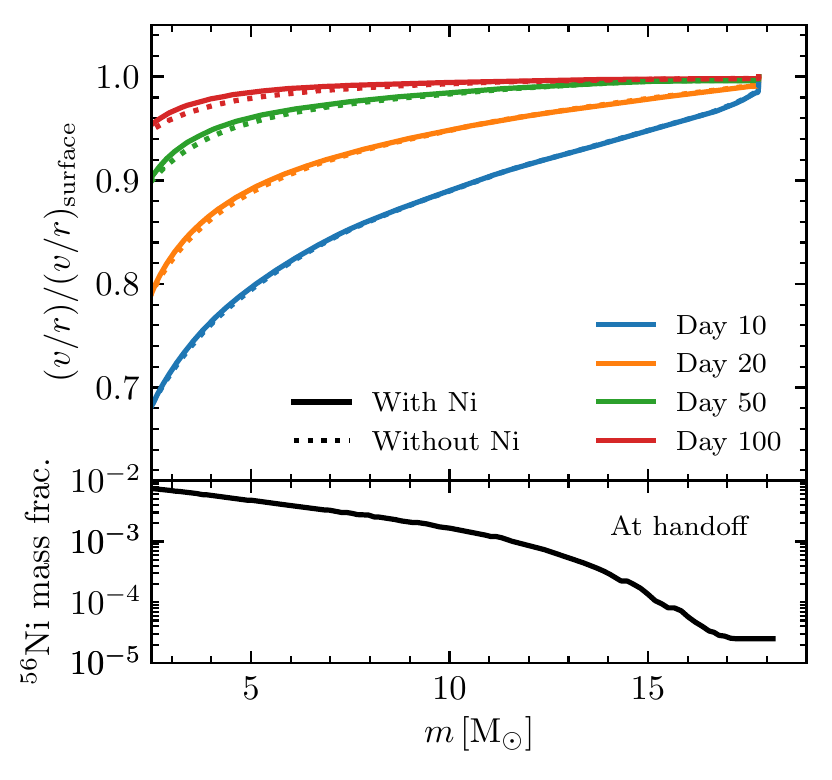}
\caption{Profile of $v/r$ throughout the model, normalized to the
  values at the surface, at a number of epochs.  In homologous
  expansion, the profiles would be constant and equal to unity.  The
  models are still significantly non-homologous at 20 days. The bottom
  panel shows the $^{56}$Ni mass fraction for this model at
  \MESA-to-\STELLA\ handoff (near shock breakout).  The extended
  $^{56}$Ni profile (total nickel mass $\approx$\,$0.04\,\Msun$)
  indicates that it has been mixed by RTI effects out to near the
  surface.  However, as illustrated by the dotted lines in the upper panel,
  the decay energy does not have a significant effect on the approach
  to homology.}
  \label{f.homology}
\end{figure}
 
\begin{figure}[!htb]
  \centering
\includegraphics[width=0.5\textwidth]{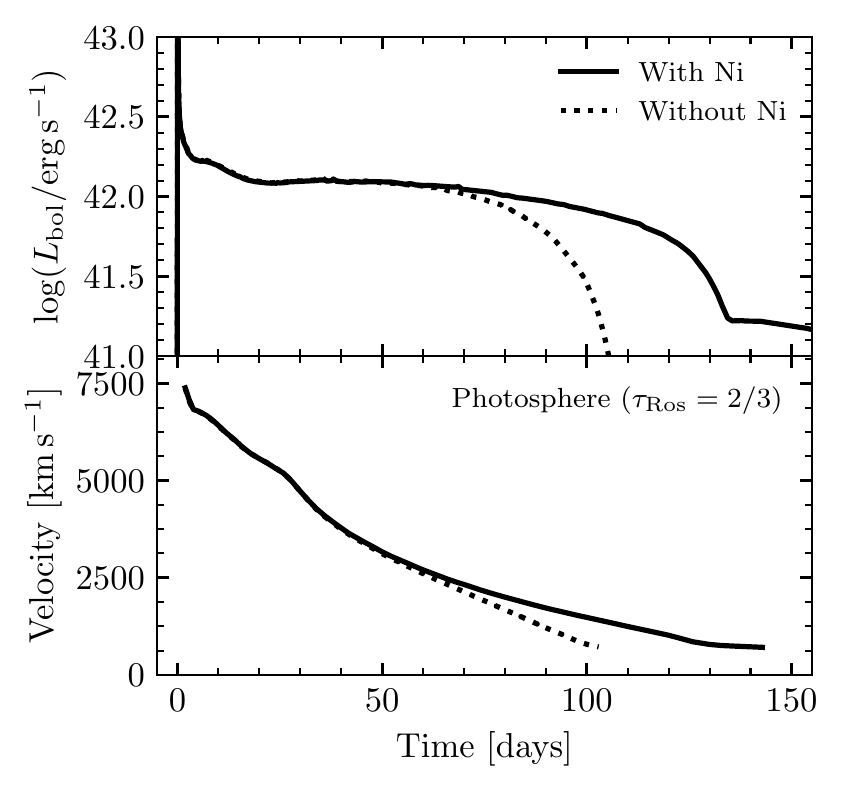}
\caption{Lightcurves and velocities for a model similar to SN 1999em
  with $\approx 0.04\,\Msun$ $^{56}$Ni and without $^{56}$Ni.  The
  main effect of the radioactive decay is to prolong the plateau.}
  \label{f.L-and-v-w-and-wo-Ni}
\end{figure}

In contrast, \STELLA\ does not assume homologous expansion, so early
handoffs are fine; it can handle the effects of remaining pressure
gradients as the model moves toward homologous expansion. This is
important, as the time it takes to reach homology in these models can
be quite long.  Figure~\ref{f.homology} shows velocity evolution
results for a model similar to SN 1999em (see discussion in
Section~\ref{s.specificsne}). Homologous expansion would imply that
$v/r$ is flat, whereas a 20\% variation from simple homology is
evident at 20 days.  An additional way in which homology can be
violated long after shock breakout is from $^{56}$Ni decay, especially
in Type Ia SNe \citep{Woosley2007}. As is evident in Figure
\ref{f.homology}, the much smaller mass fractions of $^{56}$Ni in Type
IIP SNe do not cause such a problem. The contrasting light curves with
and without $^{56}$Ni are shown in Figure~\ref{f.L-and-v-w-and-wo-Ni},
exhibiting the prolonging of the plateau due to radioactive decay
\citep{Kasen2009, Sukhbold2016}.

\subsection{Connecting to Observations: Photospheric Properties from \STELLA}
\label{s.obs-photo}

\begin{figure}[!htb]
  \centering
  \includegraphics[width=0.5\textwidth]{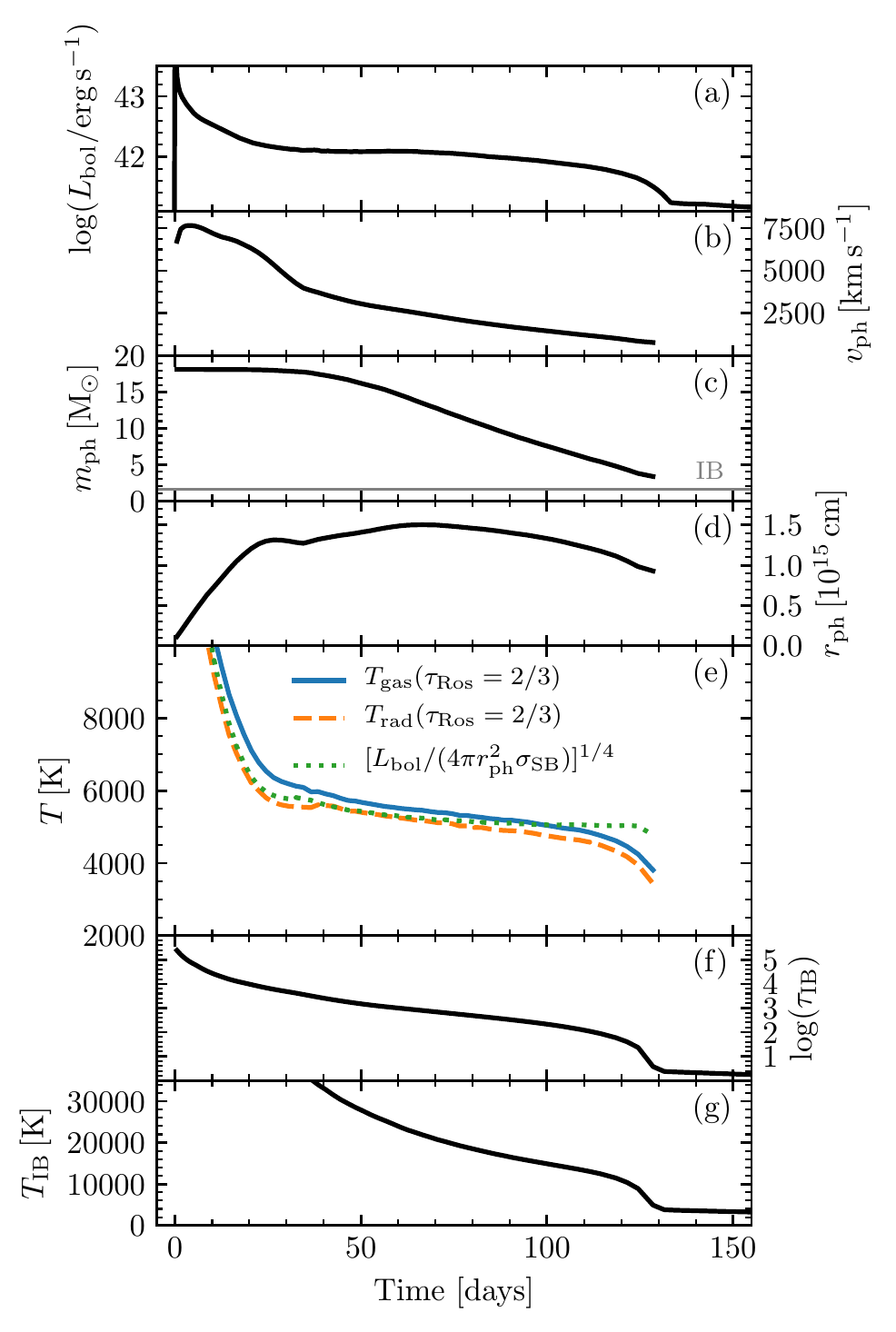}
\caption{Photosphere and IB properties of a model (99em\_19 in Table~\ref{tab:models}),
  similar to the Type IIP SN 1999em, as a function of time. From
  top-to-bottom, the figure shows the bolometric luminosity, the
  velocity, mass coordinate, and radius coordinate of the photosphere,
  three temperatures (gas, radiation, and ``effective'') at the
  photosphere, the optical depth to the IB, and the IB temperature.
  \added{The gray line in panel (c) shows the Lagrangian mass coordinate of the IB.}}
  \label{f.center-and-photosphere-99em}
\end{figure}

To set the stage for the rest of this section, we describe a
particular model in detail. Figure~\ref{f.center-and-photosphere-99em}
shows the evolution of a model we construct to be
similar to the Type IIP SN 1999em (99em\_19 in Table~\ref{tab:models}).  The quantities shown are those
generated during the \STELLA\ phase of the evolution.  Panel~(a)
is the bolometric luminosity, while panel~(b) shows velocity at
the location of the photosphere (where $\tau_{\rm  Ros}=2/3$) and
panels~(c) and~(d) show the mass and radius
coordinate of this location. This illustrates the familiar result that the photosphere only
reaches the deeper parts of the ejecta after about day 50. The
radiation and gas temperatures at the photosphere are shown in
panel~(e),
as is an effective temperature defined by the
bolometric luminosity leaving the photosphere. Panel~(f)
shows the optical depth to the IB, highlighting that
the radiative diffusion approximation is excellent (since
$\tau_{\rm IB} \gg 1$) until day 120, at which point the plateau ends and the
IB temperature (panel~g) approaches that of the
photosphere. \added{(Curves showing photospheric quantities stop once $\tau_{\rm IB} < 3$.)}
Meanwhile, the photospheric radius (panel~d)
stays remarkably constant throughout the plateau.

\begin{figure}[!htb]
  \centering
\includegraphics[width=0.5\textwidth]{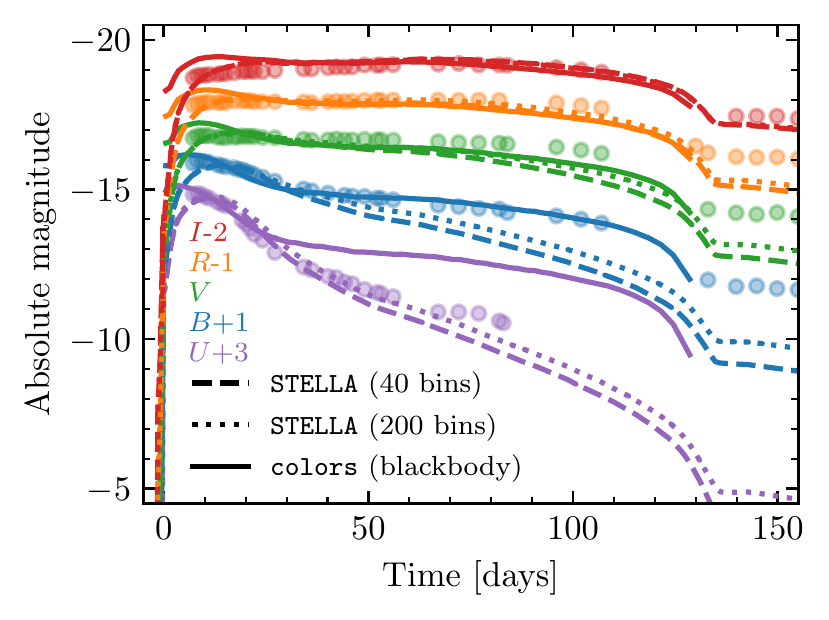}
\caption{Comparison of model 99em\_19 with the multicolor lightcurve
  of SN 1999em, showing colors from \STELLA\ and blackbody colors from
  \MESA. Circles indicate observational data. 
  This demonstrates the effect of the number of \STELLA\
  frequency bins on the predicted colors.}
  \label{f.multicolor-99em}
\end{figure}

Our emphasis is on bolometric luminosities, where 40
\STELLA\ frequency bins is adequate. However, broadband light curves
are also reported by \STELLA.  Figure \ref{f.multicolor-99em} shows
how the \STELLA\ colors change as one goes from 40 to 200 frequency
bins in a model approximately matching the bolometric luminosity of SN 1999em
(99em\_19 in Table~\ref{tab:models}).
This reflects that a given band is spanned by only a small number of
frequency bins.  The non-public research version of \STELLA\ can opt
to use many more frequency bins to address under-resolution issues.
There are no current plans to include that capability in \MESA.  We
also show what a blackbody would predict using the \MESA\
\texttt{colors} module (see Appendix~\ref{s.colors}).  This makes it
clear that the line-blanketing in the $U$ band is well handled by
\STELLA.  We do not include colors in our subsequent discussions, but
we expect they may be useful to users who have access to observations
in one or two bands, but not enough data to produce a
bolometric light curve from observations.

\subsection{Connecting to Observations: \ion{Fe}{2} Line Velocities}
\label{s.sobolev}

It is important to be able to interpret the
ejecta velocities measured by observers, which are often inferred from
the absorption minimum in the \ion{Fe}{2} 5169\,{\AA} line.
Modeling these absorption features requires
more detailed radiative transfer than available in \STELLA.
However,
rather than assume the photospheric velocity reported by \STELLA\ is
identical to the \ion{Fe}{2} 5169\,{\AA} line, we have added the capability of
finding the location (and hence the velocity) of material above the
photosphere where the Sobolev optical depth in the \ion{Fe}{2}
5169\,{\AA} line is a specified value.\footnote{This approach arose
  through the efforts of Dan Kasen, who also provided important data
  needed to complete the calculation.} This will prove to be most
important after day 30 or so, when the photosphere has started to move
inward in mass coordinate into ejecta with a shallow density profile.

The strength of a line in a
homologously expanding atmosphere is quantified by the Sobolev optical
depth \citep{Sobolev1960, Castor1970, Mihalas1978, Kasen2006}, which
for the \ion{Fe}{2} line at any position is
\begin{equation}
\tau_{\rm Sob}={\pi \qe^2\over m_e c} n_{\rm Fe}\eta_i f t_{\rm exp} \lambda_0,
\label{e.tausob}
\end{equation}
where $\lambda_0=5169$\,{\AA} is the line center wavelength for the
\ion{Fe}{2} line, $f=0.023$ is its oscillator strength, $n_{\rm Fe}$
is the number density of iron atoms, and $t_{\rm exp}$ is the time
since breakout. The quantity $\eta_i$ is the fraction of iron atoms
that are in the lower level of the transition of interest and depends
on the properties of the gas. D.~Kasen (2017, private communication)
provided an $\eta_i(\rho, T)$ table for post-processing to produce the
\ion{Fe}{2} line velocities, calculated under the
assumption of LTE and covering
$\logRho =-16 $ to $-8$ and $\logT= 3.3$ to $4.3$.

\begin{figure}[!htb]
  \centering
\includegraphics[width=0.5\textwidth]{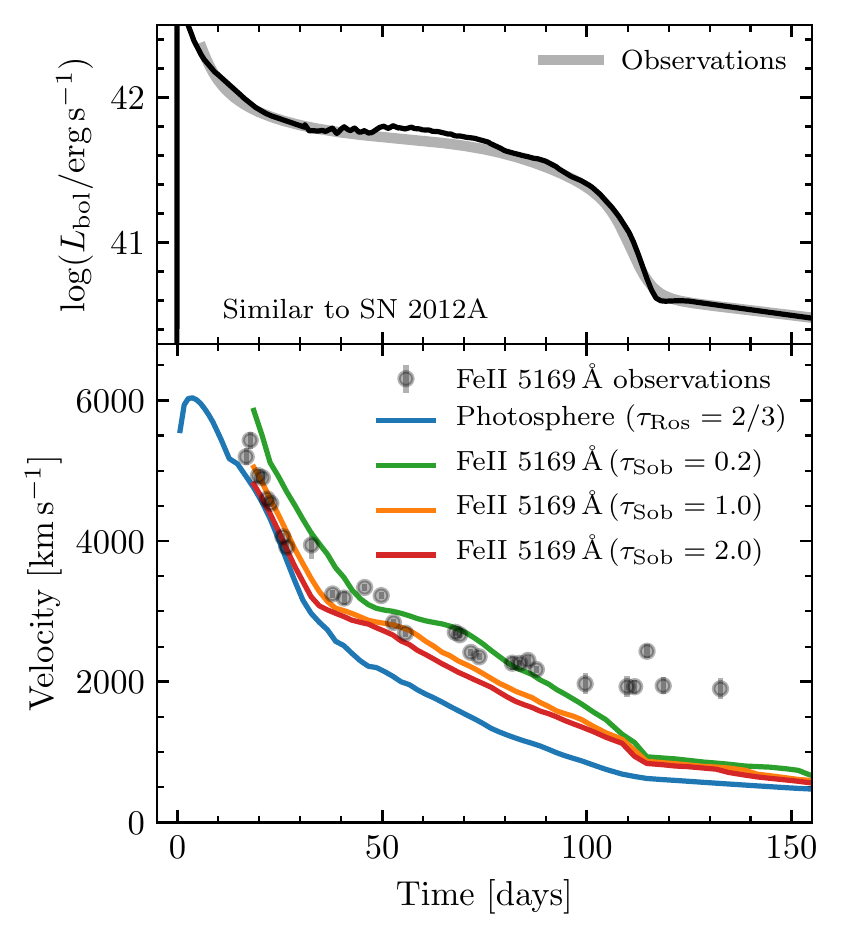}
\caption{Comparison of different definitions of velocity for a model (12A)
  similar to SN 2012A.  The upper panel shows the data and the model for
  the bolometric luminosity.  The lower panel shows the velocity of
  a few different locations depending on the Sobolev optical depth in
  the \ion{Fe}{2} line.}
  \label{f.velocity-comparison}
\end{figure}

We use Equation~\eqref{e.tausob} after the \STELLA\ run to provide the
velocity of material that satisfies a chosen value of
$\tau_{\rm Sob}$. This yields a velocity that can be compared to the
measured \ion{Fe}{2} line velocities.
Figure~\ref{f.velocity-comparison} shows the resulting comparisons for
various choices of $\tau_{\rm Sob}$ for a model similar to the Type IIP
SN 2012A found solely by matching the bolometric luminosity
(upper panel).
The lower panel displays the \ion{Fe}{2} 5169\,{\AA}
data and the velocities derived from the photosphere and for a range
of values of $\tau_{\rm Sob}$. At early times,
there is little difference between the
photospheric velocity and that of the Fe line. However, as the
photosphere moves deeper into the ejecta, the 
two velocities substantially diverge.  The velocity
inferred from the Sobolev argument gives a much better match to
observations than the photospheric velocity.  Motivated by this
comparison, we choose $\tau_{\rm Sob}=1$ for our later plots, a
parameter that the user is free to adjust.

\subsection{The Impact of Pre-breakout RTI Mixing}
\label{s.rtilight}

We have previously outlined the inclusion of a method for RTI mixing
in \MESA\ (the Duffell scheme; Section~\ref{s.RTI}), the use of \MESA\
to evolve models pre-breakout (Section~\ref{s.corecoll}), the use of
\STELLA\ to evolve models post-breakout (Section~\ref{s.postbreakout})
and described how to connect the models to observations
(Sections~\ref{s.obs-photo} and \ref{s.sobolev}).

In this way, \MESA\ plus \STELLA\ allows users to explore the impact of
RTI mixing on Type IIP light curves and velocities.  Prior work in
this direction
\citep{Eastman1994,Utrobin2007,Dessart2010,Dessart2011,Morozova2015}
focused on the impact of compositional mixing, often with
averaging approaches to achieve various levels of mixing.  Only the
recent work of \citet{Utrobin2017} incorporated compositional mixing
from a 3D model and also included the modified density and velocity
structures, also seen in the 1D RTI mixing (\duffellinprepalt).

\begin{figure}[!htb]
  \centering
\includegraphics[width=0.5\textwidth]{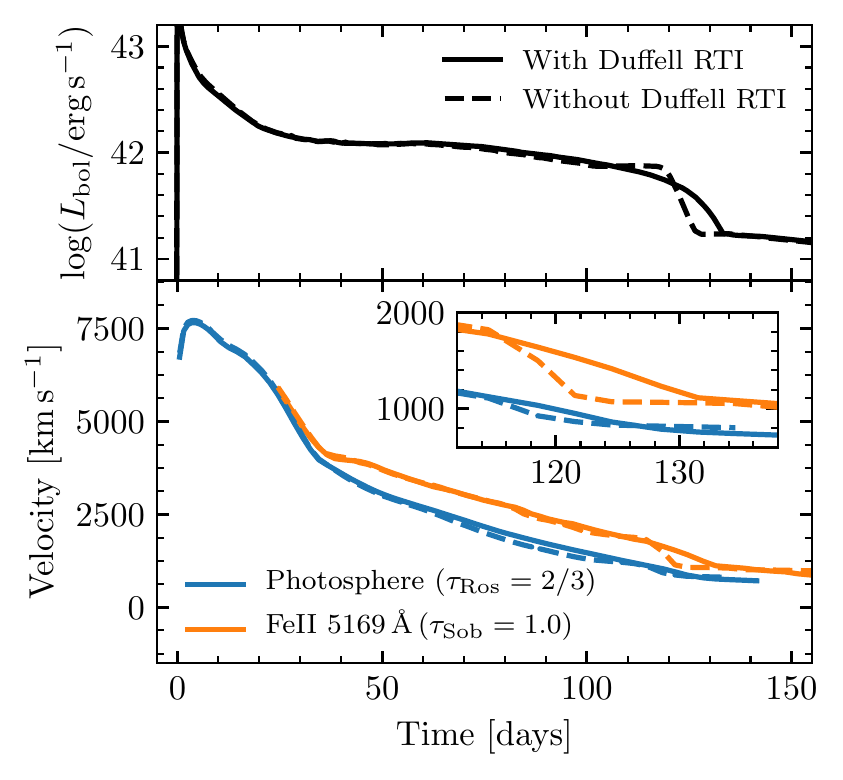}
\caption{Effect of Duffell RTI on lightcurves (upper panel) and velocities
  (lower panel) of a Type IIP SN model (99em\_19) similar to SN 1999em.  The inset shows the
  time near day 120 where the altered density structure causes a
  significant difference in the \ion{Fe}{2} line velocities.}
  \label{f.effect_of_RTI}
\end{figure}

Figure \ref{f.effect_of_RTI} shows the lightcurve and velocities of
model 99em\_19.  The luminosity without RTI mixing has a distinctive
rise just before the plateau as shown by \citet{Eastman1994} and
\citet{Utrobin2007}.  As RTI causes many associated changes in
composition, density, velocity, and energy density for the innermost
material, we cannot specifically identify the immediate cause of the
lengthening of the plateau phase when RTI is incorporated without
further experiments. These are now possible using \MESA\ and \STELLA\
but are beyond the scope of this paper.  The lower panel shows the
photospheric and \ion{Fe}{2} line velocities with and without RTI
mixing.  The most evident change is at the end of the plateau, when
the material that was near the H/He boundary in the red supergiant is
approaching the SN photosphere.  That material is strongly
affected by RTI mixing as shown in Figure \ref{f.L15-rho-and-v} and
discussed in \duffellinprep.

\begin{figure}[!htb]
  \centering
\includegraphics[width=0.5\textwidth]{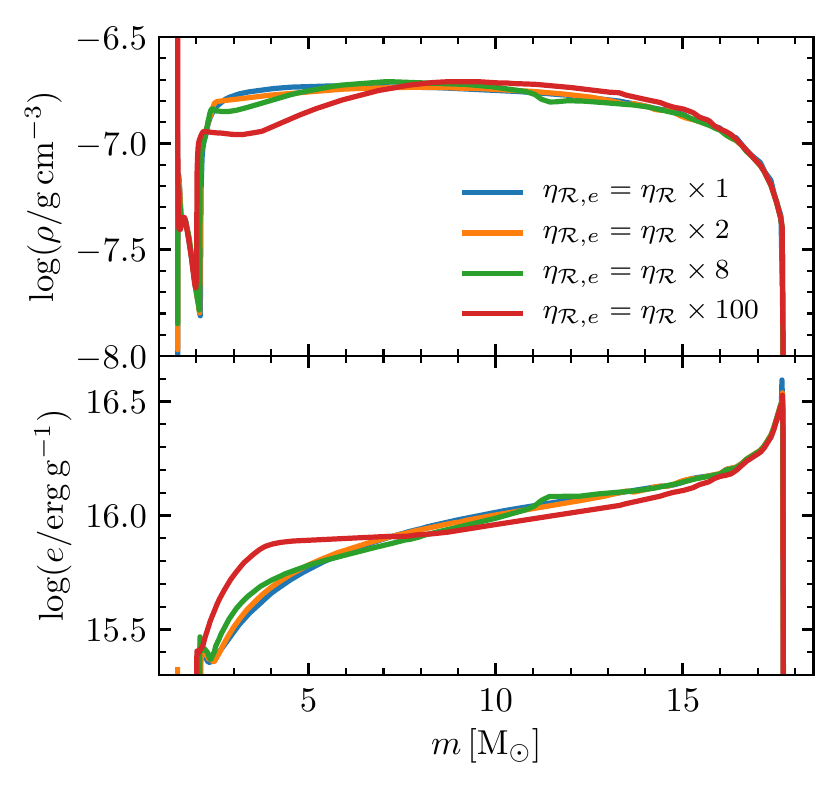}
\caption{Structural effect of the RTI energy diffusion coefficient
  $\etaRTI{,e}$.  Other parameters are the
  same as model 99em\_19.  The profiles are shown just before shock
  breakout.  The effects on the lightcurve are shown in
  Figure~\ref{f.explosion-knobs}.}
  \label{f.rti-rho-and-e}
\end{figure}

\begin{figure}[!htb]
  \centering
  \includegraphics[width=\columnwidth]{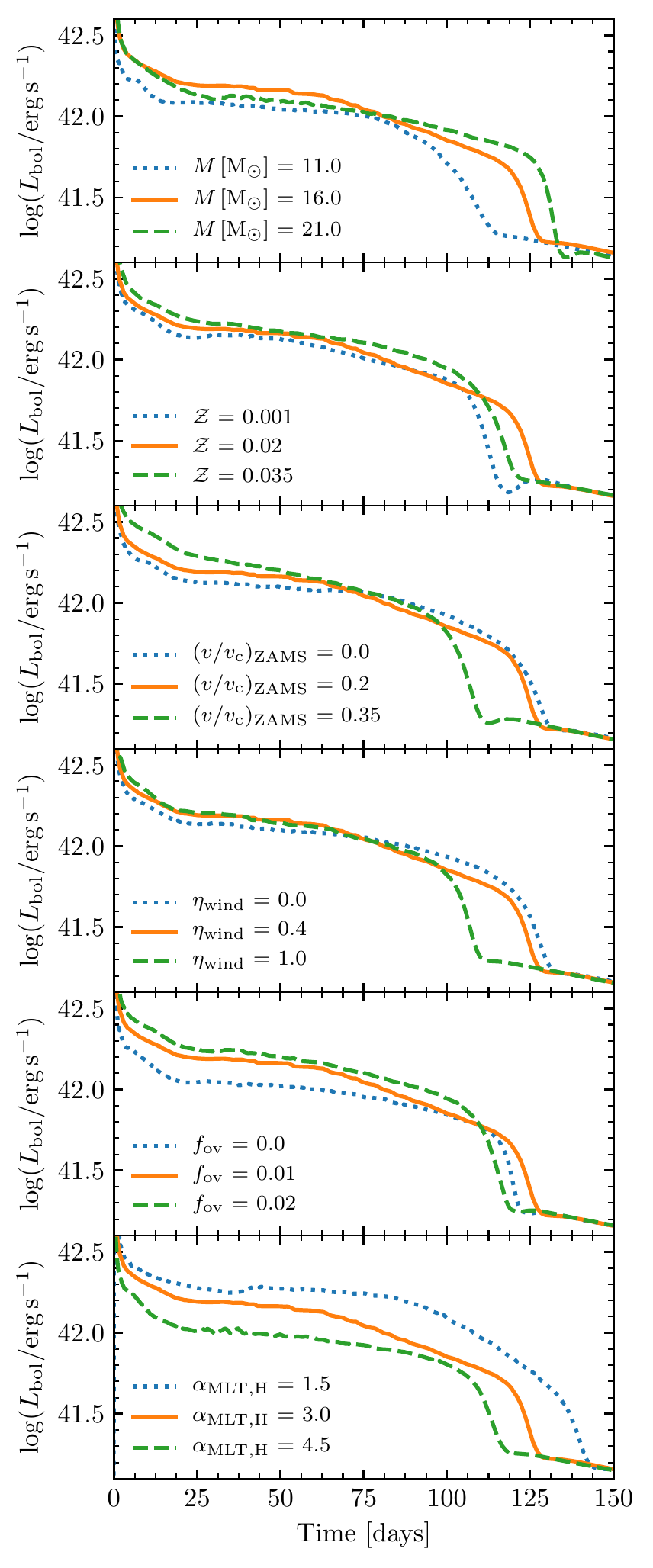}
  \caption{Effect of single-parameter variations to the progenitor std\_16.
    The upper three panels vary initial properties of the
    star; the lower three vary modeling assumptions during
    evolution to core-collapse.}
  \label{f.presn-knobs}
\end{figure}

\begin{figure}[!htb]
  \centering
  \includegraphics[width=\columnwidth]{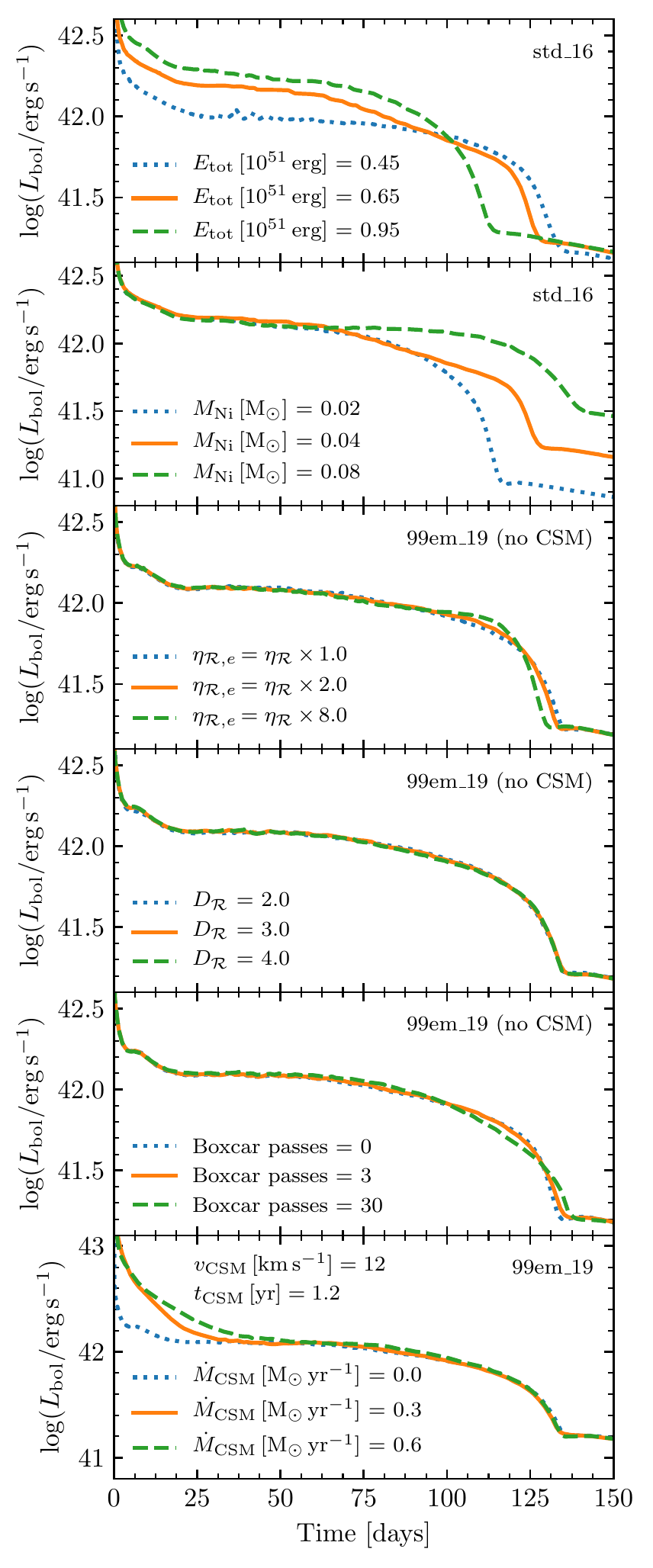}
  \caption{Effect of single-parameter variations associated with the SN
    explosion.  Unvaried parameters have the values of the 
    model listed in the upper right of that panel.
  }
  \label{f.explosion-knobs}
\end{figure}

To enable exploration of the impact of various components of the
RTI mixing, we explicitly allow for the diffusion coefficients for
density, momentum, energy, and composition to be scaled by independent
constant factors relative to the value $\etaRTI{}$ given in
Equation~\eqref{eq:eta-rti}.  We show in
Figure~\ref{f.rti-rho-and-e} the impact of varying the
coefficient in the internal energy \replaced{fluence}{flux} in
Equation~\eqref{e.energy-flux}, which we refer to as
$\etaRTI{,e}$.
These plots show the energy density and density
of the ejecta just before shock breakout in one of our models
(99em\_19) in Table~\ref{tab:models}).  The blue line is for the fiducial value, whereas the red
line is for an extreme increase of a factor of 100. The only locations
that are sensitive to these changes are the innermost mass coordinates
where RTI was most active, the same regions where the lightcurve and
velocities seem to be sensitive to changes related to RTI mixing.
The variable $\etaRTI{,e}$ was found to be a useful ``knob'' to vary for modeling
of specific SNe.

\subsection{Exploring the Explosive Landscape}
\label{s.knobs}

A strength of the new \MESA\ plus \STELLA\ capabilities is their ease
of use.  This enables detailed quantitative studies of large numbers
of core collapse SNe.  The open source nature of \MESA, the
inclusion of \STELLA\ in the \MESA\ distribution, and the repository of
examples contained within the \MESAstar\ test suite allow a user to
obtain models that can be compared directly to observations. Indeed,
with minimal manual intervention, a user can take a star from the
pre-MS to a SN light curve within a few hours of computer runtime.
To emphasize this point, we describe here how this might be done
(Section~\ref{s.howto}).  To demonstrate how parameter choices affect
lightcurves, we show a large sample of variations of a standard case
for ``high-middle-low'' settings of some of the main parameters
(Section~\ref{s.knob-results}). In Section~\ref{s.specificsne} we will
exhibit a few specific models created to be roughly similar to known
Type IIP SNe.  The potential is clear for an extensive database of
such SNe models created using \MESA\ and \STELLA; its actualization is
beyond the scope of this paper.

\subsubsection{Generating Models with \MESA\ plus \STELLA}
\label{s.howto}

The first step in generating a core collapse SN lightcurve is to use
\MESA\ to make a pre-SN stellar model that is undergoing core
collapse. The test case \texttt{example\_make\_pre\_ccsn} can serve as
a useful template.  As part of the required
inlists, the user must select values for the main variables: initial
mass ($M_{\rm ZAMS}$), initial metallicity ($\Zmet$), initial rotation
($(v/v_{\rm c})_{\rm ZAMS}$), overshooting parameter ($f_{\rm ov}$),
wind scaling factor ($\eta_{\rm wind}$), and the mixing length for
MLT in the H envelope $(\alpha_{\rm MLT,H}$).  Of
course, the user may tune other \MESA\ parameters of interest.  The
run from pre-MS to Fe core infall runs automatically given these
parameters and, depending on the case, takes roughly an hour on a
modern multicore desktop workstation.
Users interested in details of pre-SN models may
require settings that lead to significant additional runtime
\citep[e.g.,][]{farmer_2016_aa, renzo_2017_aa}.

The second step loads the model at core infall into \MESA, emulates
the core collapse explosion by excising the core and injecting energy
and Ni (as described in Section~\ref{s.corecoll}) and evolves until
near shock breakout.
The test case \texttt{example\_ccsn\_IIp} can serve as a useful template.
Again, the user must set
the value of the various ``knobs'' controlling the properties of the
explosion such as the total energy $E$ and the $^{56}$Ni mass
$M_{\rm Ni}$. Early ($t<20$ days)
lightcurves of core collapse SNe are better fit when large
amounts of CSM are placed outside the
conventional photosphere \citep{Morozova2016, Dessart2017,
  Morozova2017, Morozova2017b}.
We provide an option to include CSM.
We also provide
the option for ``boxcar'' smoothing of the model abundances before
the handoff from \MESA\ to \STELLA\
\citep{Kasen2009,Dessart2013a,Morozova2015}.
The end result
of this step is a model suitable for input into \STELLA, so one must
also indicate the number of \STELLA\ zones to be used.  This \MESA\
phase from after explosion to near breakout typically
takes about 30 minutes on a modern multicore desktop workstation.

The final step uses the results produced in the previous step as input
to \STELLA\ and evolves the model through shock break-out to the
post-plateau phase.  A script to execute \STELLA\ is provided.
This stage takes about an additional 30 minutes on a modern multicore desktop
workstation for typical cases.  When \STELLA\ finishes, a
post-processing step produces data for comparison to observational results.

\subsubsection{Sensitivity to Variations in Key Parameters}
\label{s.knob-results}

Figure \ref{f.presn-knobs} exhibits the std\_16 model lightcurves as
progenitor parameters are varied.  Many variations behave as expected from previous
analytical and numerical scalings \citep{Popov1993, Kasen2009,
  Sukhbold2016}.  For example, the decrease in the plateau duration
with lower ZAMS masses or higher mass loss (increased
$\eta_{\rm wind}$) is as expected. The increase in plateau luminosity
with decreasing $\alpha_{\rm MLT,H}$ is because those stars with lower
$\alpha_{\rm MLT,H}$ have a larger stellar radius at time of
explosion. However, other variations in these figures are not as
easily diagnosed.

Figure \ref{f.explosion-knobs} exhibits model lightcurves as
explosion parameters are varied.
Again, many cases lead to the
expected outcomes, such as the increase in the plateau
luminosity with increasing explosion energy and the increased duration
of the plateau with increasing nickel mass. The changes caused by
varying the RTI parameters are slight for the compositional mixing and
boxcars, though, as we discussed in Section~\ref{s.rtilight},
modifying the diffusion of energy density during RTI does impact the
shape at the end of the plateau. The impact of the CSM
is similar to that shown by \citet{Morozova2017} and \citet{Dessart2017}.

With experience in the effects of varying the parameters (knobs)
shown in Figures~\ref{f.presn-knobs} and~\ref{f.explosion-knobs},
it is sometimes possible to get a rough match between model and
observations after a dozen or so attempts.  That is about the
amount of effort we undertook to get the models similar to various
observed SNe presented in Section~\ref{s.specificsne}.  Of course the
effects of the various knobs do not combine in any simple manner, so
it can be a nontrivial challenge to find a combination that gives a
good match for both velocities and lightcurve.  Our experience
suggests that it is a good strategy to match velocities before
lightcurves since there are few ways available to shift velocities and
many ways to change lightcurves.  It is important to include
velocities in judging potential matches because of the multiple
degeneracies, as will be seen below where we show two models similar
to SN 1999em with quite different ejecta masses and explosion energies.
Even when using both velocities and lightcurves, it remains a
challenge to find a unique ``best'' match.

\begin{splitdeluxetable*}{lllllllBlllllllBllllllllll}
  \tablecaption{Key properties and parameters associated with the
    core-collapse SN models.  The column ``case'' identifies
    the model.  The ``progenitor parameters'' sub-table lists input
    parameters used during the \MESA\ evolution of the models to core
    infall: initial mass ($M_{\rm ZAMS}$), initial metallicity
    ($\Zmet$), initial rotation ($(v/v_{\rm c})_{\rm ZAMS}$),
    overshooting parameter ($f_{\rm ov}$), wind scaling factor
    ($\eta_{\rm wind}$), and the mixing length for MLT in the H
    envelope $(\alpha_{\rm MLT,H}$).  The ``stellar properties at the
    time of explosion'' sub-table lists physical quantities evaluated in the
    \MESA\ model at the time the Fe core begins to infall: mass
    ($M_{\rm exp}$), effective temperature (\Teff), radius
    ($R_{\rm exp}$), luminosity ($L_{\rm exp}$), mass of the He
    core ($M_{\rm He}$), and initial mass of the Fe core that will be
    excised ($M_{\rm c,i}$).  The ``explosion properties and
    parameters'' sub-table lists input parameters like the
    total energy after explosion $E_{\rm exp}$ and the $^{56}$Ni mass $M_{\rm Ni}$ as
    well as properties of the model including the final core mass
    after fallback ($M_{\rm c,f}$) and the total ejecta mass
    ($M_{\rm ej}$).  This sub-table also lists input parameters used
    in the \MESA\ plus \STELLA\ modeling such as the RTI parameter
    ($\etaRTI{,e}$) and the number of boxcar smoothing passes
    (``boxcar'').  Parameters controlling the extent of the CSM are
    also needed; for a wind profile this includes the wind duration
    ($t_{\rm CSM}$), mass loss rate ($\dot{M}_{\rm CSM}$), and
    velocity ($v_{\rm CSM}$).  Many properties are omitted for the
    stripped case because this is an ensemble of models with a range
    of envelope stripping (see Section~\ref{s.stripped}).}
    \tablehead{ & \multicolumn{6}{c}{progenitor parameters}
    & & \multicolumn{6}{c}{stellar properties at time of explosion} &
    &
    \multicolumn{9}{c}{explosion parameters and properties} \\
    \colhead{case} &
    \colhead{$M_{\rm ZAMS}\,[\Msun]$} &
    \colhead{$\Zmet$} &
    \colhead{$(v/v_{\rm c})_{\rm ZAMS}$} &
    \colhead{$\eta_{\rm wind}$} &
    \colhead{$f_{\rm ov}$} &
    \colhead{$\alpha_{\rm MLT,H}$} &
    \colhead{case} &
    \colhead{$M_{\rm exp}\,[\Msun]$} &
    \colhead{$\Teff\,[\rm K]$} &
    \colhead{$R_{\rm exp}\,[\Rsun]$} &
    \colhead{$\log(L_{\rm exp}/\Lsun)$} &
    \colhead{$M_{\rm He}\,[\Msun]$} &
    \colhead{$M_{\rm c,i}\,[\Msun]$} &
    \colhead{case} &
    \colhead{$E_{\rm exp}\,[10^{51}\,\erg]$} &
    \colhead{$M_{\rm c,f}\,[\Msun]$} &
    \colhead{$M_{\rm ej}\,[\Msun]$} &
    \colhead{$M_{\rm Ni}\,[\Msun]$} &
    \colhead{$\etaRTI{,e}/\etaRTI{}$} &
    \colhead{boxcar} &
    \colhead{$t_{\rm CSM}\,[\rm y]$} &
    \colhead{$\dot{M}_{\rm CSM}\,[\Msun\rm\,yr^{-1}]$} &
    \colhead{$v_{\rm CSM}\,[\rm km\,s^{-1}]$}
  }
  \startdata
  std\_16  & 16.0 & 0.02  & 0.2 & 0.4 & 0.01  & 3.0 &std\_16  & 14.5 & 3960  & 759 & 5.11 & 5.58 & 1.58 &std\_16  & 0.65 & 1.58 & 12.9 & 0.04  & 1.0 & 0  & 0.0 & 0.0 & 0.0 \\
  99em\_16 & 16.0 & 0.02  & 0.2 & 0.4 & 0.01  & 3.0 &99em\_16 & 14.5 & 3960  & 759 & 5.11 & 5.58 & 1.58 &99em\_16 & 0.60 & 1.58 & 12.9 & 0.042 & 2.0  & 3  & 1.0 & 0.25 & 10 \\
  99em\_19 & 19.0 & 0.02  & 0.2 & 0.4 & 0.00  & 3.0 &99em\_19 & 17.8 & 4490  & 603 & 5.13 & 6.58 & 1.50 &99em\_19 & 0.78 & 1.50 & 16.3 & 0.042 & 1.0  & 3  & 1.2 & 0.30 & 12 \\
  05cs     & 13.0 & 0.006 & 0.0 & 0.1 & 0.01  & 3.0 &05cs     & 12.9 & 4280  & 537 & 4.95 & 4.37 & 1.57 &05cs     & 0.16 & 2.51 & 10.4 & 0.009 & 7.0  & 1  & 1.0 & 0.30 & 10 \\
  09N      & 13.0 & 0.006 & 0.0 & 1.0 & 0.01  & 3.0 &09N      & 11.6 & 4290  & 549 & 4.96 & 4.34 & 1.67 &09N      & 0.36 & 1.67 & 9.9 & 0.028 & 30.0 & 3  & 1.4 & 0.30 & 10 \\
  12A      & 11.8 & 0.02  & 0.2 & 0.1 & 0.002 & 3.0 &12A      & 11.6 & 4300  & 525 & 4.94 & 4.08 & 1.49 &12A      & 0.28 & 1.49 & 10.1 & 0.009 & 3.0  & 2  & 0.9 & 0.30 & 10 \\
  13bvn    & 11.0 & 0.02  & 0.0 & 0.0 & 0.01  & 2.0 &13bvn    & 3.4  & 26520 & 7.24 & 4.37 & 3.40 & 1.57 &13bvn    & 0.95 & 1.57 & 1.8  & 0.110 & 1.0  & 5  & 0.0 & 0.0  & 0  \\
  stripped & 17.0 & 0.02  & 0.3 & 0.0 & 0.01  & 3.0 &stripped & ---  & ---   & ---  & ---  & ---  & ---  &stripped & 0.63 & ---  & ---  & 0.037 & 1.0  & 20 & 0.0 & 0.0  & 0  \\
  \enddata
\label{tab:models}  
\end{splitdeluxetable*}

%%% Local Variables:
%%% mode: latex
%%% TeX-master: "paper"
%%% End:

\subsection{Applications to a Few Type IIP SNe}
\label{s.specificsne}

To show examples of what can be accomplished with these new capabilities, we
have modeled four Type IIP SNe: 1999em, 2005cs, 2009N and
2012A. These cover a range of luminosities, plateau durations and
nickel masses and have readily available data
\citep{Pejcha2015a,Pejcha2015b} for bolometric luminosities and
\ion{Fe}{2} velocities.\footnote{We especially thank Ond\v{r}ej Pejcha
  and Stefano Valenti for providing the necessary data.} We follow
the steps described in Section~\ref{s.knobs}, iterating to reach the
matches shown.
The models are not intended to demonstrate the best matches that can
be achieved using \MESA\ and \STELLA.
An investment of more effort could produce better matches,
but is beyond the scope of this paper.  The parameters we choose are
shown in Table~\ref{tab:models}.

\begin{figure}[!htb]
  \centering
\includegraphics[width=0.5\textwidth]{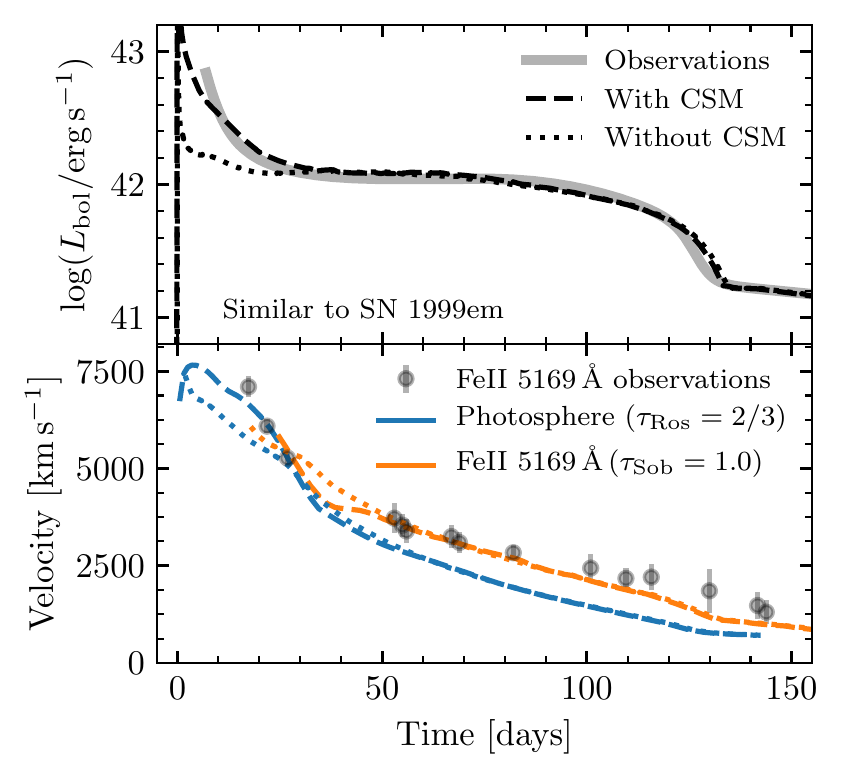}
\caption{Effect of CSM on a model (99em\_19) similar to SN 1999em.
  The influence is particularly apparent in the early time lightcurves
  and velocities.}
  \label{f.csm}
\end{figure}

\begin{figure}[!htb]
  \centering
\includegraphics[width=0.5\textwidth]{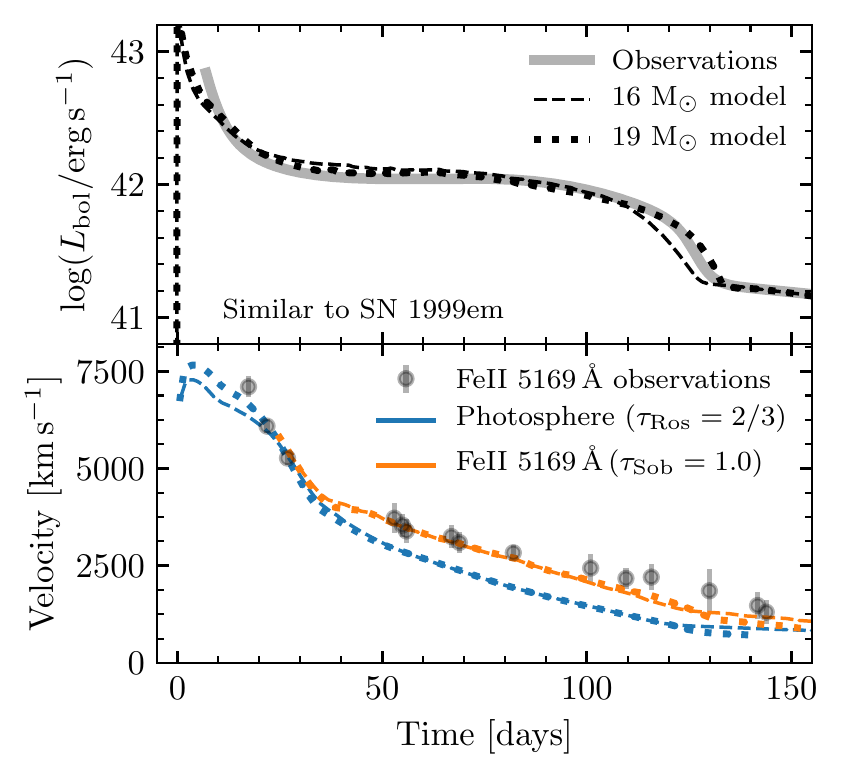}
\caption{Two models similar to SN 1999em (99em\_16 and 99em\_19) with
  significantly different ejecta masses and total energies.}
  \label{f.similar_to_99em}
\end{figure}

\begin{figure}[!htb]
  \centering
\includegraphics[width=0.5\textwidth]{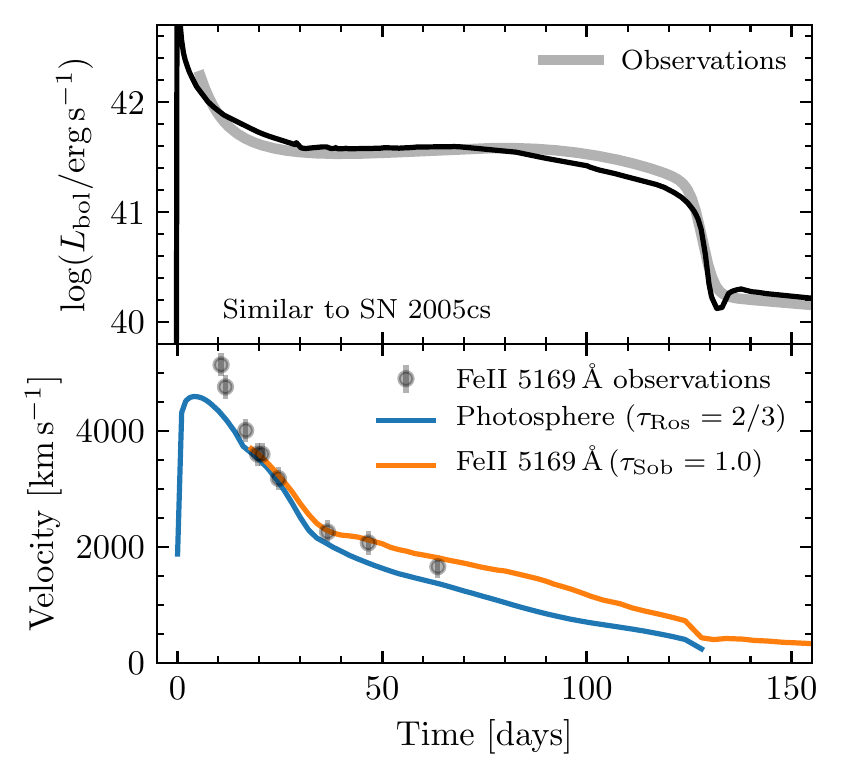}
\caption{Model (05cs) similar to the low luminosity example SN 2005cs.}
  \label{f.similar_to_05cs}
\end{figure}

\begin{figure}[!htb]
  \centering
\includegraphics[width=0.5\textwidth]{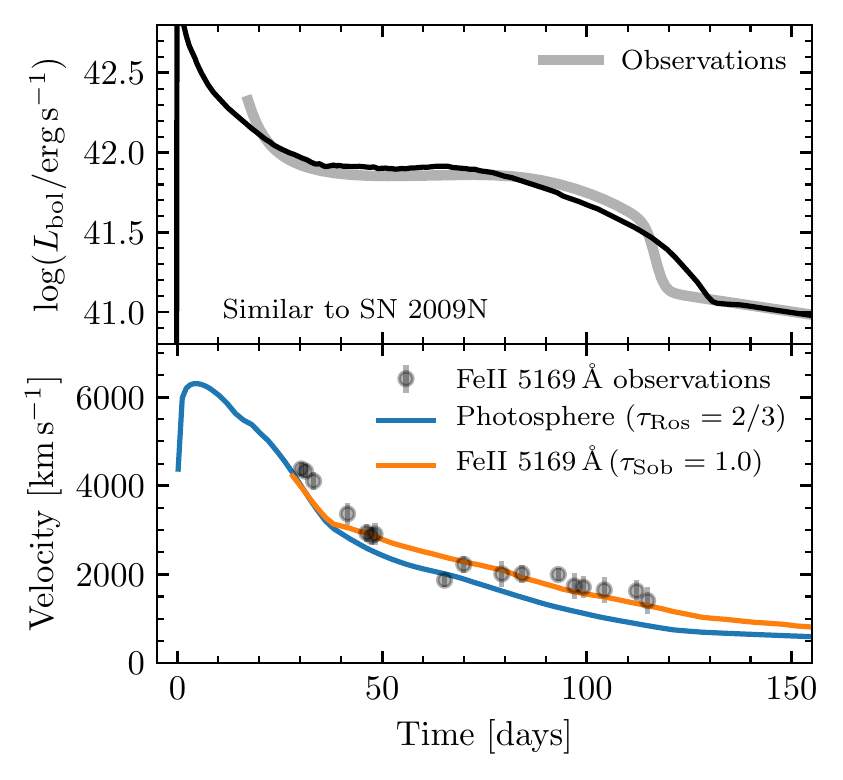}
\caption{Model (09N) similar to SN 2009N. Further experiments might produce a
  model with a better match to the drop at the end of the plateau.
Alternatively, this model might be a useful start when looking for a
match to an observed light curve with a slow decline from the plateau.}
  \label{f.similar_to_09N}
\end{figure}

\begin{figure}[!htb]
  \centering
\includegraphics[width=0.5\textwidth]{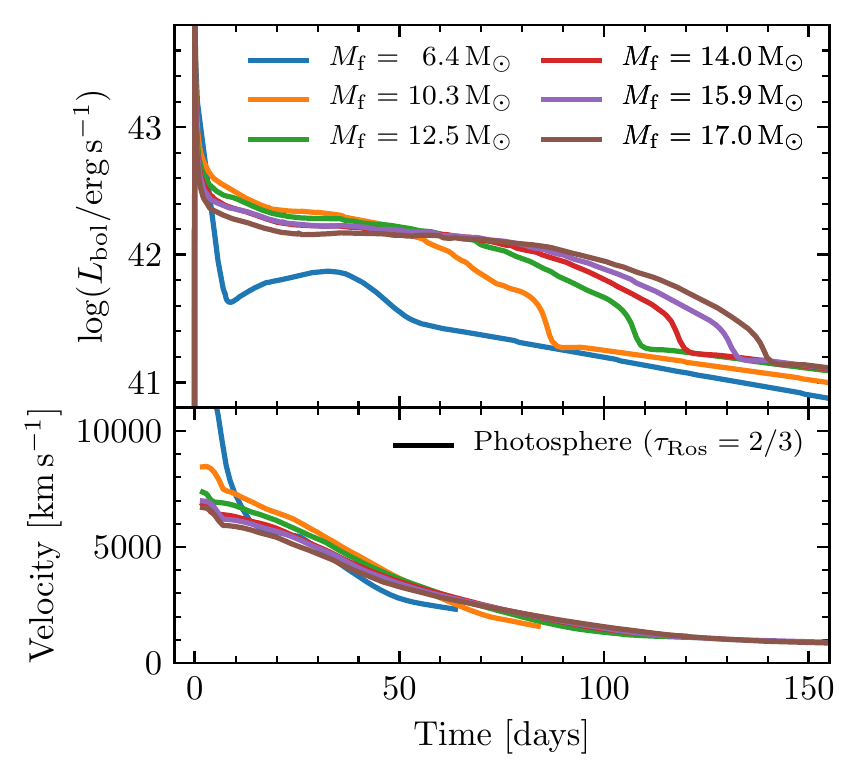}
\caption{Models of partially stripped SNe.  These $M_{\rm ZAMS}
  = 17 \, \Msun$ models have a range of 
  H envelope masses giving rise to a range of plateau durations.  The
  upper panel shows the bolometric luminosity while the lower panel
  shows the velocity.  All models have the same total energy post-explosion of
  $0.65 \times 10^{51}\,\erg$ and a $^{56}$Ni mass of $0.037\,\Msun$.
  Other model parameters are indicated in Table~\ref{tab:models}
  (case ``stripped'').  The lowest mass case has about $0.1\,\Msun$ of
  the H envelope remaining, similar to that of a Type~IIb SN \citep{Ergon2015}.}
  \label{f.partial-stripping}
\end{figure}

We note a few general insights gained from our modeling.  We found that the
radii of red supergiant models from \MESA\ were too large for these Type IIP
SNe models unless we set $\alpha_{\rm MLT,H}=3$. All models
benefited at early times by having some CSM present.
Figure~\ref{f.csm} shows how the early 1999em model predictions change
as CSM is added to the value shown in Table~\ref{tab:models}. The luminosity at early
times is a far better match, as are the earliest velocity data. As
expected, by day 50 and beyond there is no impact of the CSM on the
model predictions.  Comparisons of how the luminosity collapsed at the
end of the plateau drove us to prefer an enhancement in $\etaRTI{,e}$
in several cases.

To exhibit some of the possible degeneracies, we constructed two
distinct models for 1999em.  As shown in
Figure~\ref{f.similar_to_99em}, they are both reasonable models for
the bolometric luminosity and \ion{Fe}{2} velocities. However, their
ejected masses and radii differ significantly, one has $12.9\,\Msun$
and $770\,\Rsun$, whereas the other has $15.9\,\Msun$ and
$600\,\Rsun$.  \citet{Utrobin2007} gave an ejected
mass of $19.0 \pm 1.2\,\Msun$, a radius of $500\,\Rsun$, and an
explosion energy of $1.3\times 10^{51}$\,erg.
\citet{Bersten2011} gave an ejecta mass of
$17.6\,\Msun$, radius of $800\,\Rsun$, and explosion energy of
$1.25\times10^{51}$\,erg.
\citet{Utrobin2017} model this event with
a 3D simulation from explosion to shock breakout,
similar to the Munich L15 model we discuss in
Section~\ref{s.RTI}, but with an explosion energy of about
$0.5 \times 10^{51}$\,erg.  For comparison, the \MESA\ models for 1999em
have total energies after explosion of $0.60\times 10^{51}$\,erg for
the case with $12.9\,\Msun$ ejected mass, and $0.78\times 10^{51}$\,erg
for the case with \replaced{$15.9\,\Msun$}{$16.3\,\Msun$} ejected mass.

We previously showed 2012A in Figure~\ref{f.velocity-comparison}. Our
model had an ejected mass of $10.1\,\Msun$ to compare with
$7.8\,\Msun$ from \citet{Morozova2017b}, $12.5\,\Msun$ from
\citet{Tomasella2013} and $13.1\pm 0.7\,\Msun$ from
\citet{Utrobin2015}.  \citet{Tomasella2013} also reported a progenitor
luminosity of $\log(L/\Lsun)=4.73\pm 0.13$, just a bit fainter than
our model's value. Figure \ref{f.similar_to_05cs} shows our model for
2005cs. Our model has an ejected mass of $10.4\,\Msun$, slightly
higher than the $9.5\,\Msun$ reported by \citet{Spiro2014} and the
$7.8\,\Msun$ reported by \citet{Morozova2017b}. Figure
\ref{f.similar_to_09N} shows our model for 2009N, which has an ejected
mass of $9.9\,\Msun$, whereas \citet{Morozova2017b} found $9.3\,\Msun$
and \citet{Takats2014} found $11.5\,\Msun$.

\subsection{Partially Stripped Core Collapse SNe}
\label{s.stripped}

There is a well-defined class of core collapse SNe where either
much (Type IIb) or nearly all (Type Ib and Ic) of the H envelope was
lost prior to the core collapse event.  \citet{Dessart2015} performed
detailed radiative transfer models for a large set of progenitors from
binary evolution, while \citet{Morozova2015} carried out diffusive
calculations
with varying amounts of mass loss.
\citet{Yoon2017} explored \MESA\ models constructed
from binary transfer scenarios and applied them to a set of well
observed Type IIb events.  We have not yet been able to deal
successfully with Ic models because of numerical problems related to
the extreme ejecta velocities that occur at shock breakout.  However,
it is possible to do both IIb and Ib models as shown here.

\begin{figure}[!htb]
  \centering
\includegraphics[width=0.5\textwidth]{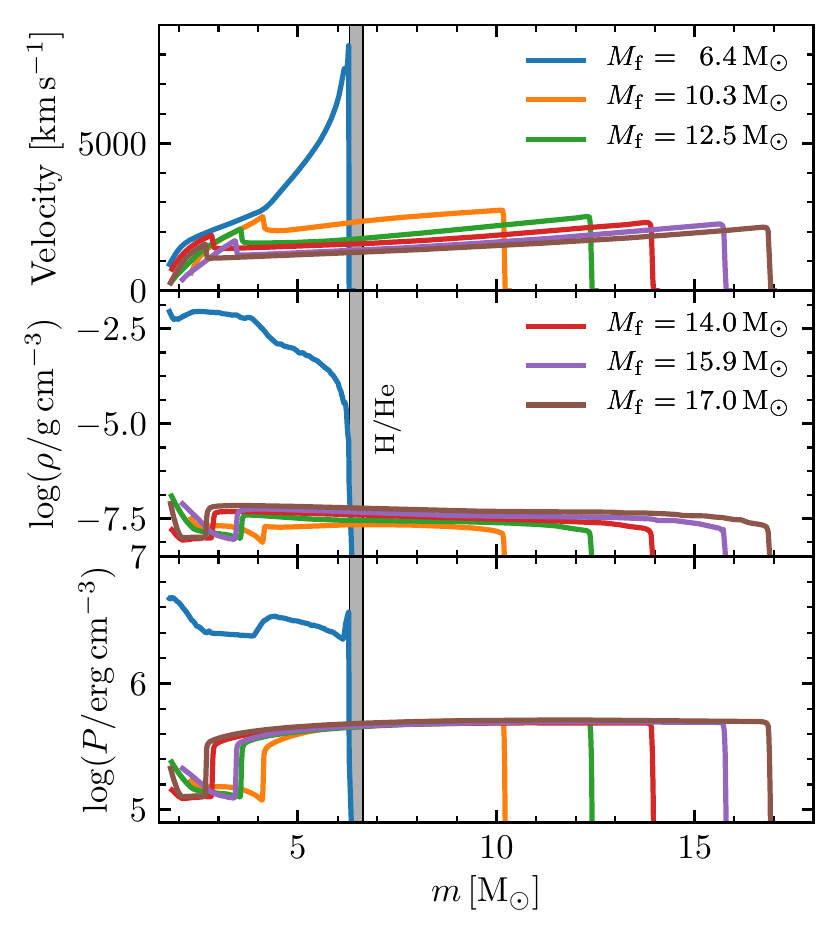}
\caption{Models of $M_{\rm ZAMS} = 17\,\Msun$ stars (case ``stripped'') that
  have experienced a range of stripping.  The density, velocity and
  pressure profiles are shown at the time of handoff from \MESA\ to
  \STELLA, very close to shock breakout.  The gray band shows the
  range of locations of the H/He boundary at the time of explosion.}
  \label{f.reverse-shock}
\end{figure}

In Figure \ref{f.partial-stripping}, we show the \MESA\ plus \STELLA\
predictions for luminosities and photospheric velocities for a range
of models with varying amounts of mass stripped from a
$17\,\Msun$ ZAMS model, ranging from the entire initial H envelope still
remaining down to only $0.1\,\Msun$ of the H envelope left at the time
of explosion.
Similar to Figure~7 of \citet{Morozova2015}, the plateau
period
becomes shorter as the residual H shell mass declines.
Our smallest mass model has an H envelope
mass comparable to typical
models of Type~IIb SNe and generates a light curve  
comparable to observed Type~IIb SNe \citep{Ergon2015}.
Figure~\ref{f.reverse-shock} shows the interior properties
of these same models near the moment of shock breakout.
For models which have been stripped,
the reverse shock has not reached the IB at the time the forward
shock reaches the surface. Since RTI mixing does not occur in \STELLA,
these models would incompletely include the effects of the RTI.

\begin{figure}[!htb]
  \centering
\includegraphics[width=0.5\textwidth]{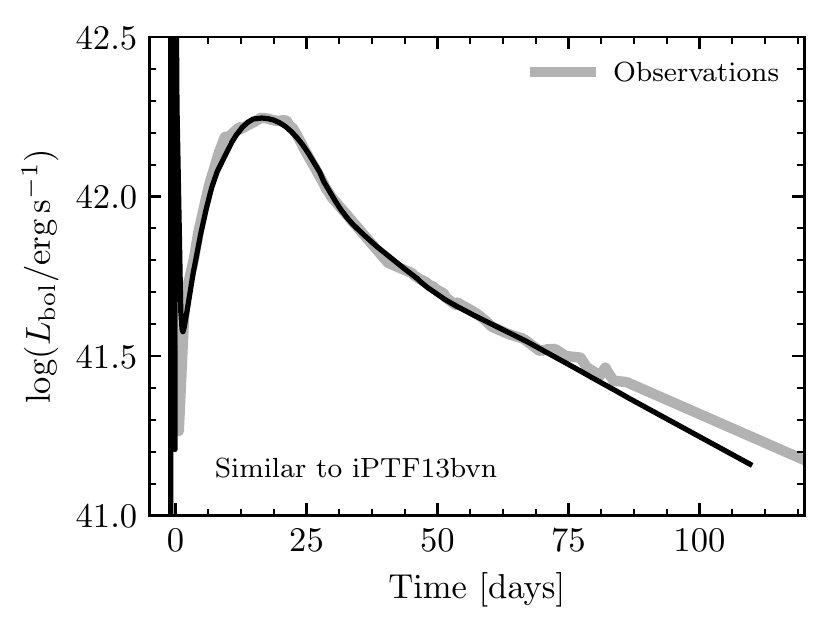}
\caption{Comparison of model (13bvn) similar to the stripped Type Ib
  SN iPTF13bvn with the observed bolometric lightcurve.}
  \label{f.iptf13bvn}
\end{figure}

\citet{Cao2013} discovered the fully stripped Type Ib SNe
iPTF13bvn in the nearby spiral galaxy NGC 5806 with the intermediate
Palomar Transient Factory \citep{Law2009}. This is one of only a few
stripped SNe with a progenitor detection.  Using data from
\citet{Cao2013} and \citet{Fremling2014}, we show in Figure
\ref{f.iptf13bvn} our model that approximately matches the iPTF13bvn
light curve. The model is derived from an $11\,\Msun$ ZAMS model
and has a remaining
mass of only $3.4\,\Msun$ at the time of explosion with total energy
after explosion of $0.95\times 10^{51}$ ergs and a $^{56}$Ni mass of
$0.11\,\Msun$ distributed throughout the remaining star (ejecta mass
1.8\,\Msun). \citet{Fremling2014} also modeled this lightcurve,
finding the total energy to be $0.85^{+0.5}_{-0.4}\times 10^{51}$
ergs, with a a $^{56}$Ni mass of $0.049^{+0.02}_{-0.012}\,\Msun$, and
total ejecta mass of $1.94^{+0.50}_{-0.58}\,\Msun$. Our parameters are
similar, falling within the range of the quoted uncertainties, except
for the $^{56}$Ni mass.

%%% Local Variables:
%%% mode: latex
%%% TeX-master: "paper"
%%% End:

\section{Black Hole Formation}\label{s.bhform}

Compact objects are a natural product of the evolution of massive
stars.  A broad consensus on which 
massive stars produce black holes (BH) has not yet been reached 
\citep{timmes_1996_ac, fryer_2001_ab, heger_2003_aa, eldridge_2004_aa,
zhang_2008_ab, ugliano_2012_aa, clausen_2015_aa, sukhbold_2016_aa,
muller_2016_ab, limongi_2017_aa}.

The lack of consensus is due to a variety of
differences in the modeling, including
stellar wind treatments during the pre-supernova stage \citep{renzo_2017_aa};
shellular rotation prescriptions \citep[e.g.,][]{limongi_2017_aa}; 
sensitivity to the initial metallicity \citep[e.g.,][]{oconnor_2011_aa},
number of isotopes in the reaction network \citep{farmer_2016_aa},
adopted values of critical reaction rates \citep{deboer_2017_aa,fields_2017_aa},
and ignition of core carbon burning \citep{farmer_2015_aa,cristini_2017_aa, petermann_2017_aa};
variations from spatial and temporal resolution \citep{farmer_2016_aa};
convection during core-collapse \citep[e.g.,][]{couch_2015_aa}; 
and effects from binary partners \citep[e.g.,][]{marchant_2016_aa, batta_2017_aa}.  
In addition, current estimates of the neutron star and BH
initial mass function chiefly rely on parameterized explosion models
and not on first principles calculations. 

This section explores \MESA\ models that can produce BHs. 
First, we consider \Mzams~$\le$~60\Msun\ models
that can form a BH without encountering dynamical instability due to
$e^+e^-$ pair production.
Second, we survey \Mzams~$\ge$~60\Msun\ models that encounter
dynamical instability,
either entering the $\gammaone$\,$\le$\,4/3 regime once to
produce a pair-instability supernova (PISN) 
\citep{fowler_1964_aa, rakavy_1967_aa, rakavy_1967_ab, barkat_1967_aa, fraley_1968_aa,
ober_1983_aa, fryer_2001_aa, scannapieco_2005_aa, kasen_2011_aa, chatzopoulos_2013_ab},
or multiple times to produce a pulsational pair-instability supernova
(PPISN) and a BH remnant
\citep{barkat_1967_aa, woosley_2007_aa, chatzopoulos_2012_aa, Woosley2017}.

\begin{figure}[!htb]
\includegraphics[width=\columnwidth]{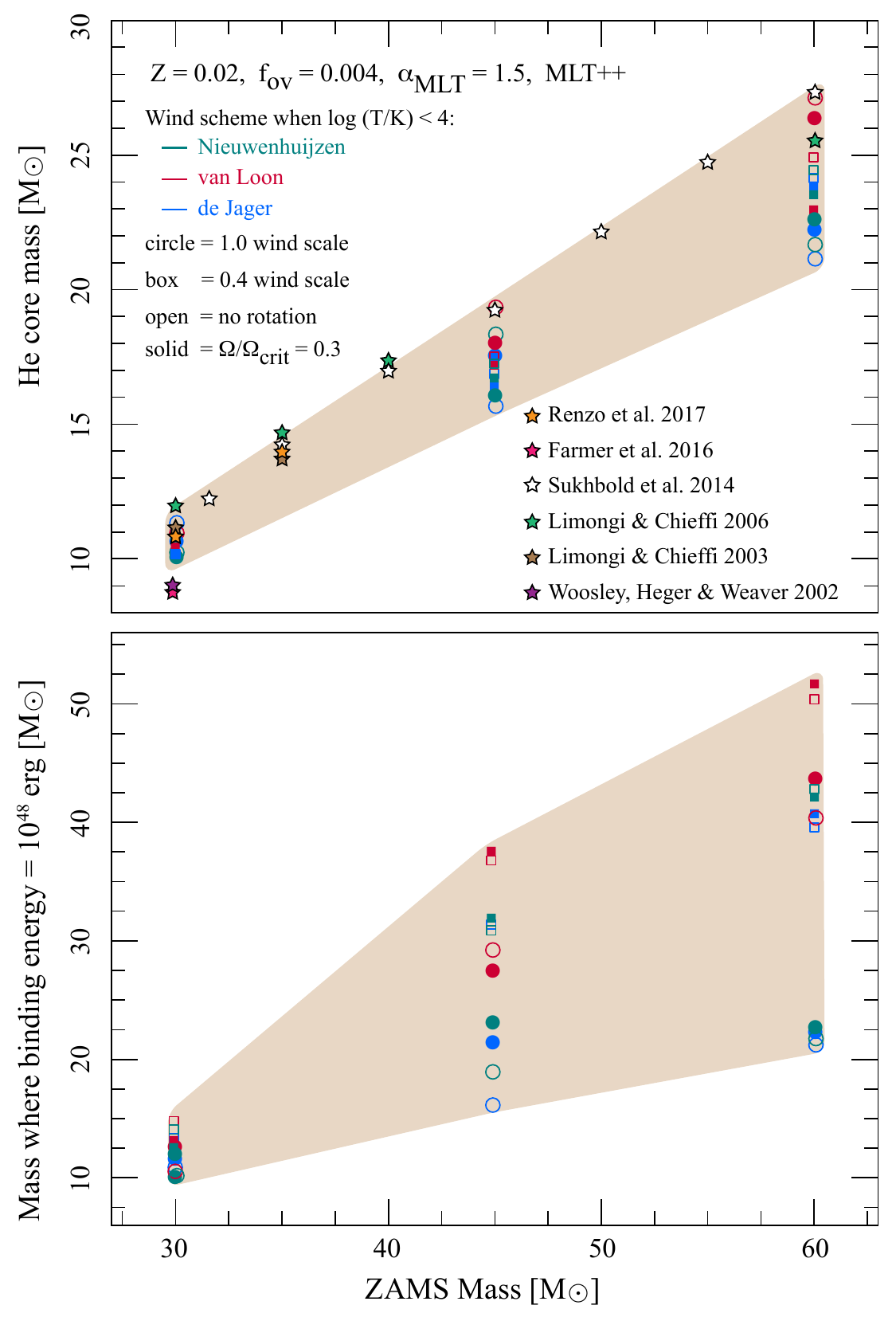}
\caption{
He core mass (upper panel) and mass location where the gravitational binding energy
is equal to $10^{48}\, \rm erg$ (lower panel) for ZAMS masses of 30, 45, and 60 \Msun.
Three stellar wind treatments, two wind scaling factors, and two rotation
rates are shown for each ZAMS mass. The variation\added{, illustrated by the tan band,} induced by these modelling choices
increases with ZAMS masses.
\added{Also shown are models from the literature \citep{renzo_2017_aa,farmer_2016_aa,sukhbold_2014_aa,limongi_2003_aa,limongi_2006_aa,woosley_2002_aa},
although each adopts different
modelling choices and definitions of the He core mass.}
}
\label{f.bh_hecore}
\centering
\end{figure}

\subsection{Progenitors that Do Not Pulse}\label{s.nopulse}

The upper panel of Figure \ref{f.bh_hecore} shows the He core mass
from $\Zmet$=0.02 \Mzams =30, 45, and 60 \Msun\ models.  The lower
panel shows the mass location where the binding energy is
$10^{48}\, \rm erg$. Each ZAMS mass uses an exponential convective overshoot
parameter $\fov$=0.004 applied at all convective boundaries, a mixing
length $\alphaMLT$=1.5, MLT++ enabled (see \mesatwo), and is run to
the onset of core-collapse (infall velocity $\ge$ 1000 km s$^{-1}$).
We illustrate the variation in the He core mass and mass location
where the binding energy is $10^{48}\, \rm erg$ from the effects of
rotation, wind strength, and the wind schemes of
\citet{nieuwenhuijzen_1990_aa}, \citet{van-loon_2005_aa}, and \citet{de-jager_1988_aa}.

To estimate a BH mass from the structure at core collapse, we
use the mass location where the binding energy integrated from the surface exceeds
10$^{48}$\,erg.  This is motivated by neutrinos removing $\approx$\,10$^{53}$\,erg
during core-collapse, reducing the gravitational mass of the core by
$\approx$\,0.3\,\Msun.
The outer part of the star responds to the sudden decrease in the gravitational
field by driving a sound wave that steepens into a shock that unbinds
some of the outer envelope \replaced{\citep{quataert_2016_aa}}{\citep{Coughlin2017}}.  Mass with
binding energy $\lesssim$\,10$^{47}$\,erg is likely to be ejected
\citep{Nadezhin1980, Lovegrove2013} while mass that is not ejected
will likely become part of the BH. 
Figure \ref{f.bh_hecore} suggests that BH masses estimated in
this simple way can be significantly larger than the final He core
mass, and more sensitive to the assumed model parameters.  For example, there is wide variation in the expected
BH mass for the 60\,\Msun\ progenitor depending on choice of
wind scheme and scaling factor, whereas modest rotation has a smaller
effect.

\subsection{Pulsational Pair-instability Supernovae}\label{s:ppisn}

Stars with \Mzams\,$\gtrsim$\,60\,\Msun\ are expected to become dynamically
unstable before core O depletion as $e^+e^-$ pair production leads to
regions where the adiabatic index $\gammaone \le$~4/3
\citep{fowler_1964_aa,rakavy_1967_ab}.  The ensuing collapse results
in explosive O burning, with a variety of possible outcomes.
Stars can produce PISNe where the energy injected from
explosive O burning completely unbinds the star without leaving a
compact remnant.
Alternatively, stars can undergo a cyclic pattern of
entering the pair instability region, contracting, burning,
and expanding, leading to PPISNe.

Individual pulses in a PPISN can remove a large fraction of the mass of the star
at velocities of several thousand km\,s$^{-1}$, with the remaining
material settling down into hydrostatic equilibrium at a lower central
temperature than before the pulse. The star then contracts as it loses energy due to radiation and
neutrino emission until it undergoes an additional pulse or collapses to form
a BH. Depending on its initial mass, the time between pulses varies from a fraction
of a year to millennia, with the outer ejected layers expanding to very low densities and
becoming optically thin. 

{\MESA} currently cannot simultaneously follow the long term
evolution of the bound core and the ejecta, making it necessary to remove the
unbound layers from the stellar model. To do this we model individual
pulses using both the Riemann solver hydrodynamics (Section~\ref{s.hydro}), as well
as the 1D treatment of the Rayleigh-Taylor instability (Section~\ref{s.RTI}),
until the star is approximately in hydrostatic equilibrium.
We then relax a new stellar
model using the methods described in Appendix \ref{app.relax}, such that it has
the same mass, entropy, and composition profiles as the layers that remained
bound in the hydrodynamical model. This model
is then evolved assuming hydrostatic equilibrium until the onset of another
pulse or the final core-collapse to a BH.

\begin{figure}%[!htb]
\includegraphics[width=\columnwidth]{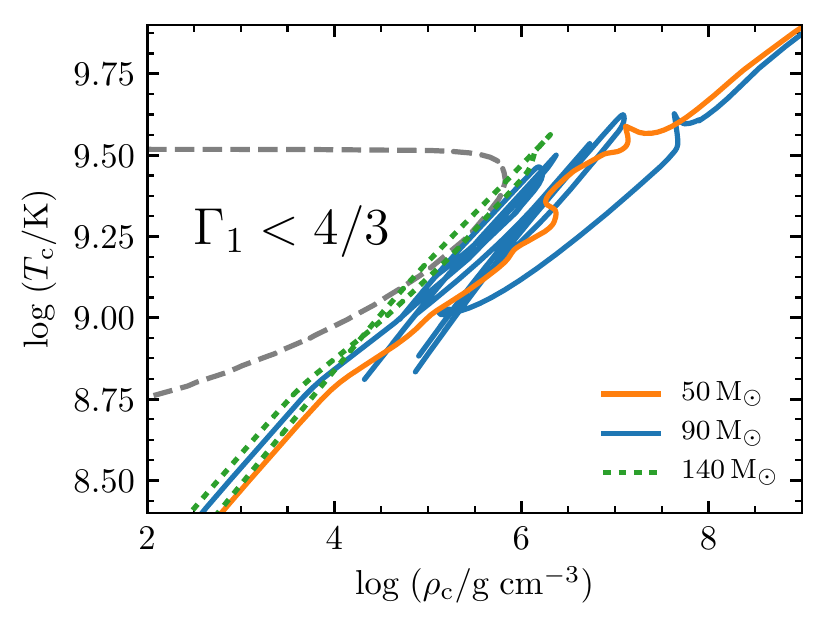}
\caption{
Central temperature and density for models with metallicity $\Zmet=0.001$ undergoing core
   collapse (50\,\Msun), PPISN (90\,\Msun), and PISN (140\,\Msun).
}
\label{f.bh_ppisn1}
\centering
\end{figure}

As an example, we compute models at a metallicity $\Zmet=0.001$, using
similar parameters as in Section~\ref{s.nopulse}.
The \texttt{van Loon} scheme is used for low-temperature winds with a scaling
factor of $0.4$. In addition, convection is modeled as a time-dependent process
by limiting changes in convective velocities as in \citet{Arnett1969} and
\citet{Wood1974}.

\begin{figure}%[!htb]
\includegraphics[width=\columnwidth]{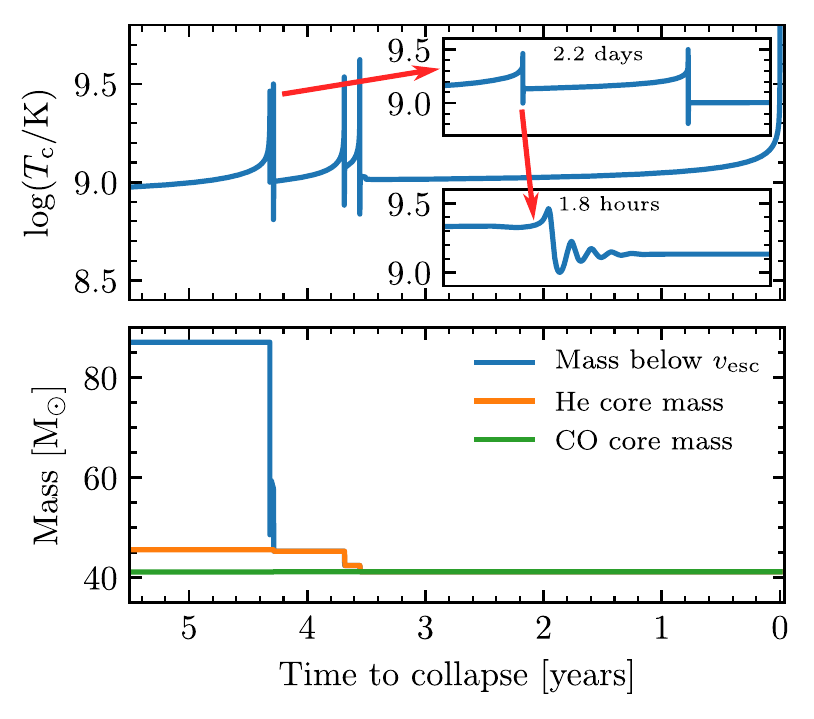}
   \caption{Late-time evolution of a $M_{\rm ZAMS}=90\,\Msun$ star with a
   metallicity $\Zmet=0.001$ undergoing a PPISN. (upper panel) Evolution of the central
   temperature showing a zoomed-in region covering 2.2 days which contains the
   first two pulses, as well as an additional zoom-in covering 1.8 hours which
   shows the first pulse and its ring-down into hydrostatic equilibrium.
   (lower panel) Total mass of the star below the escape velocity, He core mass
   and CO core mass during pulsations. The \href{http://mesa.sourceforge.net/mesa4.html}{online animated Figure} shows the time
   evolution of these quantities and the interior structure of the star.}
\label{f.bh_ppisn2}
\centering
\end{figure}

Figure \ref{f.bh_ppisn1} shows the evolution in the $\rho_{\rm c}$-$T_{\rm c}$ plane
during late burning stages for a $90\,\Msun$ model undergoing a PPISN,
a $50\,\Msun$ model experiencing iron-core collapse,
and a $140\,\Msun$ model producing a PISN. Although the
center of the $90\,\Msun$ star does not evolve into the region where
$\gammaone<4/3$, the outer layers of the CO core do.
Coupled with enhanced neutrino losses from
pair-annihilation, this causes the star to collapse and undergo four distinct pulses
before finally collapsing into a BH. At the onset of the first pulse, the star has a
mass of $87.1\,\Msun$, with a He core of $45.6\,\Msun$ and a CO core of
$41.1\,\Msun$. As shown in Figure \ref{f.bh_ppisn2} the first two pulses happen
within two days of each other and they remove the entire H envelope. The remaining two
pulses remove almost the entire He envelope, resulting in a final mass of
$41.2\,\Msun$ when the star collapses into a BH.

\begin{figure}%[!htb]
\includegraphics[width=\columnwidth]{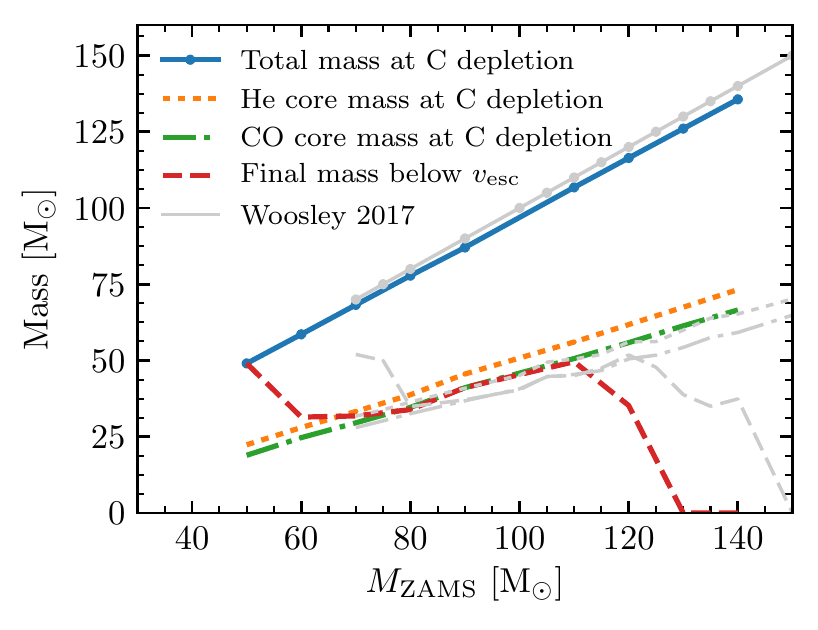}
\caption{
Total mass, He core mass, and CO core mass at carbon depletion for single
   stars at a metallicity ${\Zmet=0.001}$.
   The $50\,\Msun$ model undergoes iron-core
   collapse, while the $140\,\Msun$ model experiences complete
   disruption through a PISN. All other models experience PISNe. \added{For comparison,
   the models with no mass loss from \citet{Woosley2017} at a metallicity
   ${\Zmet=0.0016}$ are also shown.}
}
\label{f.bh_ppisn3}
\centering
\end{figure}

Figure~\ref{f.bh_ppisn3} shows key masses on a grid encompassing ZAMS
masses for which PPISNe occur under our model assumptions.
Our PPISN progenitors have He core masses in the
range of $28\,\Msun-67\,\Msun$, and no BHs with masses above
$50\,\Msun$ are formed. These results are in broad agreement with
\citet{Woosley2017}. \added{However, Figure \ref{f.bh_ppisn3} shows that the range of
ZAMS masses that result in a PPISN is significantly different to the one
computed by \citet{Woosley2017}. This can be attributed to a different choice of input
physics such as core overshooting, as well as a different initial metallicity.}

%%% Local Variables:
%%% mode: latex
%%% TeX-master: "paper"
%%% End:

\section{Energy Accounting in Stellar Evolution}\label{s.epsgrav}

\mesaone\ describes the stellar structure equations and their
implementation in \MESA.  In order to provide physically and
numerically accurate solutions of these equations, it is often necessary
to evaluate them in different ways depending on the details of the
star being simulated.  In particular, there are a number of different
ways to formulate and evaluate the equations solved by \MESA\ that
encode local and global energy conservation.  The goal of this section
is to clarify the available options, discuss when and why they are
used, and describe how various forms of energy are tracked and
accounted for in stellar evolution. While in places this section reads
like a tutorial, it is in fact the first time we have presented a
detailed description of a complex and critical aspect of how \MESA\ works,
information that is important for intelligent use of this software
tool. 

In Section~\ref{s.eps_fund} we describe the fundamental equations we
are solving, and in Section~\ref{sec:implementation} we describe choices
associated with their numerical implementation.
In Section~\ref{s.implicit_hydro} we describe the connection between
the form of the energy equation typically used in stellar evolution
calculations and the version used when the hydrodynamics options
discussed in Section~\ref{s.hydro} are enabled.
In Section~\ref{sec:ionization} we clarify how the energy associated
with ionization is included in \MESA.
In Section~\ref{sec:latent-heat} we describe the numerical approach
necessary to ensure that the latent heat associated with crystallization in a white dwarf (WD)
is included in \MESA.
In Section~\ref{sec:eos-blending} we discuss the difficulties
introduced by the necessity to blend between different equations of
state (EOS) as the thermodynamic conditions in the stellar interior change,
and how \MESA\ minimizes artifacts associated with these
blends.
In Section~\ref{sec:grav-settl} we discuss the energy associated with
gravitational settling.

\subsection{Fundamental Equations}\label{s.eps_fund}

In the stellar structure equations \citep[e.g.,][]{Cox68,Kipp12}, energy
conservation is typically formulated by considering the energy flow in
and out of a fluid parcel.  In this Lagrangian picture, to
understand how the energy of a fluid parcel is changing, we account
for the specific (i.e.,~per unit mass) rate of energy injection into the
parcel, $\epsilon$, and the specific rate of energy flow through the
boundaries ($\partial L/\partial m$; $L(m)$ is the luminosity profile and
$m$ the Lagrangian mass coordinate).  The specific heating rate
($\Dif q/\Dif t$) for the parcel must then satisfy
\begin{equation}
  \label{eq:local-energy}
\DDt{q} = \epsilon - \ddm{L},
\end{equation}
where $\Dif /\Dif t$ is the Lagrangian time derivative.
Except in the case of hydrodynamics described in
Section~\ref{s.hydro} (where a total energy equation is solved; see
Section~\ref{s.implicit_hydro}), the basic equation to be solved is
always some form of Equation~(\ref{eq:local-energy}).
By tradition, the negative of the left-hand side of
Equation~\eqref{eq:local-energy} is called $\epsgrav$.

Thermodynamics relates the heating of material to the changes in its
properties.  The first law of thermodynamics states that the
total heat added $\delta Q$ for a parcel is
\begin{equation}
  \label{eq:first-law}
  \delta Q \equiv \dif E + P \dif V,
\end{equation}
where $E$ is the internal energy, $P$ is the pressure, and $V$ is the volume.
Let $N_i$ be the number of particles of species $i$ in the parcel. Then
expanding $E$ in terms of the independent thermodynamic basis
variables $(S,V,N_i)$ yields the following thermodynamic identity:
\begin{equation}
  \label{eq:thermo-id}
  \dif E + P \dif V = T \dif S  + {\sum_i \mu_i \dif N_i}~,
\end{equation}
where $S$ is the entropy, and $T$ is the temperature.
The sum runs over all species present, and
\begin{equation}
\mu_i \equiv \left( \frac{\partial E}{\partial N_i} \right)_{S,V}
\end{equation} 
is the chemical potential for species $i$.

The number abundance of every
species is defined with reference to the total number of
baryons $N_{\rm B}$ as $Y_i \equiv N_i/N_{\rm B}$. Denoting Avogadro's
number by $N_{\rm A}$, the atomic mass unit is  $m_{\rm amu} = 1\,{\rm g}/
N_{\rm A}$. The specific (i.e.,~per unit mass)
form of Equation~\eqref{eq:thermo-id} is then given by multiplying by
the invariant $N_{\rm A}/N_{\rm B}$ to find
\begin{equation}
  \label{eq:specific-id}
  \delta q \equiv
  \dif e + P \dif \left( \frac 1 \rho \right)
  = T \dif s  + {\sum_i \left( \frac{\partial e}{\partial Y_i} \right)_{s,\rho} \dif Y_i}~.
\end{equation}
The total baryonic mass density is $\rho$, so
that $1/\rho$ is the specific volume, and $e$ and $s$ are specific
energy and entropy respectively.
Local thermodynamic equilibrium (LTE)
determines a unique solution for the ionization state of each isotope.
Thus, composition is completely specified by a set of number
abundances $\{Y_i\}$ for all nuclear isotopes.

Equation~\eqref{eq:specific-id} is relativistically correct when the
rest mass is included in the energy and the chemical potentials.
Therefore, in principle, changes in nuclear rest masses
due to nuclear reactions could be
accounted for via this equation.  However, in \MESA, the energetic
effects associated with composition changes due to nuclear reactions
are not included in \epsgrav.  Instead, these important terms are
accounted for via \epsnuc\ (the specific energy generation rate of
nuclear reactions) which is evaluated separately and included as part
of the local source term $\epsilon$ in Equation~\eqref{eq:local-energy}
(see \mesaone).

It is often convenient to specify compositions in terms of the
baryonic mass fractions $\{X_i\}$ via the relation ${X_i = A_i Y_i}$,
where $A_i$ is the mass number for isotope $i$.
Since rest-mass changes due to nuclear reactions are handled
separately from $\epsgrav$, $\rho$ and $\{X_i\}$ can be treated as
independent basis variables without introducing any ambiguity into the chemical
potential term in Equation~\eqref{eq:specific-id}.
Some EOS options express the composition dependence in terms of aggregate quantities;
examples include
hydrogen abundance $\Xhyd$, helium abundance $\Yhel$, metallicity
$\Zmet$, average mass number $\bar{A}$, and average atomic
number $\bar{Z}$.

The value
for $\epsgrav$ can be computed beginning from either
the left or right hand side of the equals sign in Equation~\eqref{eq:specific-id}.
Usually, some form of the left hand side is used, but in
Section~\ref{sec:latent-heat} we will describe a case where it is
more convenient to use the right hand side.

\subsection{Implementation}
\label{sec:implementation}

Basic variables are those quantities directly calculated by \MESAstar's solver. Examples
include velocity, radius, and the thermodynamic variables.
\MESA\ offers options for selecting ${(\rho,T,\{X_i\})}$ or
${(P_{\rm gas},T,\{X_i\})}$ as the thermodynamic variables.
The EOS routines
calculate other thermodynamic quantities as a function of
the chosen variables, e.g.,~${e=e(\rho,T,\{X_i\})}$.
\MESA\ solves the stellar structure equations implicitly, thus it is possible to
approximate total time derivatives of any quantity calculated in the
stellar model simply by differencing its value at the start and end of
a timestep.  Therefore, one way to evaluate \epsgrav\ would be to
directly calculate the time derivatives in
Equation~\eqref{eq:specific-id}.
Two possible versions of \epsgrav\ would then be
\begin{equation}
-\epsgrav = T\DDt{s} + \sum_i \frac{\partial e}{\partial Y_i} \DDt{Y_i}~,
\label{eq:epsfirst}
\end{equation}
and
\begin{equation}
-\epsgrav= \DDt{e} + P \DDt{} \left( \frac 1 \rho \right)~.
\label{eq:epssecond}
\end{equation}
While simple to construct, the finite differences necessary to
calculate these equations are often numerically problematic.

To see the potential numerical issues, consider the 
implementation of Equation~\eqref{eq:local-energy} using
Equation~\eqref{eq:epssecond} in cell~$k$ with mass~$dm_k$
over a timestep~$\dt$.  The derivative of a quantity $\Dif y/\Dif t$ is
typically constructed as a finite difference of $y$ over the timestep,
so after integrating over the mass of zone $k$, we have
\begin{equation}
\begin{aligned}
\label{eq:eps_numerical1}
0 =
\bigg( &\epsilon_{k} - \frac{e_{k, \rm end} - e_{k,\rm start}}{\dt} \\
& - P_k \frac{1/\rho_{k,\rm end} - 1/\rho_{k,\rm start} }{\dt}
\bigg) dm_k \\
- &(L_{k} - L_{k+1})~.
\end{aligned}
\end{equation}
The implicit solver scheme in \MESA\ attempts to reduce the residual
from evaluating the right hand side of this equation below some
tolerance.

While the implicit scheme in \MESA\ may sometimes find acceptable results for
an equation such as Equation~\eqref{eq:eps_numerical1}, finite
numerical precision can result in troublesome behavior for the time
derivatives involving subtractions. In particular, over a small
timestep where the change in $e_k$ or $\rho_k$ is small compared to
the overall magnitude of these quantities, floating point arithmetic
can suffer significant loss of precision. When energy scales arising
from these types of finite difference derivatives are comparable to
$\epsilon_k$, the implicit solver may be unable to converge to an
acceptable solution.

To avoid these problems, the equations can be cast in terms of
derivatives that are not evaluated using subtractions. 
Such derivatives are available only for the basic
variables, since the Jacobian matrix for an evolution step
satisfying the equations of stellar structure in \mesa\ is written in
terms of the basic variables and their derivatives (see
\mesaone, Section~6.2 and \mesatwo, Section B.2 and Figure~47).
For \mesa, $\rho$ and $T$ serve as default variables.

Modifying Equation~\eqref{eq:epssecond} to take advantage of $\rho$ as
a basic variable yields
\begin{equation}
\label{eq:epsthird}
-\epsgrav = \DDt{e} - \frac{P}{\rho} \DDt{ \ln \rho}~,
\end{equation}
but the change in $e$ is still evaluated using subtraction.
Another related form, obtained by application of mass continuity, is
\begin{equation}
\label{eq:usededt}
-\epsgrav= \DDt{e} + P \frac{\partial}{\partial m} (v\area) ~,
\end{equation}
where $v$ is the cell velocity and $\area$ is the area of the cell
face.
This is the form used in the artificial viscosity based hydrodynamics
options described in \mesathree.

Expanding the total derivative of energy and thus eliminating the
subtraction motivates the following alternative forms. Expanding $e$
in terms of its dependence on the basic variables $\rho$ and $T$ and
dropping the dependence on composition gives
\begin{equation}
\label{eq:epsfifth}
-\epsgrav = c_V T \DDt{ \ln T}
+ \left[ \rho \left(\frac{\partial e}{\partial \rho}\right)_T - \frac P \rho \right] \DDt{ \ln \rho}~,
\end{equation}
where $c_V \equiv (\partial q/\partial T)_\rho = (\partial e/\partial T)_\rho$. 
One can also choose to expand $e$ in terms of its dependence on $P$
and $T$ (dropping composition dependence) and then convert to
a form given in terms of $\rho$ instead of $P$ to obtain
\begin{equation}
\label{eq:epslast}
-\epsgrav = c_P T \left[ \left( 1 - \nabla_{\rm ad} \chi_T \right) \DDt{ \ln T} - \nabla_{\rm ad} \chi_\rho \DDt{ \ln \rho}  \right]~,
\end{equation}
where $c_P \equiv (\partial q/\partial T)_P$ and $\grada \equiv (\partial
\ln T /\partial \ln P)_s$.
The derivation for this expression in terms of $P$ and $T$ is given in
Chapter~4 of \cite{Kipp12}, from which it is straightforward to obtain
Equation~\eqref{eq:epslast} using
$\chi_T \equiv (\partial \ln P/ \partial \ln T)_\rho$ and
$\chi_\rho \equiv (\partial \ln P/ \partial \ln \rho)_T$. 

Since $\rho$ and $T$ are basic variables, the time derivatives appearing
in Equations~\eqref{eq:epsfifth} or~\eqref{eq:epslast} involve no
subtractions. Hence, solving Equation~(\ref{eq:local-energy})
with $\epsgrav$ as defined by those two equations will not be
susceptible to the same losses of numerical precision as other
forms, at the cost of dropping the composition terms. 
Similarly, Equation~(4.47) of \cite{Kipp12} will yield the same
stability when $P$ and $T$ are used as basic variables.
When $P_{\rm gas}$ and $T$ are selected as basic variables,
the identification ${P = P_{\rm gas} + aT^4/3}$ allows writing
\begin{equation}
\label{eq:eps_Pgas}
\begin{aligned}
-\epsgrav = c_P T
\bigg[
\left( 1 - 4 \grada \frac{P_{\rm rad}}{P} \right) &\DDt{\ln T} \\
- \grada \frac{P_{\rm gas}}{P} &\DDt{\ln P_{\rm gas}}
\bigg]~.
\end{aligned}
\end{equation}
Section~4.5 in \cite{Kipp12} also shows how this local
energy treatment of \epsgrav\ results in global energy conservation,
including total gravitational potential energy from which the name
\epsgrav\ is derived.

The superior numerical stability of
Equations~\eqref{eq:epsfifth}--\eqref{eq:eps_Pgas} comes at the cost
of using derivative quantities such as  $c_V$ and $\chi_{\rho}$. 
The Jacobian matrix of an implicit method thus requires
the partial derivatives of $c_V$ and $\chi_{\rho}$. An EOS must therefore be capable 
of returning the state functions $P$, $e$, and $s$ along with their
first derivatives (e.g.,  $c_V$ and $\chi_{\rho}$)  and 
second derivatives  (e.g.,  $\partial c_V / \partial T$).

\begin{figure*}
\begin{center}
% Couldn't get tikz to cooperate with this larger tex document,
% so it's rendered externally in another tex document, and just
% loaded as a pdf here.
\includegraphics{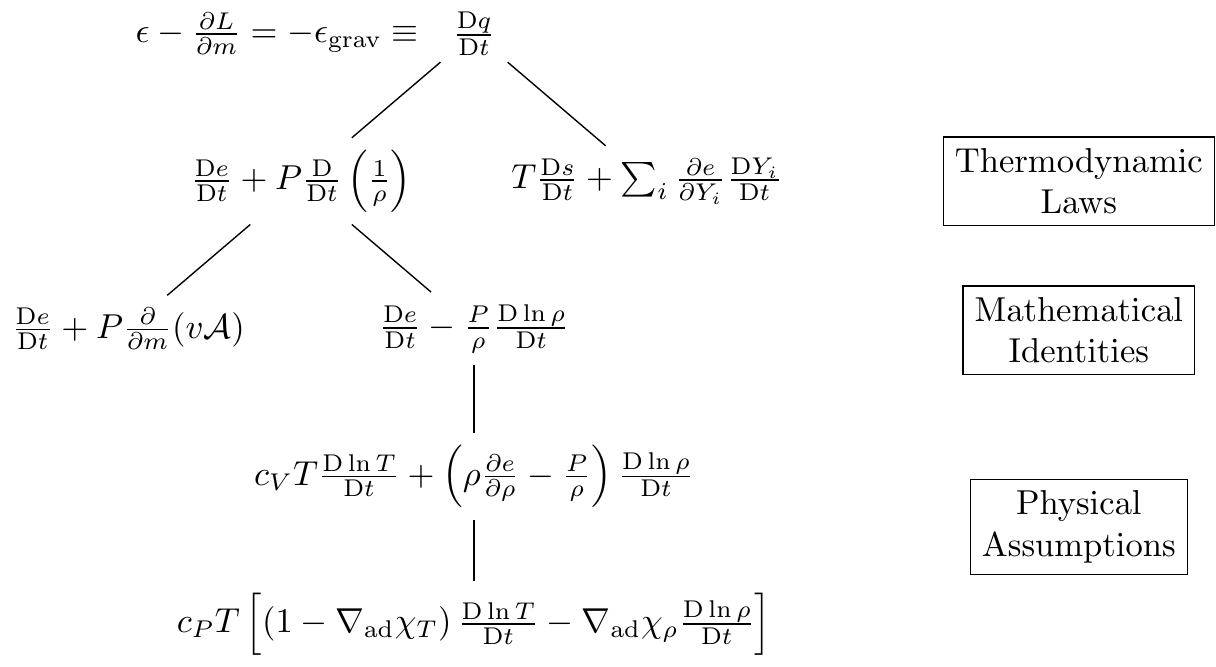}
\caption{Schematic showing the relationships in Equations~(\ref{eq:epsfirst})--(\ref{eq:epslast}).}
\label{fig:scheme}
\end{center}
\end{figure*}

\begin{table}
\caption{Summary of \epsgrav\ Options}
\begin{center}
\begin{tabular}{l | c } %| c}
Inlist Option & $\epsgrav$ \\ % & Equation Solved \\ 
\hline
\verb|use_PdVdt_form_for_eps_grav| & (\ref{eq:epssecond}) \\ % & (\ref{eq:luminosity}) \\
\verb|use_dlnd_dt_form_for_eps_grav| & (\ref{eq:epsthird}) \\ %& (\ref{eq:luminosity}) \\
\verb|use_dedt_form_of_energy_eqn| & (\ref{eq:usededt}) \\ %& (\ref{eq:dedt}) \\
\verb|use_dEdRho_form_for_eps_grav| & (\ref{eq:epsfifth}) \\ %& (\ref{eq:luminosity}) \\
\mesa\ default (all other inlist options \verb|.false.|) & (\ref{eq:epslast}) \\ %& (\ref{eq:luminosity}) \\
\verb|lnPgas_flag| (and other inlist options \verb|.false.|) & (\ref{eq:eps_Pgas}) \\ %& (\ref{eq:luminosity}) \\
\verb|use_lnS_for_eps_grav| & (\ref{eq:eps_S}) \\ % & (\ref{eq:luminosity}) \\
\end{tabular}
\label{tab:summary}
\end{center}
\end{table}

As noted above, Equations~(\ref{eq:epsfifth})--(\ref{eq:eps_Pgas}) drop
the composition terms, which is justifiable if the
derivatives $(\partial e / \partial X_i)(\Dif X_i/\Dif t)$ are negligible for each $X_i$. 
Dropping composition terms is
often justified in stellar evolution scenarios where timescales for
these changes are very slow or their associated energies are
negligible, such as MS burning where energy release from
nuclear burning dominates any small changes in internal energy due to
composition evolution over a single step \citep{Kipp65,Berro08}.
Making this assumption, \MESA\ also offers an option for calculating
$\epsgrav$ in terms of a simplified form of
Equation~\eqref{eq:epsfirst}:
\begin{equation}
\label{eq:eps_S}
- \epsgrav = T \DDt{s}~,
\end{equation}
which drops composition dependence to offer an expression that is more
convenient to evaluate.

However, even after composition dependence related
to nuclear burning is accounted for with a separate $\epsnuc$ term as discussed in
Section~\ref{s.eps_fund}, other processes that change abundances
(e.g.,~mixing) may be important. In cases where dropping these terms is
not justifiable, it may be necessary to add a compensating local
source term $\epsilon$ in Equation~\eqref{eq:local-energy}.

In summary, \mesa\ currently offers options for solving
Equation~\eqref{eq:local-energy} with $\epsgrav$ defined in any of the
ways given in
Equations~\eqref{eq:epssecond}--\eqref{eq:eps_S}. Figure~\ref{fig:scheme}
schematically summarizes the relationships between these equations and
Table~\ref{tab:summary} shows the inlist commands necessary for
invoking each of these options.  Usually, the superior numerical
stability gained by using Equation~\eqref{eq:epslast} is to be
preferred, and hence it is the \mesa\ default, but users should be
aware of the possibility that other forms may be necessary to capture
important physics.  One such case for Equation~\eqref{eq:eps_S} is described in 
Section~\ref{sec:latent-heat}.
Another is the artificial viscosity-based implicit hydrodynamics described in
\mesathree\ (see Section 4, Equation 41), where choosing Equation~\eqref{eq:usededt} 
helps ensure intrinsic energy conservation.

\subsection{Relationship to the Riemann Solver-Based Hydrodynamics Implementation}\label{s.implicit_hydro}

When using the Riemann solver-based hydrodynamics capabilities
described in Section~\ref{s.hydro}, \MESA\ does not cast the stellar
structure equations in terms of local heating as in
Equation~\eqref{eq:local-energy}. Instead, it combines
Equation~(\ref{eq:local-energy}) with the constraint of fluid momentum
conservation to form a local \textit{total} energy equation.

We begin with the mass continuity equation,
\begin{equation}
  \label{eq:mass-continuity}
  \DDt{\rho} = -\frac{\rho }{r^2}\frac{\partial}{\partial r}\left( r^2 u\right),
\end{equation}
and the momentum equation,
\begin{equation}
  \label{eq:momentum}
   \DDt{u} = -\frac{1}{\rho} \ddr{P} - \ddr{\Phi}~,
\end{equation}
written in Lagrangian form and assuming spherical symmetry.  
The variable $u$ is the radial velocity and $\Phi$ is the gravitational potential.
The Lagrangian derivative operator is
$\Dif/\Dif t = \partial/\partial t+  u \partial/\partial r$.

Multiplying Equation~\eqref{eq:momentum} by $u$ gives
\begin{equation}
  \label{eq:dot-v}
  \DDt{}\left(\frac{1}{2}u^2\right) =  -\frac{u}{\rho}\ddr{P} - u\ddr{\Phi}~.
\end{equation}
The gravitational potential does not explicitly depend on time
($\partial \Phi/ \partial t = 0$), so
$\Dif\Phi/\Dif t = u \partial \Phi/\partial r$.  
This implies
\begin{equation}
  \label{eq:dot-v-rearranged}
  \DDt{}\left(\frac{1}{2}u^2 + \Phi \right) =  -\frac{u}{\rho}\ddr{P} ~.
\end{equation}
Using Equations~\eqref{eq:local-energy} and
\eqref{eq:epssecond} we have
\begin{equation}
  \label{eq:eps-grav-first-law}
  \DDt{e} - \frac{P}{\rho^2} \DDt{\rho}  =  \epsilon - \ddm{L} ~.
\end{equation}
Adding Equations~(\ref{eq:dot-v-rearranged}) and
(\ref{eq:eps-grav-first-law}) gives
\begin{equation}
   \DDt{}\left(e + \frac{1}{2}u^2 + \Phi \right)
   = \frac{P}{\rho^2} \DDt{\rho} - \frac{u}{\rho}\ddr{P} +  \epsilon - \ddm{L} ~.
\end{equation}
Using mass continuity (Equation~\ref{eq:mass-continuity}) this becomes
\begin{equation}
  \label{eq:1}
     \DDt{}\left(e + \frac{1}{2}u^2 + \Phi \right)
     =   -\frac{1}{\rho r^2}\ddr{}\left(Pur^2\right) + \epsilon - \ddm{L} ~.
\end{equation}
In spherical coordinates
\begin{equation}
   \frac{1}{\rho r^2}\ddr{}\left(r^2f\right) = \ddm{(\area f)}~,
\end{equation}
where $\area = 4\pi r^2$. Thus we arrive at the equation that \MESA\ solves,
\begin{equation}
  \label{eq:total_conservation}
  \DDt{}\left(e + \frac{1}{2}u^2 + \Phi \right)
  =   \epsilon - \ddm{}\left(L + P\area\,u\right) ~.
\end{equation}

\subsection{Ionization}
\label{sec:ionization}

The internal energy reported the EOS should include the energy
associated with ionization\footnote{Since this energy is released upon
  recombination, it is also often referred to as ``recombination
  energy''.} and molecular
dissociation. The assumption of LTE specifies the ionization state given
$(\rho,T, \{X_i\})$.
Since \mesa\ does not regard a change in ionization as a change in composition, it is not
necessary to include separate composition derivatives in \epsgrav\ in
order to account for the energetic effects of changes in ionization
state.

To demonstrate a specific scenario where \MESA\ 
accounts for ionization energy,
we evolve a $1\,\Msun$ pre-MS model composed of pure H. We compare quantities calculated by \MESA\
with other, simpler estimates.  We
calculate the thermal energy assuming a monatomic ideal gas,
\begin{equation}
  e_{\rm thermal} = \frac{3 N_{\rm A} \kB T}{2 \mu}~.
\end{equation}
We calculate the ionization energy for pure H as
\begin{equation}
  e_{\rm ion} = (1-f_{\rm H}) N_{\rm A} E_{\rm H} + \frac{N_{\rm A} E_{\rm H2}}{2} ~,
  \label{e.eion}
\end{equation}
where we assume the ionization fraction of
H is given by the Saha equation.  The variable $f_{\rm H}$
represents the neutral fraction of H.  The H ionization energy
is $E_{\rm H} = 13.6\,\eV$ and Equation~\eqref{e.eion} also includes the
dissociation energy of molecular H ($E_{\rm H2} = 4.52\,\eV$)
assuming that no H is in the molecular state.

\begin{figure}
  \centering
  \includegraphics[]{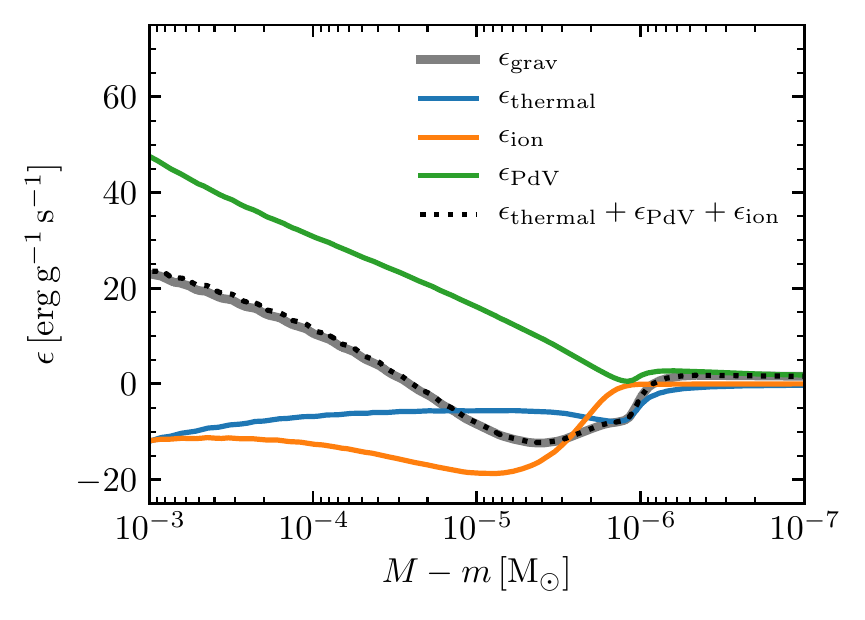}
  \caption{The value of \epsgrav\ in the pure-H pre-MS model, evaluated over
    a region near the stellar surface that includes an ionization zone (where
    $\epsilon_{\rm ion} \ne 0$).  The solid colored lines indicate the
    individual energy terms.  Their sum (dotted black line) agrees
     with the value calculated by \MESA\ (solid \replaced{grey}{gray} line).}
  \label{fig:ionization}
\end{figure}

During the evolution, we record $\epsgrav$ calculated by
\MESA\ using Equation~\eqref{eq:epslast}.  We also evaluate the quantity
\begin{equation*}
  \underbrace{\DDt{}\left(e_{\rm thermal}\right)}_{\epsilon_{\rm thermal}} +
    \underbrace{\DDt{}\left(e_{\rm ion}\right)}_{\epsilon_{\rm ion}} +
    \underbrace{P\DDt{}\left(\frac{1}{\rho}\right)}_{\epsilon_{\rm PdV}}
  \label{eq:epsgrav-split}
\end{equation*}
that separates out the thermal and ionization energy.  In
Figure~\ref{fig:ionization} we compare these two approaches, making it
clear that all three terms in the above expression play
an important role. Their sum agrees with the \MESA\ \epsgrav,
indicating that each of these terms is accounted for in the
\MESA\ calculation.

\begin{figure}
  \centering
  \includegraphics[]{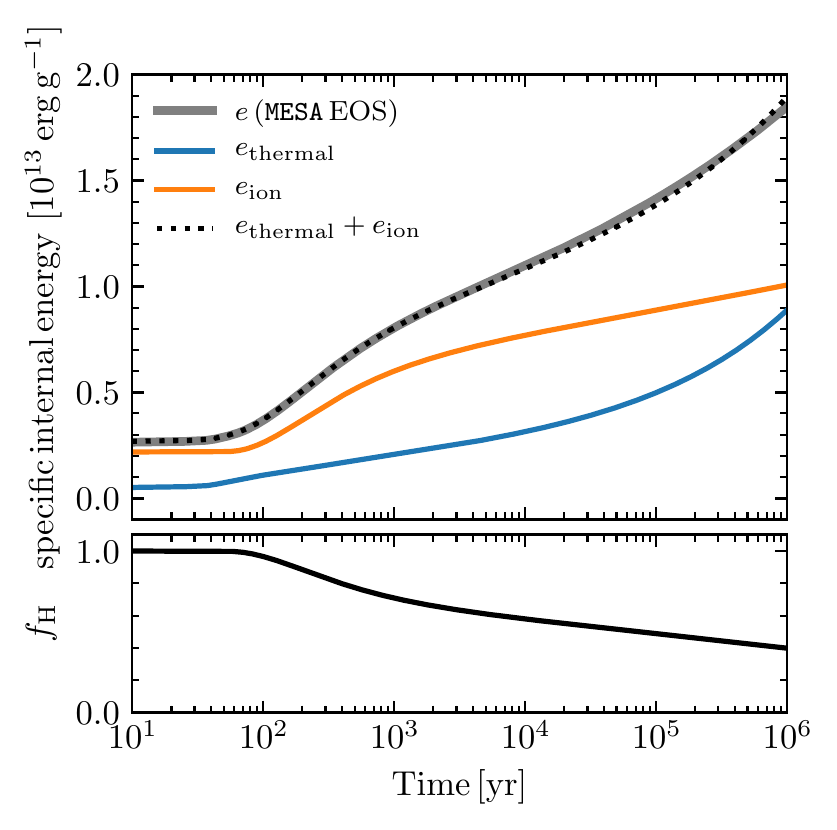}
  \caption{Specific internal energy at a fixed Lagrangian coordinate
    in the pre-MS model (upper panel).  The solid colored curves indicate the
    individual energy terms.  The internal energy reported by \MESA\
    (solid \replaced{grey}{gray} curve) exceeds the thermal energy because of the
    ionization energy.  The lower panel shows the neutral fraction of the H.}
  \label{fig:ionization-energy}
\end{figure}

Figure~\ref{fig:ionization-energy} shows the history of the material at the
Lagrangian coordinate $(M-m)/\Msun = 10^{-5}$.
We plot $e$ reported by the \MESA\ EOS along with $e_{\rm
  thermal}$ and $e_{\rm ion}$ (calculated in same manner as above).
At this location, the specific internal energy is dominated
by the ionization energy. The lower panel of this figure shows the neutral fraction of H; 
towards the left of the plot, the H is fully neutral. In this region the ionization
energy plateaus at the
dissociation energy of molecular H (see Equation~\ref{e.eion}).

\begin{figure}
  \centering
  \includegraphics[]{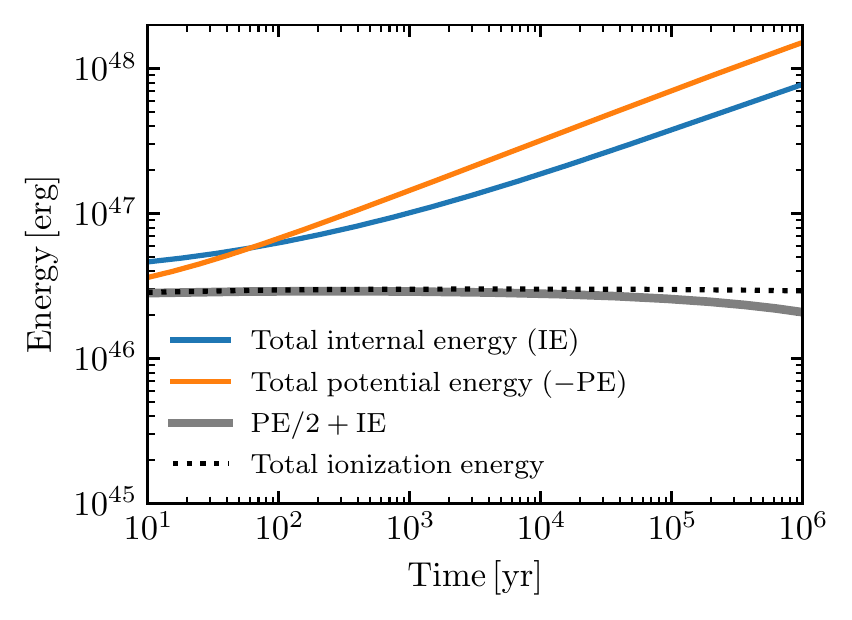}
  \caption{Total potential and internal energy in the pure-H pre-MS model.
    The sum of half the total potential energy plus the internal
    energy (solid \replaced{grey}{gray} curve), which by the virial theorem should be
    the non-thermal internal energy, agrees well with our estimate of
    the ionization energy (dashed black curve).  The deviation at
    $\ga 10^5$ yr is caused by non-ideal gas effects.}
  \label{fig:total-energy}
\end{figure}

\begin{figure}
  \centering
  \includegraphics[]{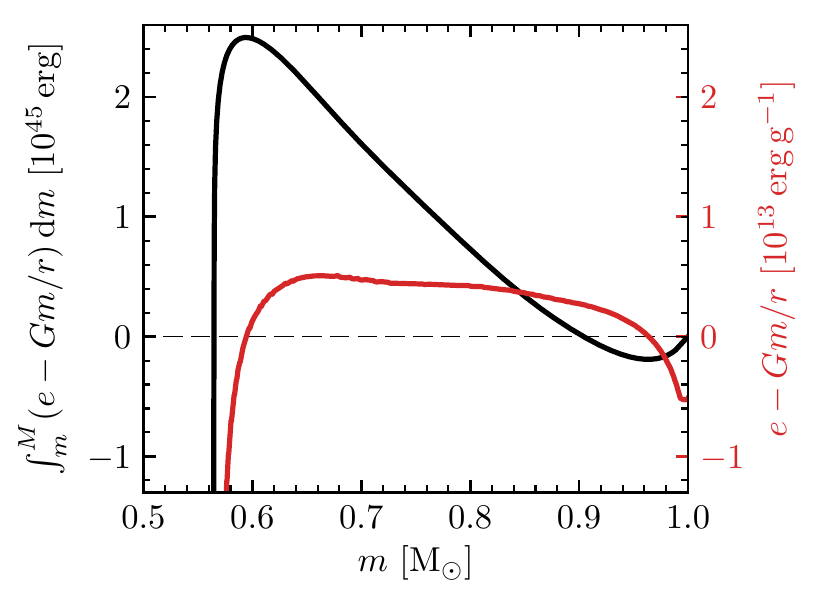}
  \caption{Specific (red) and cumulative (black)
    total energy (IE + PE) in the envelope of an
    AGB model ($M=1.0 \, \Msun$, $L = 4.97 \times 10^3 \, \Lsun$, $T_{\rm
      eff} = 2,920 \, {\rm K}$, $R = 276 \, \Rsun$).
    This energy is positive in the envelope
    due to the inclusion of ionization energy in the internal energy
    reported by \mesa.
  }
  \label{fig:pos_env}
\end{figure}

For a star in hydrostatic equilibrium, the virial theorem states
that
\begin{equation}
  -\frac{1}{2}\int_{0}^{M} \frac{Gm}{r} dm + \int_{0}^{M}\frac{3 P}{2 \rho} dm = 0 ~.
\end{equation}
The right term's integrand, $3P/(2\rho)$, is the specific thermal
energy of an ideal monatomic ideal gas.  Figure~\ref{fig:total-energy} shows
the total internal energy and gravitational potential energy reported
by \MESA\ for the pure-H pre-MS model.  On the same scale we show half the total potential energy
plus the internal energy.  This quantity is not zero; rather, by the
virial theorem, it should sum to the non-thermal and non-ideal internal energy
(e.g., the ionization energy).
This value, recorded from the \MESA\
model, agrees well with our estimate of the ionization energy.  Also
note that at early times the total energy of the star (internal +
potential, not shown) is positive.
The phenomenon of positive total energy when ionization energy is
included also occurs for envelopes of stars on the asymptotic giant branch
\citep[AGB;][]{Pac68}. Figure~\ref{fig:pos_env} shows
the total energy in the
envelope of a $1.0 \,\Msun$ \mesa\ model on the AGB.
This confirms that the ionization energy is included when
\MESA\ reports the total energy of a model.

\subsection{Latent Heat}
\label{sec:latent-heat}

\begin{figure*}[ht]
  \centering
  \includegraphics{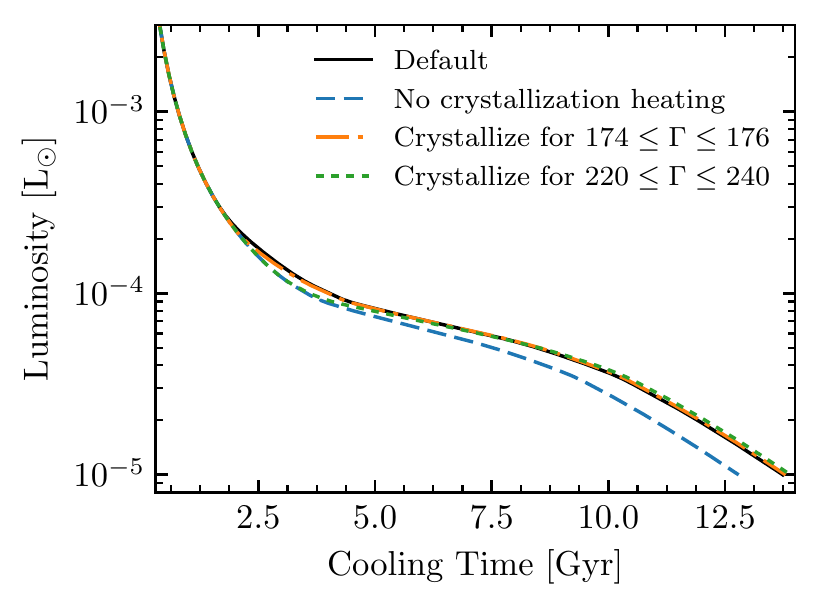}
  \includegraphics{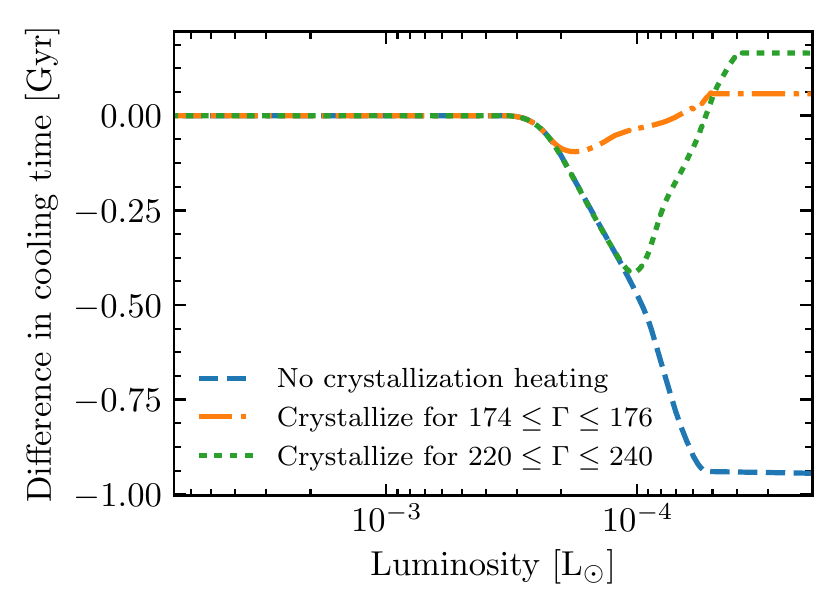}
  \caption{Cooling for a $0.6 \, \Msun$ C/O WD ($M_{\rm H} = 2.7
    \times 10^{-5} \, \Msun$, $M_{\rm He} = 1.5 \times 10^{-2} \,
    \Msun$) with different treatments of the latent heat of
    crystallization. The default treatment smoothly
    injects the latent heat over the range $150 \leq \Gamma \leq
    175$. The right panel shows differences in cooling time (relative to the default
    shown in the left panel) required to
    reach a given luminosity for other treatments.}
\label{fig:wd_cool}
\end{figure*}

\mesatwo\ discusses the inclusion of the latent heat of
crystallization for long term WD cooling. Crystallization is a
first-order phase transition that manifests in the PC EOS 
\citep{PC2010} as an entropy discontinuity at a plasma coupling
parameter of $\Gamma = 175$, and can be captured in stellar evolution
with $\epsgrav$ in the form of Equation~\eqref{eq:eps_S}.
Since the publication of \mesatwo, controls have been
added to \mesa\ to allow smoothing out the injection of latent heat in
$\epsgrav$ over a user-specified range of $\Gamma$.
By default, the range for crystallization is softened to $150 \leq
\Gamma \leq 175$ to avoid numerical difficulties with sudden energy
injection associated with a sharp transition at $\Gamma = 175$.
The controls allow for tightening this range for more precise timing
on the occurrence of crystallization if necessary.
Figure~\ref{fig:wd_cool} shows the small impact on cooling time for a
$0.6 \, \Msun$ WD from spreading the latent heat over this broader
range of $\Gamma$ relative to a tighter phase transition for $174 \leq
\Gamma \leq 176$.

The spreading of the phase transition is accomplished by calculating both
the liquid and solid solutions  within the PC EOS and linearly
blending the entropy $s$ and internal energy $e$ over the specified
range of $\Gamma$. With $\epsgrav$ expressed in the form of
Equation~\eqref{eq:eps_S}, the energy of the phase transition is
captured as fluid elements smoothly traverse from liquid-phase to solid-phase.
By default \mesa\ automatically switches to using $\epsgrav$ in the form of
Equation~\eqref{eq:eps_S} for $\Gamma > 150$.
This choice ensures the capture of latent heat release.

Theoretical and observational work has suggested that crystallization
in C/O mixtures may occur at higher $\Gamma$ than the classical one
component plasma value of $\Gamma = 175$
\citep{Horowitz07,Winget09,Medin10,Althaus12}. Our updated crystallization
controls allow for investigating the effect on stellar evolution of
crystallization at $\Gamma \approx 240$. 
Figure~\ref{fig:wd_cool} shows the potential effects on WD cooling times of varying the
$\Gamma$ for crystallization.
Because the heating from crystallization is
released very late in the WD evolution, its effects on cooling times
are on the order of $\rm Gyr$, and variations in crystallization
treatment can lead to changes that are a significant fraction of this
timescale.

The composition terms in Equation~\eqref{eq:epsfirst} that were dropped to
form Equation~\eqref{eq:eps_S} are negligible as long as there is no
mixing in the crystallization region.
Phase separation may violate this assumption and require a
modified treatment, but we do not consider this process here.
Detailed phase-diagrams for crystallization and
the possible associated phase-separation effects are not currently
supported in \mesa, so our investigation here is limited to the
effects of crystallization as a function of a fixed $\Gamma$ range.

\subsection{EOS Blending}
\label{sec:eos-blending}

\begin{figure*}[ht]
  \centering
  \includegraphics[width=0.495\textwidth]{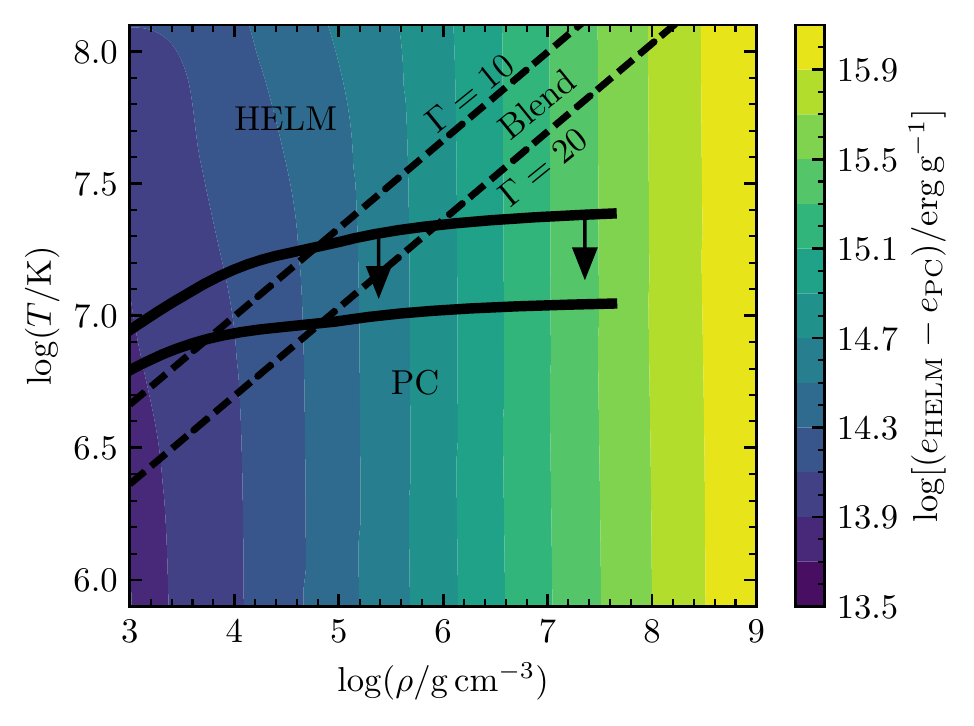}
  \includegraphics[width=0.495\textwidth]{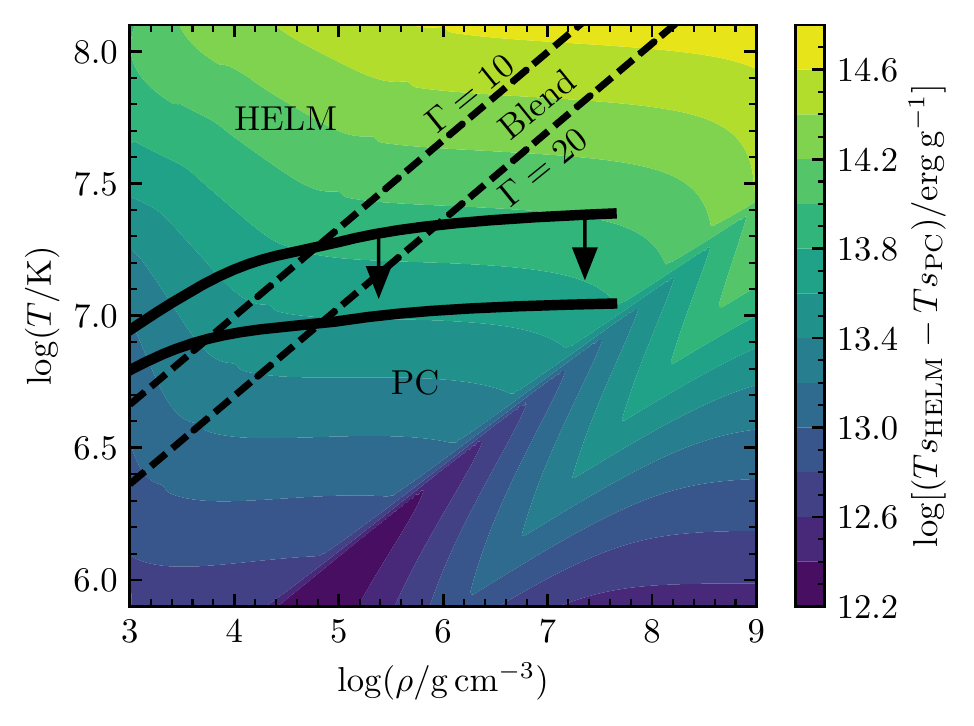}
  \caption{Magnitude of the energy differences between the HELM and PC
    EOS for specific internal energy $e$ (left) and entropy $s$
    (right) in a 50/50 C/O mixture. Dashed lines show
    the EOS blending boundaries for $10 \leq \Gamma \leq 20$,
    and the solid black lines show representative profiles for
    a $1.0 \, \Msun$ WD cooling from 
    $T_{\rm eff} = 26,000 \, \rm K$ to $T_{\rm eff} = 17,000 \, \rm K$.}
  \label{fig:eos_energy}
\end{figure*}

\begin{figure*}[ht]
  \centering
  \includegraphics[width=0.495\textwidth]{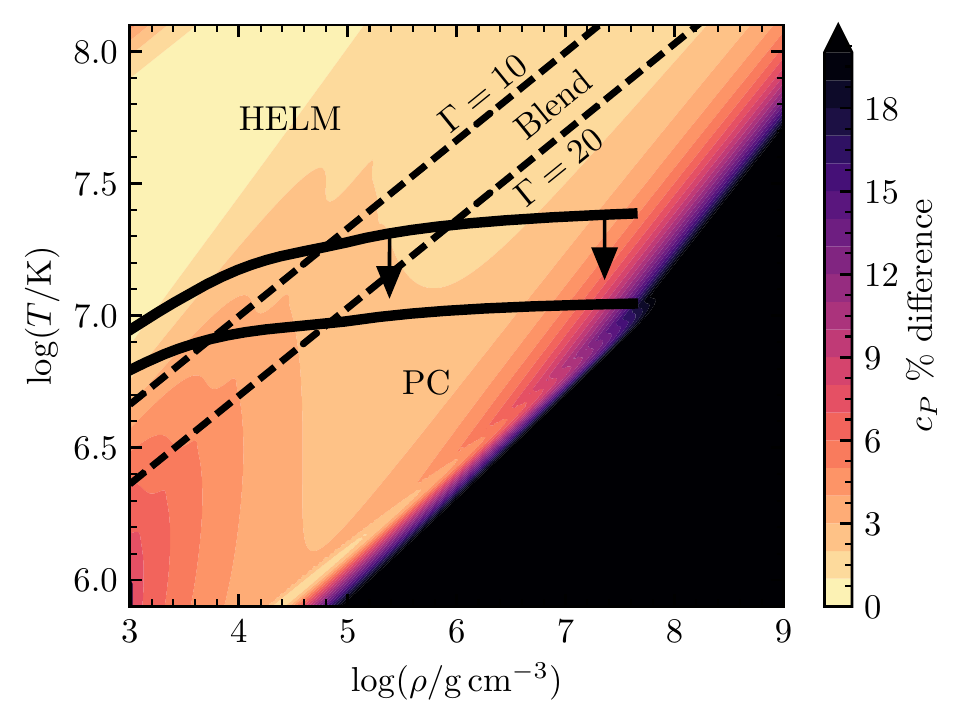}
  \includegraphics[width=0.495\textwidth]{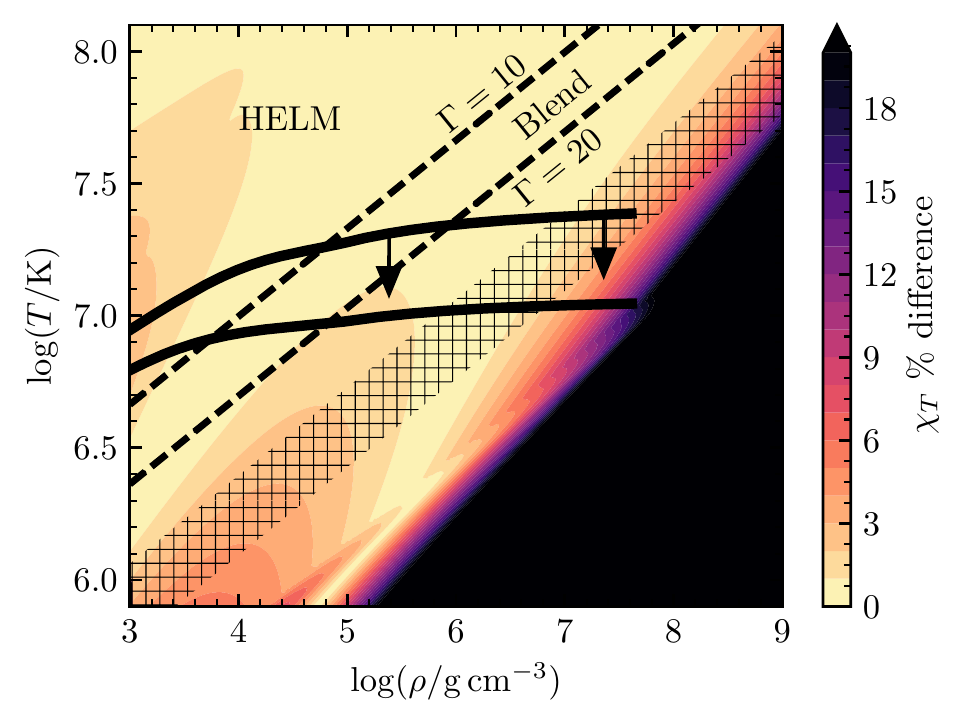}
  \caption{Percent difference between HELM and PC EOS for $c_P$
    (left) and $\chi_T$ (right). Dashed lines show
    the EOS blending boundaries for $10 \leq \Gamma \leq 20$, and
    the hatched region in the right panel shows the previous default
    blending range $40 \leq \Gamma \leq 80$. The new EOS blending region
    disagree by only a few percent for these quantities.}
  \label{fig:eos_derivs}
\end{figure*}

As shown in Figure~1 of \mesaone, \mesa\ employs a patchwork of
several EOSs to provide coverage of a
maximal amount of $\rho-T$ space. When blending from one EOS region
into another, care is required to avoid introducing spurious
contributions into $\epsgrav$. At high density, \mesa\ blends from the
Helmholtz EOS \citep[HELM,][]{Timmes00} for $\Gamma < 10$ to the PC EOS
\citep{PC2010} for $\Gamma > 20$ by
default. This default has been changed from the original default of
$40 \leq \Gamma \leq 80$ given in \mesaone\ due to the optimal agreement
between relevant quantities shown in Figure~\ref{fig:eos_derivs}, as
explained below.
Overall, the two EOSs agree well on thermodynamic quantities
in the blending region ($\sim 1 \%$ for $e$ and $s$), but
Figure~\ref{fig:eos_energy} shows that the absolute magnitude of the
disagreement can still be large enough to influence $\epsgrav$ for a cooling
WD when $\epsgrav$ is expressed in the form of
Equations~\eqref{eq:epsfirst}--\eqref{eq:usededt}.

The left panel of Figure~\ref{fig:eos_energy} indicates that typically
the internal energy difference is ${\Delta e \sim 10^{15} \,
\rm erg\,g^{-1}}$, while $c_P T \sim 10^{14} - 10^{15} \, \rm erg\,g^{-1}$ in the
region of the blend. As a WD model cools, most of its $\sim 10^{33} \,
\rm g$ of mass must eventually pass through this transition.
If the energy equation is being solved in the form of
Equation~\eqref{eq:eps_numerical1}, $\sim 10^{48} \,
\rm erg$ of spurious energy would be introduced into the model by EOS
blending. Since much of this blending happens after the WD model has
cooled to a luminosity of $L \lesssim 0.1 \, \Lsun$, this extra
energy corresponds to $\Delta t \gtrsim 100 \, \rm Myr$ of extra WD cooling time.

The default form of $\epsgrav$ given in
Equation~\eqref{eq:epslast} does not suffer from this
spurious heating, since it is expressed in terms of thermodynamic
derivatives from the EOS rather than $e$ and $s$. For this form of
$\epsgrav$, the differences between $e$ or $s$ do not directly enter
the equations. Instead, changes in $e$ with $\Dif \ln \rho/\Dif t$ and $\Dif \ln
T/\Dif t$ are tracked with quantities such as $c_P$ and $\chi_T$, and
Figure~\ref{fig:eos_derivs} shows that these agree well for the EOS
blend region. Since the implementation of Equation~\eqref{eq:epslast} does
not involve any derivatives constructed as finite differences, the
fact that quantities such as $c_P$ agree to within a few percent
guarantees that $\epsgrav$ will be consistent across the blend, with
no significant spurious energy injected due to blending.
Crucially, the release of latent heat described in
Section~\ref{sec:latent-heat} requires switching to $\epsgrav$ in the
form of Equation~\eqref{eq:eps_S} only for zones with $\Gamma >
150$, so both EOS blending and crystallization simultaneously receive
appropriate treatments with different forms of $\epsgrav$ in different
stellar regions.

\subsection{Gravitational Settling}
\label{sec:grav-settl}

Equation~\eqref{eq:epslast} for $\epsgrav$ ignores changes in internal
energy $e$ due to composition changes. \cite{Berro08} point out
that a self-consistent evolutionary approach to WD cooling including
the effects of $^{22}{\rm Ne}$ settling requires accounting for
composition changes due to element diffusion in $\epsgrav$.
They adopt pure $^{12}\rm C$ or
$^{16} \rm O$ core compositions with trace $^{22}\rm Ne$ and
no other isotopes. While this approach is useful for rigorous study of
self-consistent WD evolution with diffusion fully coupled to
evolution, it is not well suited for a general treatment of realistic
mixed core compositions.

\mesa\ splits element diffusion into a
separate step before the main structural solve, and
hence diffusive
effects are not included in $\epsgrav$. 
We ensure that the energy associated with $^{22}
\rm Ne$ settling is not included in $\epsgrav$ by using
Equation~\eqref{eq:epslast}, and we compensate by including an
 extra heating term $\epsilon_{22}$ in
Equation~\eqref{eq:local-energy}. This term is calculated using
velocities saved from the element diffusion step as described
in Section~\ref{s.heating}. Our results for the effects of $^{22}\rm
Ne$ settling on WD cooling agree well with
\cite{Berro08} and with \cite{DeloyeBildsten} who adopt a
 heating term similar to our approach.

%%% Local Variables:
%%% mode: latex
%%% TeX-master: "paper"
%%% End:

% !TEX root = ./paper.tex
\section{Summary}\label{s.conclusions}

We explain significant new capabilities and improvements 
implemented in \MESA \ since the publication of \mesaone, \mesatwo, and \mesathree.
Progress in the treatment of \deleted{energy conservation (Section \ref{s.epsgrav}),}
convective boundaries (Section \ref{s.convect}) and element diffusion (Section \ref{s.diffusion} and Appendix \ref{app.diffusion})
will improve studies of their impact on stellar evolution.
Advances to \MESA\ in implicit hydrodynamics (Section \ref{s.hydro}),
approximation of 3D RTI effects (Section \ref{s.RTI}),
and coupling with a public version of the \STELLA \ radiative transfer instrument 
will enhance the modeling of Type~IIP SN light curves from post-explosion to post-plateau
(Section \ref{s.sneiip}). We integrate these improvements with an exploration of PPISN 
and black hole formation models (Section \ref{s.bhform}).
\added{We describe
energy conservation in \MESA\ and demonstrate improvements relevant to WD cooling (Section \ref{s.epsgrav}).}
Upgrades to estimating the absolute magnitude of a model in a chosen passband
(Appendix \ref{s.colors}), guidance on importing multi-dimensional models into \MESA \ (Appendix  \ref{app.relax}),
and new \MESA-based software tools (Section \ref{s.software}) will strengthen research and education.
Input files and related materials for all the figures are available at \url{http://mesastar.org}.

%%% Local Variables:
%%% mode: latex
%%% TeX-master: "paper"
%%% End:

% !TEX root = ./paper.tex

\acknowledgements

% people and places shout-outs

We thank Dan Kasen for many discussions, \texttt{SEDONA} calculations and
interactions that led to the implementation of the Sobolev optical
depth approach to estimating the \ion{Fe}{2} 5169 \AA\ line velocities in Type
IIP SNe in \MESA. Both Ond\v{r}ej Pejcha and Stefano Valenti
were generous in sharing and making available photometry, bolometric
luminosities and \ion{Fe}{2} line velocities of core collapse SNe.
We acknowledge Annop Wongwathanarat for both communications regarding
his 3D Rayleigh Taylor simulations of core collapse SNe and for
providing detailed simulation results that enabled our comparison of
the \MESA\ implementation of the Duffell RTI modeling to 3D data.
S.B.~is grateful to L.B.~and B.P.~for hospitality at KITP where the
public version of \STELLA\ was finalized.  Contributions to \STELLA-tools
by Petr Baklanov, Ulrich Noebauer, and Marat Potashov are gratefully
acknowledged.
We thank Carl Fields for initiating and developing \MESAWeb.
It is a pleasure to thank
Conny Aerts, 
\added{Warrick Ball,}
Matteo Cantiello, 
\added{Barbara Castanheira,}
Sean Couch, 
Luc Dessart,
Pavel Denisenkov, 
Jim Fuller, 
Falk Herwig,
Raphael Hirschi,
Christian Iliadis, 
Thomas Janka,  
Sam Jones,
\added{Alexandra Kozyreva,}
% Dan Kasen, 
Takashi Moriya, 
% Ond\v{r}ej Pejcha
\added{Marat Potashov,}
Eliot Quataert,
Ken Shen,
Matt Turk, 
%Stefano Valenti, 
%and
Michael Wiescher,
\added{and Matt Wood}
for discussions.
We also thank the participants of the 2016 and 2017 \MESA \ Summer
Schools for their willingness to experiment with new capabilities,
as well as Mitchell Lachat and Jamie Lombardi for providing us with
SPH simulations of stellar mergers.

% funding acknowledgements

This project was supported by NSF under the Software Infrastructure
for Sustained Innovation (SI$^2$) program grants 
(ACI-1339581, ACI-1339600, ACI-1339606, 
 ACI-1663684, ACI-1663688, ACI-1663696), and 
by NASA under the Theoretical and
Computational Astrophysics Networks (TCAN) program grants (NNX14AB53G,
NNX14AB55G, NNX12AC72G).
This research is
funded in part by the Gordon and Betty Moore Foundation
through Grant GBMF5076 and was also supported at UCSB by
the National Science Foundation under grant PHY 11-25915.
The work at Arizona State University was also supported by 
the NSF under grant PHY-1430152 for the Physics Frontier Center ``Joint Institute
for Nuclear Astrophysics - Center for the Evolution of the Elements'' (JINA-CEE). 
Support for this work was provided by NASA through Hubble Fellowship
grant \# HST-HF2-51382.001-A awarded by the Space Telescope Science
Institute, which is operated by the Association of Universities for
Research in Astronomy, Inc., for NASA, under contract NAS5-26555.
\added{This work used the Extreme Science and Engineering Discovery
Environment (XSEDE), which is supported by National Science Foundation
grant number ACI-1548562; specifically the Comet cluster at the San Diego Supercomputer
Center (SDSC) through allocation TG-AST150065.}
P.M.~acknowledges support from NSF grant AST-1517753 to Vassiliki Kalogera at Northwestern University.
A.T.~is Research Associate at the Belgian Scientific Research Fund (F.R.S-FNRS).
This research made extensive use \added{of} the SAO/NASA Astrophysics Data System (ADS).

\software{
\texttt{crlibm} \citep{de-dinechin_2007_aa},
\texttt{sedov} \citep[][\url{http://cococubed.asu.edu/research_pages/sedov.shtml}]{kamm_2007_aa},
\texttt{STARSMASHER} \citep[][\url{https://jalombar.github.io/starsmasher/}]{Gaburov+2010, Lombardi+2011},
\texttt{Python} avaliable from \href{https://www.python.org}{python.org},
\texttt{matplotlib} \citep{hunter_2007_aa},
\texttt{NumPy} \citep{der_walt_2011_aa},
\texttt{ipython/jupyter} \citep{perez2007ipython,kluyver2016jupyter},
\texttt{py\_mesa\_reader} \citep{pmr},
\MESA-Web available at \url{http://mesa-web.asu.edu},
\MESA-Docker \citep[][\url{https://doi.org/10.5281/zenodo.1002851}]{MESA-Docker}, 
\texttt{pyMESA} \citep[][\url{https://doi.org/10.5281/zenodo.846305}]{pymesa}
         }

%%% Local Variables:
%%% mode: latex
%%% TeX-master: "paper"
%%% End:

\appendix
\section{Colors}\label{s.colors}

We describe \MESA's implementation of bolometric corrections (BCs) for use in estimating the
absolute magnitude of a model in a user-chosen filter system. Note this is different than the
colors reported by \STELLA\ (Section \ref{s.sneiip}), as the \colors\ module uses pre-computed tables of BCs
while \STELLA\ solves the radiative transfer equations on-the-fly \citep{Blinnikov1998}.

The absolute bolometric magnitude ($M_{\rm{bol}}$) of a star is defined, with reference to the solar absolute bolometric magnitude, as \citep{2010AJ....140.1158T}: 
\begin{equation}
 M_{\rm{bol}} = {\rm M}_{\rm{bol,\odot}}-2.5\log_{10}\left(L/\Lsun\right)~,
\end{equation} 
where ${\rm M}_{\rm{bol,\sun}}$ is the absolute bolometric magnitude of the Sun, taken as 4.74 (2015 IAU Resolution B2). This can be transformed into the pass band dependent
absolute magnitude, $M_{X}$, for a nominal pass band $X$, via
\begin{equation}
  M_{X} = M_{\rm{bol}} - {\rm{BC}}_{X} ~,
\end{equation}
${\rm{BC}}_{X}$ is the BC for pass band band $X$ and accounts for the flux
emitted outside of the wavelength range of the filter system. The derivation of a BC
requires an atmospheric model of a star such that a stellar spectrum can be computed over all wavelengths,
a computationally costly process.
To prevent the requirement of actually having to generate a spectra at each time-step,
we make use of pre-computed BC tables.
These define the BC as a function of the stellar
photosphere; $\Teff/\K$, $\log \left(g/\cmpersecondSc\right)$ and the metallicity $[\rm{M/H}]$, and are derived from pre-computed grids of stellar atmosphere models,
\citep[see e.g.,][]{1970SAOSR.309.....K,2013A&A...553A...6H}.
Given the parameters at the stellar photosphere, we 
interpolate each set of BCs over $\log \left(\Teff/\K\right)$,
$\log \left(g/\cmpersecondSc\right)$ and $[\rm{M/H}]$
using linear interpolation over nearest neighbors and without extrapolation for
points outside of the table range.

\begin{figure}[htb!]
\centering
\includegraphics[width=0.5\columnwidth]{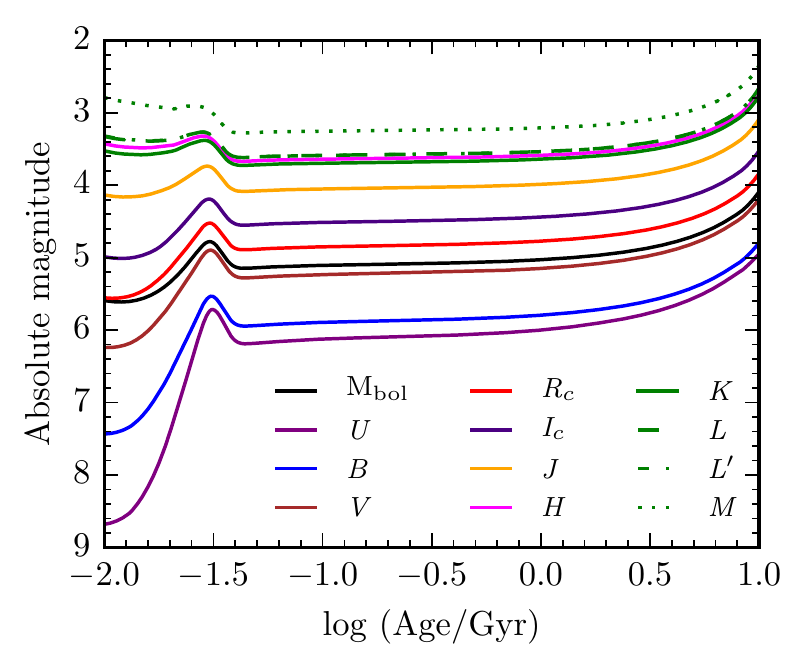}
\caption{Evolution of the absolute magnitude of a 1\,\Msun\ star for the bolometric magnitude and 
magnitude in the filter bands \textit{UBVR$_{\rm{c}}$I$_{\rm{c}}$JHKLL$^\prime$M}.\label{f.color}}
\end{figure}

We provide two sets of pre-processed tables of BCs, though a user may provide their own.
From \citet{1998A&AS..130...65L} we provide the Johnson-Cousins-Class bands \textit{UBVR$_{\rm{c}}$I$_{\rm{c}}$JHKLL$^\prime$M}. 
This table provides the
BCs over the parameter range $2,000 \leq \Teff/\K \leq  50,000$, 
$-1.02 \leq \log \left(g/\cmpersecondSc\right) \leq 5.0$ and 
$-5.0 \leq [\rm{M/H}] \leq 1.0$, with a variable sampling rate. Figure \ref{f.color} shows the time evolution
of the absolute magnitude of a 1\,\Msun\ star with the pass bands defined in \citet{1998A&AS..130...65L}.
We also provide a set of blackbody BCs for the pass bands \textit{UBVR$_{\rm{c}}$I$_{\rm{c}}$}, over the range
 $100 \leq \Teff/\K \leq 50,000$ in steps of $100\,\K$. As these are blackbody corrections there is no $g$ or $[\rm{M/H}]$
 dependence.

There are many other possibilities for other pass bands or classes of object
\citep{1996AJ....111.1748F,2002A&A...391..195G,2011PASP..123.1442B,2012PASP..124..140B}. Thus the tables we 
provide are not a definitive set, but merely a reasonable starting point for modeling stellar objects. 
Other astrophysical objects like WDs, exoplanets, or SN light curves require calculating specialized tables. Users may provide BC tables defined in terms of $\Teff/\K$, $\log
\left(g/\cmpersecondSc\right)$ and $[\rm{M/H}]$.

%%% Local Variables:
%%% mode: latex
%%% TeX-master: "paper"
%%% End:

\section{Model Relaxation}\label{app.relax}

To simplify the process of importing a model into {\MESA}, we
have developed simple relaxation routines that allow the construction of a
starting model in hydrostatic equilibrium with specified profiles for
composition, angular momentum, and entropy. Examples that motivate
importing a model into \MESA\ include multi-dimensional simulations of
stellar mergers, common envelope evolution, and the effects of SN explosions on nearby
companions.

The relaxation process inputs include 1D profiles of composition
and angular momentum. The process also requires either an entropy
profile or the profiles of pairs of values $(\rho,T)$,
$(P_{\rm gas},T)$, or $(\rho, e)$, from which {\MESA} extracts
the entropy using the \texttt{eos} module. Note that
in the case where the entropy is not provided directly,
the relaxed model will match the entropy computed by the \texttt{eos} module,
but not neccesarily the input $(\rho,T)$, $(P_{\rm gas},T)$, or $(\rho,
e)$ profiles.
A good match for the input profiles depends on the input
data corresponding to a model in hydrostatic equilibrium computed with an
EOS that is consistent with {\MESA}'s.

Relaxation is done via pseudo-evolution of a stellar model for
which mixing, angular momentum transport, and changes in composition from nuclear
burning are suppressed,
while a quantity of interest is incrementally altered until it reaches the
desired value up to a pre-defined tolerance. Throughout this relaxation process,
hydrostatic equilibrium is enforced. The starting stellar model can be
any {\MESA} model with the required mass, and for most cases a ZAMS star at $\Zmet=0.02$
works well. The first two steps in the relaxation of a model fix the composition and
angular momentum profiles. This is done by directly adjusting the
variables for composition and angular momentum of each cell until the
desired values are reached.
Since the entropy is a derived quantity in \MESA, the third step
relaxes the entropy indirectly via the energy equation.
This is achieved by adding a heating term that injects
energy in regions where the entropy is below the target value, and removes
energy in regions where the entropy is above the target value.
This specific heating rate is
\begin{eqnarray}
   \epsilon_{\rm relax}(m) = \left( 1 - \frac{s(m)}{s_{\rm
   target}(m)}\right)\frac{e(m)}{\tau},
\end{eqnarray}
where $e(m)$, $s(m)$ and $s_{\rm target}(m)$ are the specific internal energy,
current entropy, and target entropy respectively at the mass coordinate $m$.
The timescale for the relaxation process is specified by $\tau$.
The value $\tau$ should be chosen to be small enough that 
energy transport is negligible during the pseudo-evolution.
In practice, $\tau$ can be chosen to be orders
of magnitude smaller than the dynamical timescale of the system.

We verified that using the entropy, composition, and angular momentum profiles
of a model computed with {\MESA} as input, the relaxation procedure can
reproduce the original model to within $0.1\%$. An example is provided in the test
suite under the name \texttt{relax\char`_composition\char`_j\char`_entropy}.

We tested these relaxation routines using the outcome of a stellar merger
computed with the \texttt{STARSMASHER}\footnote{The
  \texttt{STARSMASHER} code is open source and freely available at
  \url{https://jalombar.github.io/starsmasher/}} SPH code
\citep{Gaburov+2010, Lombardi+2011}, configured to use the {\MESA} EOS.
Two coeval non-rotating {\MESA} models with
ZAMS masses of $20\,\Msun$ and $15\,\Msun$ are evolved until
the $20\,\Msun$ star reaches $\Xhyd_{\rm c}=0.34$.
These models are then imported into \texttt{STARSMASHER} to simulate a head-on
collision, such that the relative velocity of the two stars at infinity is zero.
We find that $2.18\,\Msun$ of material
is lost from the system due to the collision.

\begin{figure}%[htb!]
\centering
\includegraphics[width=0.5\columnwidth]{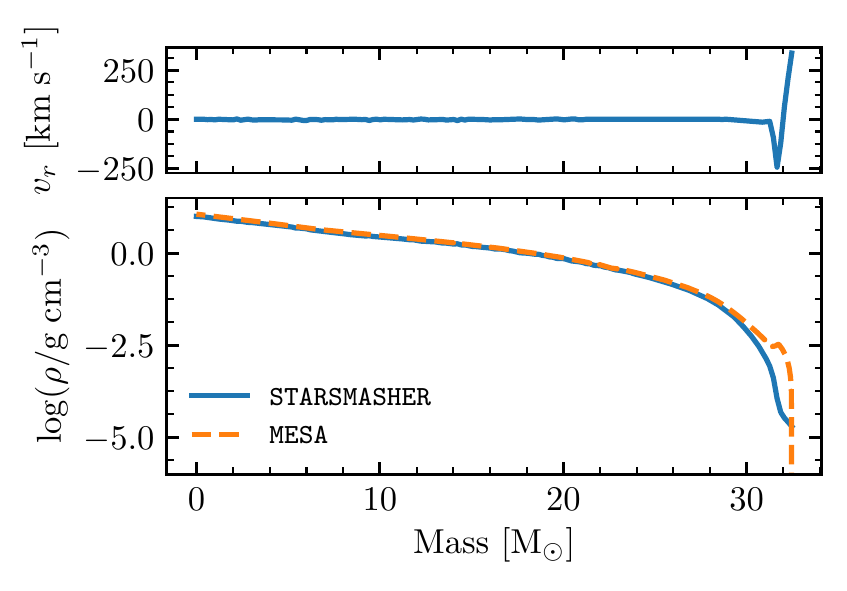}
\caption{Mass-weighted spherical averages of radial velocity and density from the
   \texttt{STARSMASHER} simulation of a head-on collision between non-rotating 
   $20\,\Msun$ and $15\,\Msun$ stars. The dashed line shows the
   resulting density profile of a {\MESA} model relaxed to the entropy and
   composition profile of the simulation.}\label{relax:merger}
\end{figure}

We compute spherical mass-weighted
averages of the composition, $\rho$, and $e$. These
profiles are input into the \MESA\ relaxation process, along with a
zero angular momentum profile
since the model is a head-on collision of non-rotating stars.
Figure~\ref{relax:merger} shows that the relaxed model closely follows
the input smoothed particle hydrodynamic (SPH) merger model in the
central regions, though densities are $\approx 10\%$ larger
throughout the inner $25\,\Msun$.
Density differences of more than an order of magnitude are present in the outer
layers. This is a consequence of these layers not being in hydrostatic
equilibrium in the input SPH simulation. The \MESA\ relaxation process
matches the entropy rather than density profile of the SPH model assuming hydrostatic
equilibrium as discussed above. The relaxed model corresponds to the
final configuration if it contracts adiabatically,
which is a good approximation as velocities in the SPH model are well below the local sound
speed \citep{Pan+2013}.

%%% Local Variables:
%%% mode: latex
%%% TeX-master: "paper"
%%% End:

\section{Element Diffusion Implementation Details}
\label{app.diffusion}

This appendix provides implementation details not contained in Section~\ref{s.diffusion}.
Equations~\eqref{eq:BurgersEnergy}--\eqref{eq:charge_flux} and \eqref{eq:BurgersMomentum} give the full set of diffusion equations
that must be solved to obtain diffusion velocities.
For $S$ total species in the plasma (including electrons),
Equation~\eqref{eq:BurgersMomentum} provides $S-1$ equations (one for
each ion species), Equation~\eqref{eq:BurgersEnergy} provides $S$
equations (one for each species including electrons), and
Equations~\eqref{eq:mass_flux} and~\eqref{eq:charge_flux} each provide
one additional equation, for a total of $2S + 1$ independent equations.
The $2S + 1$ unknowns are $S$ diffusion velocities $w_s$, $S$ heat flow vectors
$r_s$, and the electric field $E$. 

The inputs provided from the \mesa\ model are the number densities $n_s$,
temperature $T$, gradients of each of these quantities $\dif \ln n_s/\dif r$
and $\dif \ln T/\dif r$, species mass in atomic units $A_s$, species mean
charge as an average ionization state $\bar{Z}_s$, and resistance
coefficients $K_{st}$, $z_{st}$, $z_{st}'$, $z_{st}''$
(defined in Equation~86 of \mesathree).
The coefficients are calculated as described in Section~\ref{s.coefficients}.
Together with the mean ionization states, these are the key
pieces of input physics that determine the diffusion of all ions.
Extra acceleration terms $g_{{\rm rad},s}$ for radiative levitation are either set to zero
by default, or calculated as in \cite{Hu11} when the option to include
radiative levitation is enabled.

In the spirit of \cite{Thoul94},
Equations~\eqref{eq:BurgersEnergy}--\eqref{eq:charge_flux} and \eqref{eq:BurgersMomentum} are grouped
into a single matrix equation:
\begin{equation}
\label{eq:fullMatrix}
\beta_{{\rm rad},i} + \alpha_i m_{\rm p} g + \nu_i k_{\rm B} T \frac{\dif
  \ln T}{\dif r} + k_{\rm B} T \sum_j \gamma_{ij} \frac{\dif \ln n_j}{\dif r}
= \sum_{j} \Delta_{ij} W_j.
\end{equation}
The vectors capturing the driving terms are
\begin{equation}
\label{eq:alpha}
\alpha_{i} = 
\begin{cases}
n_i A_i & i = 1,\ldots,S-1,\\
0 & i = S, \ldots, 2S+1,
\end{cases}
\end{equation}
\begin{equation}
\label{eq:nu}
\nu_{i} = 
\begin{cases}
n_i & i = 1,\ldots,S-1,\\
\frac 5 2 n_i & i = S,\ldots,2S-1,\\
0 & i = 2S,2S+1,
\end{cases}
\end{equation}
\begin{equation}
\gamma_{ij} = 
\begin{cases}
n_i \delta_{ij} & i =1,\ldots, S - 1, \\
0 & i =S,\ldots,2S+1,
\end{cases}
\end{equation}
\begin{equation}
\beta_{{\rm rad},i} = 
\begin{cases}
-n_i A_i m_{\rm p} g_{{\rm rad},i} & i = 1,\ldots,S-1, \\
0 & i = S,\ldots,2S+1.
\end{cases}
\end{equation}
The vector containing the unknowns is
\begin{equation}
W_{j} = 
\begin{cases}
w_j & j = 1,\ldots,S, \\
r_j & j = S+1,\ldots,2S, \\
\qe E & j = 2S + 1.
\end{cases}
\end{equation}
For $i = 1,\ldots,S-1$, the right hand side matrix of Equation~\eqref{eq:fullMatrix} is
\begin{equation}
\Delta_{ij} = 
\begin{cases}
- \sum_{l \neq j} K_{il} & j = i,\\
K_{ij} & j = 1,\ldots,S \text{ and } j \neq i,\\
\sum_{l \neq j} K_{il} z_{il} A_l/(A_i + A_l) & j = i + S, \\
-K_{i,j-S}z_{i,j-S} A_i/(A_i + A_{j-S}) & j = S+1,\ldots,2S \text{ and } j \neq i + S, \\
n_i \bar{Z}_i & j = 2S + 1.
\end{cases}
\end{equation}
For $i = S,\ldots,2S-1$, the matrix terms are
\begin{equation}
\Delta_{ij} = 
\begin{cases}
\begin{aligned}
\frac 5 2 \sum_{l \neq j}  K_{i-S,l} z_{i-S,l} \frac{A_l}{A_{i-S} + A_l}
\end{aligned}
& j = i-S,\\
\begin{aligned}
- \frac 5 2  K_{i-S,j}z_{i-S,j} \frac{A_j}{A_{i-S} + A_j} 
\end{aligned}
& j = 1,\ldots,S \text{ and } j \neq i-S,\\
\left.
  \begin{aligned}
    -\sum_{l \neq j-S} &K_{i-S,l}
    \left[\frac{3A_{i-S}^2 + A_l^2z_{i-S,l}'}{(A_{i-S} + A_l)^2}
      + \frac 4 5  \frac{A_{i-S} A_l}{(A_{i-S} + A_l)^2}z_{i-S,l}''
    \right] \\
    -\frac 2 5 &K_{i-S,i-S} z_{i-S,i-S}''
  \end{aligned}
\right\}
& j = i, \\
\begin{aligned}
K_{i-S,j-S} \frac{A_{i-S} A_{j-S}}{(A_{i-S} + A_{j-S})^2}
\left(3 + z_{i-S,j-S}' - \frac 4 5 z_{i-S,j-S}'' \right)
\end{aligned}
& j = S+1,\ldots,2S \text{ and } j \neq i, \\
0 & j = 2S + 1.
\end{cases}
\end{equation}
For $i=2S$,
\begin{equation}
\label{eq:constraint1}
\Delta_{ij} = 
\begin{cases}
n_j A_j & j = 1,\ldots,S, \\
0 & j = S+1,\ldots,2S+1.
\end{cases}
\end{equation}
For $i=2S+1$,
\begin{equation}
\label{eq:constraint2}
\Delta_{ij} = 
\begin{cases}
n_j \bar{Z}_j & j = 1,\ldots,S, \\
0 & j = S+1,\ldots,2S+1.
\end{cases}
\end{equation}
Indices $i=1 \ldots S-1$ capture the $S-1$
equations~\eqref{eq:BurgersMomentum} for the ions. Indices
$i=S \ldots 2S-1$ capture the $S$
equations~\eqref{eq:BurgersEnergy}. Indices $i = 2S,2S+1$ capture
the two constraints in Equations~\eqref{eq:mass_flux}
and~\eqref{eq:charge_flux}. 

For a generic driving term that takes the form of an extra force $f_s$ on ions of
species $s$, a term $-n_s f_s$ appears on the left hand side of
Equation~\eqref{eq:BurgersMomentum}.  This can be accounted
for in the matrix setup by adding another vector $\beta_{f,i}$
to the left hand side of Equation~\eqref{eq:fullMatrix} with the form
\begin{equation}
\beta_{f,i} = 
\begin{cases}
- n_i f_i & i = 1,\ldots,S-1, \\
0 & S,\ldots,2S+1.
\end{cases}
\end{equation}
One such extra driving force that may be explored with \mesa\ in the
future is Coulomb separation in dense matter arising from non-ideal
corrections for the ions \citep{CBA2010,Beznogov13,Diaw16}.

The diffusion velocities are separated into two terms capturing
the distinct effects of gravitational settling and ordinary diffusion in the
tradition of Equation~(11) of \cite{Iben85}:
\begin{equation}
\label{eq:diff_separation}
w_i = w_i^g - \sum_j \sigma_{ij} \frac{\dif \ln C_j}{\dif r},
\end{equation}
where $C_j \equiv n_j/n_{\rm e}$ following the notation of \cite{Thoul94}.
These separate terms are constructed by inverting the matrix
$\Delta_{ij}$ and then solving Equation~\eqref{eq:fullMatrix} for
just one of $\alpha$, $\beta$, $\nu$, and $\gamma_{*,j}$
at a time on the left hand side.
These results can then be linearly combined
to construct $w_i^g$ and $\sigma_{ij}$ such that
the the full sum in Equation~\eqref{eq:diff_separation} gives a
solution that satisfies the complete set represented by
Equation~\eqref{eq:fullMatrix}.

When electrons become degenerate, we drop all $S$
Equations~\eqref{eq:BurgersEnergy} and set the $S$ heat flow vectors
to ${r_s = 0}$.  Equation~\eqref{eq:fullMatrix} then represents a
system of just $S+1$ equations and the vectors and matrices simplify
considerably, dropping all entries for indices $i = S \ldots 2S-1$ or
$j = S+1 \ldots 2S$ in the definitions given in
Equations~\eqref{eq:alpha}--\eqref{eq:constraint2}.  To avoid
discontinuities, we employ a blend that smoothly transitions between
the diffusion velocity
solutions
over a range in ${\eta \equiv \mue / k_{\rm B} T}$, where $\mue$ is
the electron chemical potential.  By default, the blend is centered
around $\eta \approx 1$, with user controls available to adjust the range
of this blending region.

%%% Local Variables:
%%% mode: latex
%%% TeX-master: "paper"
%%% End:

% !TEX root = ./paper.tex
\section{Software Infrastructure}\label{s.software}

Software is an integral enabler of observation, theory, and
computation and a primary modality for realizing the discoveries and
innovations expressed, for example, in the astronomy and astrophysics
decadal surveys 
\citep[e.g.,][]{national-research-council_1991,national_resource_council_2001_aa,  national-research-council_2010_aa}.  
In this appendix we describe new
software stacks at a variety of scales that enhance the 
research and education infrastructure.

\subsection{Not A Number}\label{s.nantrap}

Not a Number (NaN) is a numeric data type representing an undefined or
unrepresentable value \citep[e.g.,][]{goldberg_1991_aa, hauser_1996_aa}. 
Examples include $0/0$ and $\sqrt{-1}$ in real
arithmetic. In the IEEE 754 floating-point standard \citep{ieee_2008_aa} there are two types of
NaNs: quiet (qNaN) and signaling (sNaN).  A qNaN propagates errors
resulting from invalid operations or values without triggering a
floating point exception. An sNaN precipitates an invalid
operation exception whenever an attempt is made to use one as an
arithmetic operand. The IEEE 754 standard requires qNaN as the
default, while an sNaN can be used to support features such as filling
uninitialized memory or other extensions to floating-point arithmetic.

NaN and infinity (INF) setting and testing routines are provided within the \texttt{utils\char`_nan.f90} file.
A consistent set of interfaces allows for initializing scalars/arrays 
to NaN values, and testing for qNaN, sNaN, or INF values. Interface
overloading allows handling of single, double or quad precision
scalars or arrays of rank between 1 and 4.
This module provides four generic interfaces.
Logical function \texttt{is\char`_nan(x,signal)} returns true if \texttt{x} contains NaNs and false otherwise. 
The optional logical argument \texttt{signal} determines whether qNaN,
sNaN or both are tested for. 
Logical function \texttt{is\char`_inf(x)} returns true if \texttt{x} contains INFs and false otherwise.
Logical function \texttt{is\char`_bad(x)} returns true if \texttt{x} contains NaN or INF values and false otherwise.
Routine \texttt{set\char`_nan(x,signal)} sets a scalar or array \texttt{x} to NaN values. 
The optional logical argument \texttt{signal} determines whether a qNaN or sNaN is set.

The library framework of \MESA \ is designed to be interoperable within
other software ecosystems.  For example, these NaN and INF interfaces
are of potential interest to users of \MESA\ or developers of similar
software instruments.

\subsection{\MESAWeb}\label{s.mesaweb}

Stellar evolution software instruments can be complicated to install
and use, especially when the aim is primarily pedagogical (e.g.,
high-school or undergraduate courses).  Motivated by the
community's expressed need for a lower barrier to entry for education,
a web-based interface to \MESA\ was developed, \MESAWeb \ at
\url{http://mesa-web.asu.edu}.  \MESAWeb \ currently allows choices for
the initial mass, metallicity, rotation, mass loss, nuclear reaction network, 
custom nuclear reactions rates, spatial and temporal resolution, and
model output rate.

\MESAWeb \ sends the user an email message when their job has completed
that contains a URL of a zip file to download. The unzipped output directory
contains a \MESA \ history data file holding the time evolution of 57
quantities, as well as a series of \MESA \ profile data files containing information
on 56 quantities in each zone of the stellar model at discrete model
numbers. Also included in the output is an MP4 formatted video
containing a plot dashboard of the abundance profiles, Kippenhahn
diagram, Hertzsprung-Russell Diagram, rotational profile, and temperature, density, and
pressure profiles.

\MESAWeb\ is currently hosted on a 4-core server at Arizona State
University and allows jobs to run on a single core for 4 hours of
walltime or until the model reaches iron core collapse. Launched in
June 2015, \MESAWeb \ has presently served more than 3000 models to
over 600 different users at over 40 academic institutions. Efforts to
expand \MESAWeb's capabilities include porting the service to a host
with enhanced compute resources, simulating core-collapse
supernova explosions (see \mesathree) and light curves (see
Section \ref{s.sneiip}), and binary star evolution (see \mesathree).

\subsection{\MESA-Docker}\label{s.mesadocker}

Docker is a software technology designed to deploy and run
applications by using ``containers''. Containers provide much of the
virtualization power of traditional virtual machines while requiring
far less resource overhead. This allows efficient packaging of
an entire operating environment, with all of the necessary libraries
and other dependencies for a large software tool such as \MESA. 
The \MESA-Docker\ package \citep{MESA-Docker} provides a solution that simplifies the 
requirements for locally running a full \MESA\ installation with all 
capabilities available, with only minor overhead associated with 
running in a container. \MESA-Docker will be useful for 
new users, students with educational projects, and Windows operating system users.

\subsection{\texttt{pyMESA}}\label{s.pymesa}

\texttt{pyMESA} \citep{pymesa} allows embedding of \MESA\ modules into
\texttt{Python} projects.  \texttt{pyMESA} currently supports using
the equation of state (\eos), nuclear
reaction (\rates), neutrino (\neu), atmosphere (\atm), and opacity
(\kap) packages.  This software infrastructure will be useful for users who want
to use parts of \MESA\ in their own \texttt{Python} software projects.
As an example of these capabilities,
Figures~\ref{fig:eos_energy} and~\ref{fig:eos_derivs} were
produced using the \texttt{pyMESA} \texttt{eos} interface to make
direct calls to the \MESA\ EOS routines.

\subsection{\MESAstar\  Model Optimization}\label{s.mesastarsimplex}

The \MESAstar\ test suite contains a sample case that shows how to use the simplex
optimization algorithm \citep{NelderMead1965} to find stellar models
that minimize a specified $\chi^2$ by automatically adjusting a variety of
control parameters.\footnote{This is the same simplex algorithm that
  is used for finding matches in asteroseismology applications using
  \MESA\ (see \mesathree, Section 3).  The code reported here is a
  simplified subset of that tool and is now easier to use and adapt to
  new problems.}  The $\chi^2$ to be minimized can contain both
pre-supplied and user-defined terms.  Pre-supplied terms include
\Teff, $L$, $R$, $g$, surface $\Zmet$, surface $\Yhel$, and age.  An
easy-to-use framework allows the user to define other terms to include
in the $\chi^2$.  Control parameters include $M$, $\Zmet/\Xhyd$,
$\Yhel$, $\alpha_{\rm MLT}$, and $f_{\rm ov}$.  Other stellar
evolution parameters can be easily added from the extensive set of
controls in \MESA.  We provide a \MESA\ test suite case using this new
capability to calibrate a solar model.  This can serve as a template
for users wishing to use this method to search for models that match
the observed properties of specific stars.

\subsection{\rm{\url{http://mesastar.org}}}\label{s.mesastar}

Reproducibility is bedrock to scientific research.  Provenance, as the
term relates to software instruments \citep{buneman_2001_aa,carata_2014_aa}, 
is the ability to record the full history of a result.
Scientific research is generally held to be of good provenance when it
is documented in detail sufficient to allow reproducibility.
The \MESA \ project facilitates provenance by the research community in four ways. 
One, by curating public releases of the source code, makefiles, test suite, and how the source code was
compiled $-$ GNU compilers are redistributed in the \MESA \ Software
Development Kit (see \mesatwo) $-$ at \url{http://mesa.sourceforge.net}.
Two, by providing bit-for-bit consistency for all results across all the supported platforms (see \mesathree).
Three, by supporting a user mailing-list to openly share knowledge  (see the Manifesto in \mesaone).
Currently, over 12,000 messages are archived and searchable.
Four, by hosting a web-portal at \url{http://mesastar.org}
to share \MESA-oriented software contributions and 
reposit the \MESA \ files (\texttt{inlist}, \texttt{run\char`_star\char`_extra.f}, etc) 
that specify all the ingredients needed to reproduce a scientific result. 
Currently,  \url{http://mesastar.org} offers over 120 \MESA-oriented 
software contributions and inlist repositories.

%%% Local Variables:
%%% mode: latex
%%% TeX-master: "paper"
%%% End:

\bibliographystyle{aasjournal}
\bibliography{paper}

\listofchanges

\end{document}